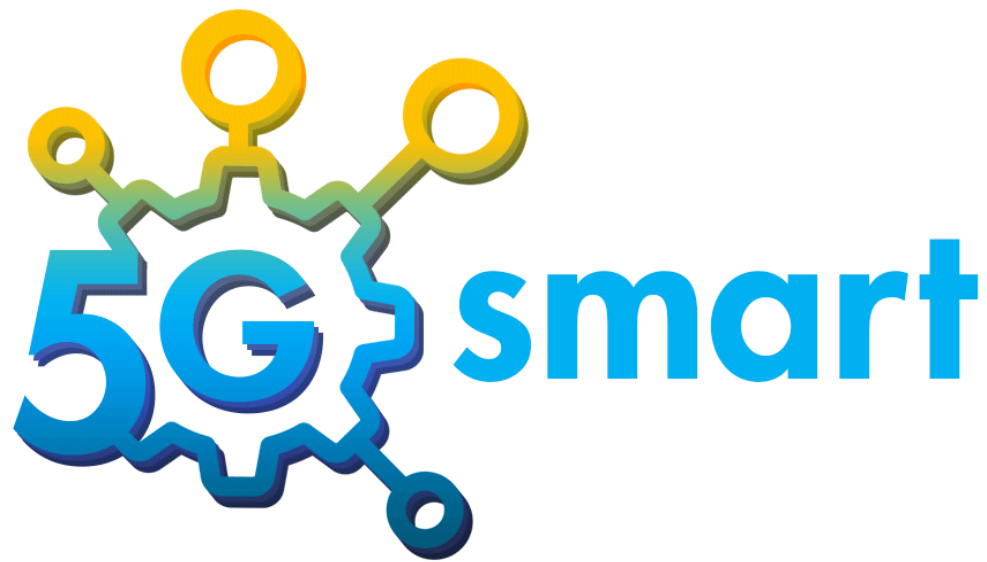

# D1.5

EVALUATION OF RADIO NETWORK DEPLOYMENT OPTIONS

The 5G-SMART project has received funding from the European Union's Horizon 2020 research and innovation programme under grant agreement no 857008.

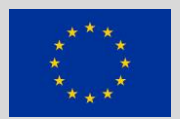

# Evaluation of radio network deployment options

| | |
|---|---|
| Grant agreement number: | 857008 |
| Project title: | 5G Smart Manufacturing |
| Project acronym: | 5G-SMART |
| Project website: | www.5gsmart.eu |
| Programme: | H2020-ICT-2018-3 |
| Deliverable type: | Report |
| Deliverable reference number: | D5 |
| Contributing workpackages: | WP1 |
| Dissemination level: | Public |
| Due date: | 2021-11-01 |
| Actual submission date: | 2021-12-21 |
| Responsible organization: | UPV |
| Editor(s): | Fedor Chernogorov, Saúl Inca |
| Version number: | V1.0 |
| Status: | Final |
| Short abstract: | This deliverable results from the work on the radio network performance analysis of the identified use cases and deployment options. Covered topics include latency reduction and mobility features of the 5G NR itself, as well as detailed analysis of the radio network KPIs, such as latency, reliability, throughput, spectral efficiency and capacity. Corresponding trade-offs for the identified deployment options and industrial use cases are quantified with an extensive set of technical results. Also, this deliverable is looking into co-channel coexistence performance analyzed through a real-life measurement campaign and considers performance optimization in presence of a special micro-exclusion zone within a factory. |
| Keywords: | Radio network deployment, radio network performance, spectrum, trade-off analysis, mobility, latency reduction, coexistence. |
| Contributor(s): | Fedor Chernogorov, Antti Ratilainen (ERI-FI) |
| | Joachim Sachs, Leefke Grosjean, Yanpeng Yang, Alexey Shapin (ERI-SE) |
| | Junaid Ansari, Jordi Biosca Caro (ERI-DE) |
| | Meriem Mhedhbi (Orange) |
| | Saúl Inca, Concepción García-Pardo, Jose F. Monserrat (UPV) |

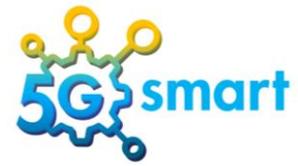

## Disclaimer

This work has been performed in the framework of the H2020 project 5G-SMART co-funded by the EU. This information reflects the consortium's view, but the consortium is not liable for any use that may be made of any of the information contained therein.

This deliverable has been submitted to the EU commission, but it has not been reviewed and it has not been accepted by the EU commission yet.

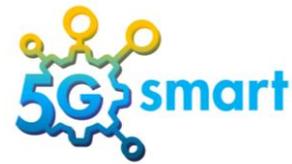

## Executive summary

This report makes an analysis of capacity and required spectrum for smart manufacturing radio network deployment options suggested in [5GS20-D14]. Evaluations are performed through performance evaluations in link-level and system-level simulations. The Key Performance Indicator (KPI) framework from D1.1 (latency, reliability, coverage, capacity, throughput, etc.) will be used as an input [5GS20-D11]. Other radio network characteristics, configurations, and models, such as the duplex mode (TDD/FDD), air interface characteristics (sub-carrier spacing, radio frame structure, repetitions, etc.), antenna models/configurations, traffic models and propagation models, are considered in alignment with the radio network deployment options developed in [5GS20-D14].

The analysis presented in this report suggests that the various radio network deployment options and features can significantly impact the performance of a non-public factory network supporting ultra-reliable low-latency communication (URLLC) services. The applied Time Division Duplex (TDD) downlink-uplink pattern defines a lower bound for the achievable latency and has a clear impact on the maximum system capacity. Packet size also has impact on the performance of the system, which marks that a proper selection of the payload is of interest for the improvement of system performance. Finally, for latency critical URLLC services the downlink is more sensitive to interference increase at high load due to inter-cell interference, as compared to the uplink.

A complete trade-off analysis has been made for different bands, bandwidths, density of users and allocation of resources in TDD mode. The system limits of capacity can come from the limitation of resources and in some cases due to interference or propagation conditions. For each of the different configurations a capacity analysis for uplink and downlink is provided.

In addition, we have empirically investigated the coexistence between a wide-area outdoor network and an indoor non-public network (NPN) both operating in the same channel. Under some conditions the transmission in the outdoor network can create interference to the indoor NPN and thereby increase the transmission latency of the indoor NPN. Such interference can in particular appear in a (practically uncommon) situation, when an outdoor User Equipment (UE) is located right outside an unshielded window of a factory. This can lead to an increased NPN latency, which may be very small for the median values but can impact the 99th and 99.9th percentile of the latency distribution, especially if the two networks use un-synchronized TDD configurations with strong cross-link interference over very short distance. We show that such cross-link interference can be avoided by an indoor TDD configuration that avoids Downlink (DL) transmission slots during Uplink (UL) transmission slots of the outdoor network, while still allowing for a separate TDD configuration for the NPN. When the outdoor UE is located further away from the factory wall, the interference on the indoor NPN becomes negligible independent from the used TDD pattern.

Finally, a realistic performance evaluation in the Bosch factory digital twin has shown the feasibility of defining micro-exclusion zones via the usage of a proper planning plus the use of selective beamforming. Results show that it is possible to isolate a certain area while maintaining on average good levels of service. However, URLLC type of service cannot be deployed together with micro-exclusion zone solution due to significant signal quality degradation. Careful planning of the radio network deployments and agreeing on appropriate emission limits, should be applied to control the level of the interference.

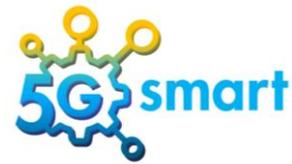

## Contents

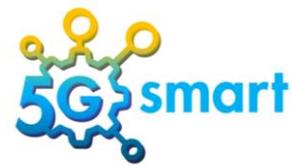

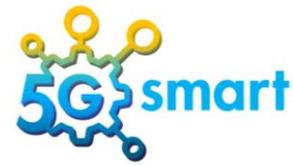

# 1        Introduction

In a first step, the project identified the main use cases and their design objectives. From there, the evaluation work consisted of setting the performance, designing the optimisation of the system and, ultimately in this report, assessing the extent to which the various parameters affect the performance of the system taking into account that some requirements push in opposite directions. The boundaries and interactions between latency, reliability and capacity are the subject of this report, which builds on previous project work to further characterise industrial 5G NR system.

## 1.1       Objective of the report

The objective of this deliverable is to evaluate the radio network deployment options identified in the project, density of the radio site/antenna placement, duplexing mode, TDD pattern, bandwidth, amount of available spectrum, quality characteristics of different industrial services as well as properties of the services themselves, e.g., packet size and bitrate per user.

An analysis of capacity and required spectrum for these radio network deployment options is carried out by performance evaluations through system-level simulations. The KPI framework defined in the project (latency, reliability, coverage, capacity, throughput, etc.) is used as an input. Other radio network characteristics, configurations and models, such as the duplex mode (TDD/FDD), air interface characteristics (sub-carrier spacing, radio frame structure, repetitions, etc.), antenna models/configurations, traffic models and propagation models, will be considered in alignment with the radio network deployment options developed in this activity.

The trade-off between decreasing latency, improving reliability and reducing spectral efficiency/control signaling overhead will also be addressed.

This report also presents the results of a large empirical measurement campaign which investigates the co-channel coexistence between an outdoor and an indoor 5G network that use the same channel. The evaluation was conducted in the 5G Industry Campus Europe in Aachen in the 5G N78 TDD band in a spectrum allocation between 3.7-3.8 GHz.

Finally, this report provides some relevant conclusions on the planning impact for the definition of indoor micro-exclusion zones in a real scenario, as the Bosch factory in Reutlingen. Results show promising capacities of 5G to allocate beams in such a way that signal is minimized in an interest area.

The overall goal of this report is to complete the radio network deployment analysis initiated in [5GS20-D14] and to conclude the detailed analysis of the radio latency and reliability as well as the trade-off analysis between the low latency, high-reliability and spectral/control signaling efficiency for the identified deployment options.

## 1.2       Relation to other documents

The radio network deployment and coexistence analysis made in 5G-SMART Deliverable 1.4 [5GS20-D14] will serve as basis for the radio network evaluations that is carried out also in this report. Moreover, this report is closely related to the other work performed within 5G-SMART Work Package 1 (WP1) - "Use Cases, Business Models and Network Design". This includes the 5G-SMART Deliverable 1.1 [5GS20-D11] discussing the forward-looking smart manufacturing use cases, requirements and key performance indicators, and the work on the common 5G terminology [5GS20-Term].

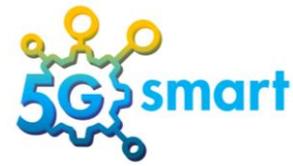

This report is also related to the work performed within 5G-SMART WP5 ("5G Optimization and Design for Manufacturing") when it comes to the different architecture options for the non-public networks and the descriptions of the different operation models (i.e., roles and stakeholders). In particular, some of the definitions presented in 5G-SMART Deliverable 5.2 [5GS20-D52] discussing 5G network architecture options and assessments are reused within this report. However, any lower-level details regarding the network architecture and the operation model are omitted from this report and can be found in [5GS20-D52].

## 1.3    Structure of the report

The rest of the report is organized as follows. Chapter 2 provides a summary of the use cases and scenarios defined in 5G-SMART that are of use in the deliverable. The relation to the other similar documents from 3GPP and 5G-ACIA is also highlighted.

Chapter 3 highlights some functions which allow 5G NR standard to support low latency, considering air-interface latency reduction features and mobility enhancements from layer 1 to layer 3 of the NR protocol stack. Also means to maintain reliability in mobility situations are discussed.

Chapter 4 discusses the detailed performance trade-off analysis for various industrial URLLC deployments and services. The impact on latency, reliability, throughput, system capacity and spectral efficiency is investigated for various frequency bands, duplexing modes, numerologies, network configurations, Quality of Service (QoS) targets and network loads. Several industrial use cases are also considered.

Then, Chapter 5 on the one hand focuses on the co-channel coexistence behavior, highlighting performance evaluated with empirical measurement campaign of an indoor non-public network in a factory and corresponding impact due to a coexisting outdoor (public) network operating in the same spectrum. On the other hand, a possibility to have a micro-exclusion zone with an optimum planning and a subsequent use of digital precoders for the isolation of a certain area within the indoor industrial facility are presented in the second part of Chapter 5. Finally, conclusions are drawn in Chapter 6.

## 2      Industrial use cases

In order to be able to evaluate radio network deployment options and their performance and capabilities, certain industrial use cases need to be considered. Within 5G-SMART several use cases have been defined [5GS20-D11]; some of them are being implemented and validated in the 5G-SMART trial activities. 3GPP has defined industrial use cases and their requirements in [TS22.104], which has been largely influenced by the work in 5G-ACIA. High-level descriptions of use cases by 5G-ACIA can be found in e.g. [5GA19-1] [5GA19-2] [5GA20].

### 2.1      QoS KPIs and translations between them

In this deliverable we follow the Quality of Service (QoS) KPI definitions in timeliness, such as latency, jitter, etc., and dependability: Communication Service Availability (CSA), survival time, network reliability, communication service reliability (CSR) [TS22.104] [TS22.261] [5GS20-Term] [5GS20-D11].

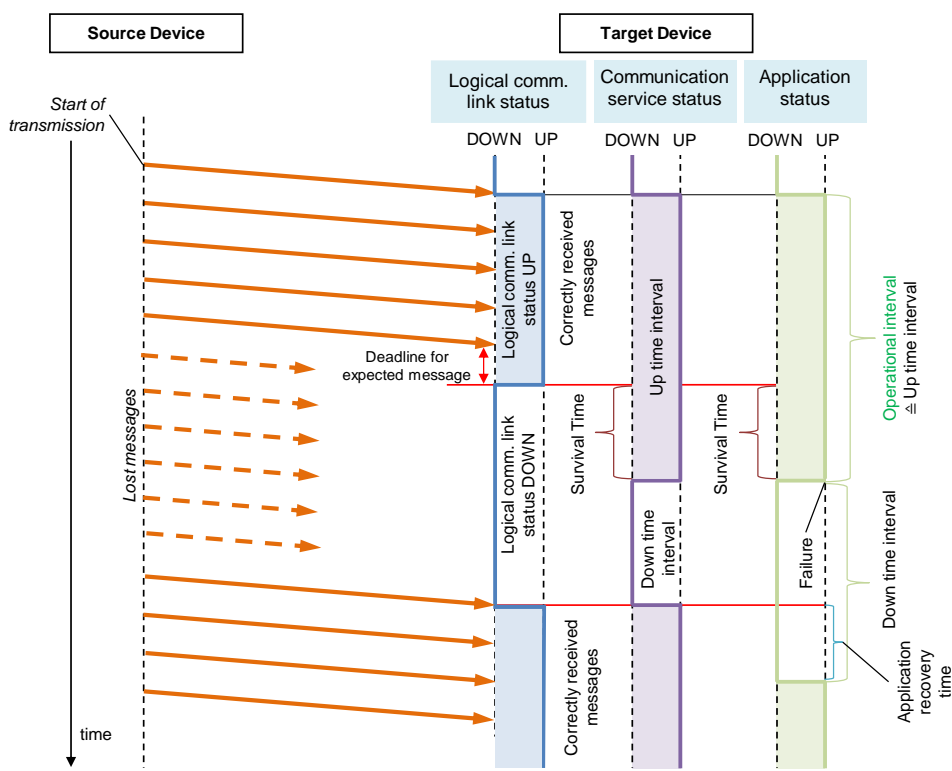

Figure 1. Relation between logical communication link, communication service and application statuses (example with lost messages). Source [TS22.104], Figure C3-1.

- The **survival time** indicates the maximum time period the communication service may not meet the application's requirements before there is a failure on the application layer, such that the communication service is deemed to be in an unavailable state. Such a situation occurs when the communication with the network is lost or the network performance is degraded, and the application stops as a result and an alarm is raised.
- In the context of network layer packet transmissions, (**network) reliability** is the percentage value of the amount of sent network layer packets successfully delivered to a given system entity within the time constraint required by the targeted service, divided by the total number of sent network layer packets [TS 22.261]. In order to well differentiate reliability from communication service reliability (defined below), we will use the term **network packet transmission reliability**.

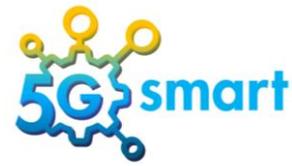

- The **communication service availability** relates to the ability to allow correct operation of the application. It is defined as the "percentage value of the amount of time the end-to-end communication service is delivered according to an agreed QoS, divided by the amount of time the system is expected to deliver the end-to-end service according to the specification in a specific area" [TS 22.104]. The service is unavailable if the messages received at the target are impaired and/or untimely (e.g., latency > stipulated maximum), resulting in survival time being exceeded.

- The **communication service reliability** relates to the ability to continuously operate as required by the application, without failure, for a given time interval and under given conditions (e.g., mode of operation, stress levels, and environment). It can be quantified using metrics such as **mean time between failures** (MTBF) or the probability of no failure within a specified period of time. MTBF is the mean value of how long the communication service is available before it becomes unavailable. For instance, a mean time between failures of one month indicates that a communication service runs without any failure for one month on the average before a failure makes the communication service unavailable. Note that the network transmission failures shorter than the survival time remain unnoticed by the application. This KPI is an end-to-end reliability metric comprising the reliability of several sub-segments, such as the equipment (including hardware and software) reliability as well as the network packet transmission reliability.

Even though mean time between failures (MTBF) is suggested as a measure of communication service reliability, this metric may be misleading in the context of telecommunications, especially when the target reliability values are expressed in long periods of time, e.g., years. MTBF is not a conventional measure of reliability of communication services, because state of each communication link is an individual statistical variable defined by the conditions of each particular link. Therefore, we consider the following way to quantify reliable communication for an industrial service. We perform a translation of the industrial communication service requirements to radio-access network requirements on the basis of CSA, cycle time and survival time, following the methodology described in [TS22.104]. According to this methodology, for example, if the required communication service availability is 99.9999% - 99.999999% and a survival time equivalent to one cycle time, the corresponding requirement of network reliability for message transmission is approx. 99.9% - 99.99% (see [TS22.104], section 5.1).

## 2.2    Considered industrial use cases

The industrial scenarios evaluated in this report focus on use cases that are defined in 5G-SMART and in 3GPP and which are listed in Table 1. Use cases marked with "22.104" are described in [TS22.104]. The use cases from 5G-SMART cover UC1, UC4 and UC7 as described in [5GS20-D11]. Those 5G-SMART use cases are also implemented in the 5G-SMART trials and are further described in [5GS20-D21] [5GS20-D32] [5GS21-D33] [5GS21-D43]. The selected use cases comprise a range of industrial use cases, from mobile robotics, different types of closed-loop control, real-time monitoring of workpieces and machining processes. One commonality is that all of these use cases are in need of time-critical communication, which requires guaranteed performance with bounded low latency according to the 5G URLLC capabilities. The use cases cover a range of service requirements in terms of message sizes, cycle times of the control loops, required data rates, latency bounds and the needed communication service availability (CSA).

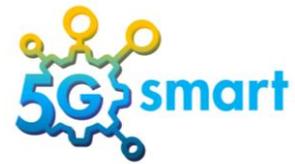

Table 1 Industrial use cases and their requirements.

| Use case | Message size [byte] | cycle time [ms] | rate [Mb/s] | Communication Service Availability (CSA) [# of nines] | survival time [cycle times] | Latency bound [ms] | Network reliability [%] |
|---|---|---|---|---|---|---|---|
| 3GPP URLLC target - modified 22.104 motion control (2) | 32 | 1 | 0.256 | 5 | 0 | 1 | 99.999% |
| 22.104 motion control (2) | 40 | 1 | 0.32 | 6 to 8 | 1 | 1 | 99.99% |
| UC1 (robotics motion planning) | 500 | 5 | 0.8 | 4 | 0 | 5 | 99.99% |
| UC4 (process monitoring) | 1024 | 5 | 1.6384 | 5 | 1 | 10 | 99.9% |
| UC7 (controller-to-controller) | 500 | 10 | 0.4 | 3 to 5 | 2 | 10 | 99% |
| 22.104 mobile robots (1) - precise cooperative robotic motion control | 40 to 250 | 1 | | > 6 | 1 | 1 | > 99.9% |
| 22.104 mobile robots (1) - machine control | 250 | 10 | 0.2 | 6 | 1 | 10 | 99.9% |
| 22.104 mobile robots (1) - co-operative driving | 40 to 250 | 10 to 50 | | > 6 | 1 | 10 to 50 | > 99.9% |
| 22.104 (controller-to-controller) (1) | 1000 | 10 | 0.8 | 6 or 8 | 1 | 10 | 99.99% |
| 22.104 (controller-to-controller) (2) | 1000 | 50 | 0.16 | 6 or 8 | 1 | 50 | 99.99% |

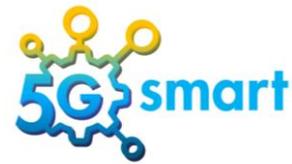

# 3      5G NR capabilities in latency reduction and mobility for URLLC services

## 3.1      Achievable 5G NR air-interface latencies

As can be seen from requirement definitions, evaluations to estimate network performance can focus on latency distributions and checking a likelihood of reaching target latency bound. In order to simplify evaluations even more the worst-case packet arrival delay can be alternatively tested.

5G NR has been standardized to support low latency and high reliability from day one (Release 15). Later in Release 16 and 17 the functionality has been extended to better support extreme latency from 0.5 ms to 1 ms and reliability from 99.999% – 99.9999%, as well as to enhance system capacity.

NR base station scheduler has full control on radio resources which allows to perform careful resource management, admission control, resource slicing for different services, etc. In addition, the scheduler can choose the most efficient strategy to allocate resources to users by knowing the link quality, traffic characteristics and required QoS.

To support stringent requirements there are numerous of tools in the radio interface supported by the NR standard. When latency requirements are extremely short, allocation granularity in time becomes important. For a typical NR mid-band spectrum around 3.5 GHz, the slot duration is 0.5 ms; and for mmWave spectrum, it is even shorter, i.e., 0.125ms. This granularity can be shortened even more by sub-slot granularity scheduling from 0.5ms to ~0.071ms for mid-band, or from 0.125 ms to ~0.02 ms for mmWave.

Furthermore, 5G NR devices can do very fast processing of transmissions taking only ~0.2 ms when encoding or decoding data. There are also numerous tools to minimize alignment delay and other delay components introduced by scheduling mechanisms, such as more frequent downlink control monitoring, pre-scheduling in UL (configured grant) instead of scheduling request-based method, frequent HARQ feedback transmission, etc.

For reliability, 5G NR has more robust Modulation and Coding Scheme (MCS) tables, multi-antenna techniques, reliable control formats and possibility of repeating transmissions to maximize robustness.

### 3.1.1      Performance analysis

We estimate potential 5G NR latencies through evaluation of the worst-case data arrival latency for a user in a system which employs the following assumptions:

- A data packet comes at the most "inconvenient" time for the scheduler, i.e., alignment delay is the longest possible.
- Resource blocking is not taken into account, meaning there is no queuing delay.
- All latency/delay values are one-way and between gNB (gNodeB) and the user equipment (UE) (User Equipment) i.e. radio interface only.
- UE is considered in connected mode and has zero ms handover interruption time, or no mobility at all, i.e. no handovers.

In addition to the assumptions above, there is an agreed methodology to evaluate latency in 5G with a list of delays which needs to be considered TR 37.910 [TR 37.910]. The evaluation is based on the procedure illustrated in Figure 2.

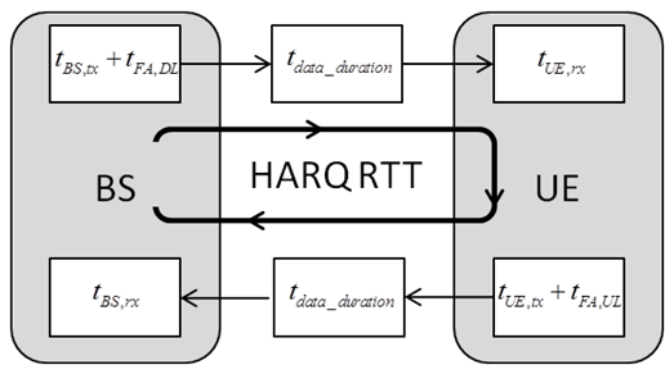

Figure 2 - User plane procedure for evaluation.

The figure represents an exchange of data and feedback between base station and a UE with delay components, where:

- $t_{BS,tx}$ and $t_{BS,rx}$ are BS processing delays
- $t_{FA,DL}$ is DL transmission alignment delay
- $t_{data\_duration}$ is a transmission time in radio interface
- $t_{UE,tx}$ and $t_{UE,rx}$ are UE processing delays
- $t_{FA,UL}$ is UL transmission alignment delay.

As explained in [SKA18], any downlink transmission to a UE starts from resource allocation signaling over the physical downlink control channel (PDCCH). If it is successfully received, the UE will know about the scheduled transmission and may receive data over the physical downlink shared channel (PDSCH). If retransmission is required according to the hybrid automatic repeat request (HARQ) scheme, a signaling of a negative acknowledgement (NACK) on the physical uplink control channel (PUCCH) is involved and PDCCH together with PDSCH transmissions (possibly with new redundancy information) are transmitted and soft-combined with previous data. Otherwise, if no valid control signaling for scheduling data is received, nothing is transmitted on PUCCH (discontinuous transmission - DTX), and the base station upon detecting DTX will retransmit the initial data.

An uplink transmission normally starts from a scheduling request (SR) – a signaling message from the UE to the base station indicating that the UE has data to transmit. Once the scheduler knows about buffered data in the UE, it sends a signaling message in the form of an UL grant through PDCCH to inform the UE about allocated resources. If the UL grant is successfully received, the UE transmits a data packet on the physical uplink shared channel (PUSCH). The transmission of SR can be omitted in UL if "semi-persistent scheduling" (SPS) is activated in a cell, when the UE periodically gets already pre-scheduled resources in UL.

In uplink and downlink, a total one-way user plane latency is calculated according to the formula:

$$T_{UP} = T_1 + n \times T_{HARQ}$$

where $n$ is the number of re-transmissions ($n \geq 0$) and $T_{HARQ} = T_1 + T_2$. Moreover, $T_1$ and $T_2$ calculation is different for uplink and downlink:

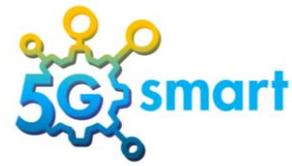

- In downlink, $T_1 = (t_{BS,tx} + t_{FA,DL}) + t_{DL\_duration} + t_{UE,rx}$ and $T_2 = (t_{UE,tx} + t_{FA,UL}) + t_{UL\_duration} + t_{BS,rx}$.
- In uplink, $T_1 = (t_{UE,tx} + t_{FA,UL}) + t_{UL\_duration} + t_{BS,rx}$ and $T_2 = (t_{BS,tx} + t_{FA,DL}) + t_{DL\_duration} + t_{UE,rx}$.

All components are summarized and well explained in Table 5.7.1.1.1-1 and Table 5.7.1.1.2-1 of TR 37.910 [TR37.910]. In general, there are several factors impacting size of each component and, as a result, the total latency:

- Periodicity of downlink control channel (PDCCH) to send resource assignment/grant : the more often the scheduler has an opportunity to send assignment/grant the less assignment/grant is awaiting to be sent in air-interface, thus, the smaller the total delay.
- Duration of a transmission (PUSCH, PDSCH, PDCCH, PUCCH) in air-interface: the shorter transmission, the smaller total delay.
- Signal processing delay in user equipment and base station: for UE there are two processing capabilities defined in standard – baseline capability 1 and more advanced capability 2. UEs supporting processing capability 2 can perform processing approximately two times faster. Base station processing is assumed to be as fast as UE processing.
- How often HARQ feedback can be sent: the more often the less HARQ feedback is awaiting to be sent in air-interface, thus, the smaller the total delay.

In Figure 3, one can get an impression of how fast NR can deliver data over the air, with and without retransmissions and for different configurations. The results are based on typical processing delay values, while actual product performance may differ depending on implementation. Widely-used in 5G networks time-division duplex (TDD) split of 4DL:1UL (DDDSU) slots is assumed for the resources into downlink (DL) and uplink (UL) in these spectrum ranges.

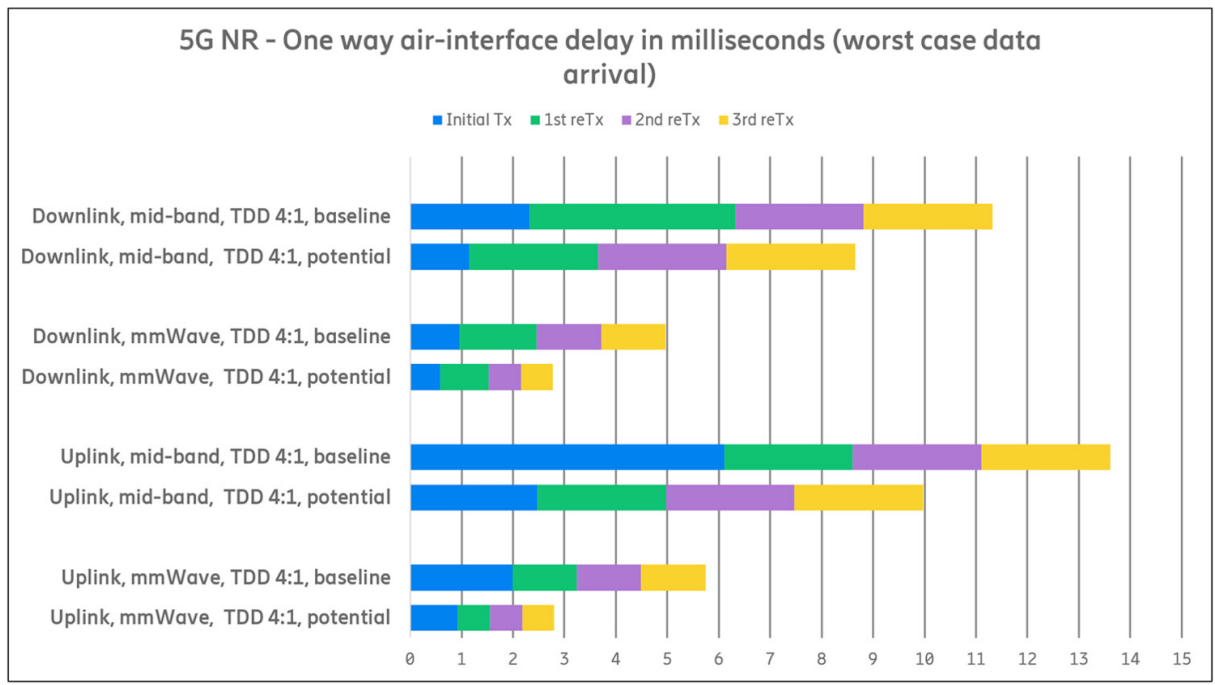

Figure 3– 5G NR one-way air-interface delay

Delays on Figure 3 have been derived with the following assumptions on the 5G NR system:

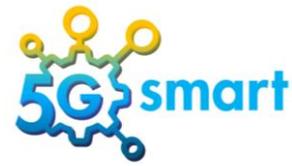

- Baseline:
    - Processing Capability 1;
    - Full slot-based scheduling (14 OFDM-symbols);
    - UL Scheduling request (SR) based scheduling, one SR opportunity per slot;
    - DL one PDCCH per slot and two SR/HARQ-ACK opportunities per slot;
    - Scheduling delay is 0.5 ms.
- Potential:
    - Processing Capability 2 (mid-band only);
    - Sub-slot-based (2 OFDM symbols resulting in 7 sub-slots per slot) scheduling;
    - UL configured grant scheduling;
    - DL seven PDCCHs per slot and seven HARQ-ACK opportunities per slot;
    - Scheduling delay is 0.17ms for mid-band and 0.08ms for mmWave.

Below we list several evaluation studies done to make sure 5G NR is compliant with stringent high-reliability and low latency requirement. The first comprehensive study called "Study on self-evaluation towards IMT-2020 submission" has been done in 3GPP and documented in TR 37.910 [TR37.910]. Study shows that 5G NR can achieve sub millisecond one-way latency at 99.999% reliability. More E2E system performance studies have been held in 3GPP during Release 16 "Enhanced URLLC" study item. The outcome is documented in TR 38.824 [TR 38.824]. Several use cases have been evaluated including industrial use cases, e.g. motion-control. It confirms that 1 ms one-way latency can be achieved at 99.9999% reliability. Finally, 5G-ACIA asked 3GPP to evaluate industry use cases considering E2E application performance metrics, i.e., availability. It has been shown that 5G NR can support more than 200 industry devices on a factory floor 120x50m having 1 ms @99.9999% reliability requirements [RP-210884].

## 3.2 URLLC Mobility

Some URLLC use cases may require mobility while still fulfilling URLLC QoS requirements. The critical moment during mobility is when the current serving cell or transmission/reception point (TRP)[1]. quality becomes bad and the neighboring cell/TRP becomes better than the current one.

During mobility, when the UE needs to change its serving point, cell or TRP, to another one, this may cause critical interruption to an ongoing URLLC service. Thus, it is important to identify scenarios when this interruption is acceptable, when it is not and what can be done to reduce the interruption caused by this mobility.

Different layer 2 and layer 3 mobility mechanisms were considered alongside with different deployment scenarios and use cases. Three deployment scenarios that were considered: Local area deployment, confined wide area deployment and wide area deployment.

- Typical to local area deployment is that the TRPs can be planned and controlled very tightly, and the network is mostly isolated from other networks.
- Wide area deployment consists of generic macro cell deployment over e.g., urban/rural area with no densified areas.

---

[1] TRP is a general term for a network node with an antenna array located at a specific geographical location able to transmit in downlink and receive in uplink user data to serve UEs in the network e.g. macro cell, pico cell, femto cell, relay node etc. A cell can consist of multiple TRPs or each TRP may form its own cell.

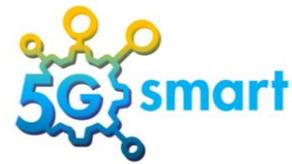

- Confined wide area deployment is a mix of the two above, with macro cell deployment and additional densified areas e.g., over a certain path or hotspot with smaller TRPs.

Many URLLC deployment environments resemble the local area deployment, where the network is dense and can be planned according to the use case, and there are no major reductions in channel quality. This deployment also typically hosts the URLLC use cases with stricter interruption and reliability requirements.

It is worth noting that "interruption-free mobility" does not mean 0 ms user plane interruption for all use cases, as the delay tolerance is use case dependent, i.e. for some use cases we need 0 ms user plane interruption at mobility to meet end user service requirements, but for others 10s of milliseconds can be acceptable without having a negative impact on end user service quality. [TS22.104] lists many examples of industrial use cases where the latency requirement is several 10's to more than 100ms, further including a survival time in a similar order. This includes examples from control-to-control communication, or mobile robots in factory automation, as well as, closed-loop control or process and asset monitoring in process automation.

By studying the 3GPP specifications on mobility, it was noted that the handover interruption time caused by NR Release-15 L3 handover is in the order of 40 ms – 60 ms for both FDD and TDD. Here, NR Release-15 L3 handover refers to a L3 handover without specific mobility optimizations. A majority of this interruption time is caused by the random access procedure, during which the time is spent for message processing either on gNB or UE side and message transmission. Theoretical random access delay during handover according to 3GPP specifications (3GPP TR 36.881 [TR36.881] and 3GPP TS 38.331 [TS38.331] (assuming 20 ms periodicity for Synchronization Signal Block (SSB)) is as follows:

1. UE processing time for handover command: 16 ms
2. Retune and synchronize to target cell: 20 ms
3. Wait for RACH (Random Access Channel) slot: 0 – 10 ms
4. RA (Random Access) procedure: 6 ms

This equals to a minimum of 42 ms until RA is completed. In actual implementation this delay is higher and varies a lot. Simplified depiction of L3 handover procedure signaling can be seen in Figure 4.

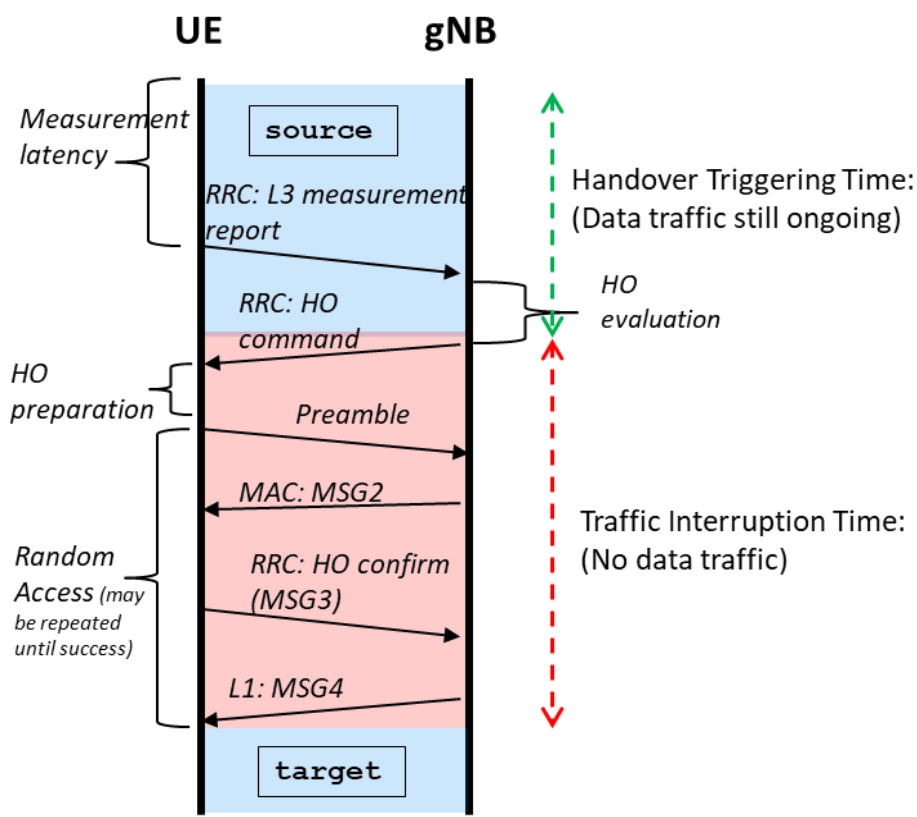

Figure 4 NR Release-15 L3 handover procedure with contention based random access.

For URLLC use cases that require very low latency (lower than the L3 handover interruption time of 40 ms – 60 ms), possibly even down to zero interruption time during mobility, there is a need for improved latency performance during mobility in order to address these use cases.

The mobility mechanisms considered here include layer 2 Multi-TRP framework with intra-cell mobility (L2 M-TRP inter-cell mobility is planned to be standardized in 3GPP Release-18), NR Release-15 L3 handover, L3 DAPS (Dual Active Protocol Stack) handover and L3 conditional handover.

Generally, L3 solutions are bound to have higher interruption times during mobility due to the two main sources that contribute to the interruption times: Random access procedure and RRC (Radio Resource Control) reconfiguration in the target cell.

Some L3 solutions aim to reduce the delay caused by these procedures but can't remove it completely. With L2 intra-cell solutions there is no need for the random access procedure and also RRC reconfiguration is less common, which leads to lower interruption times during mobility.

### 3.2.1    L3 handover solutions (DAPS, CHO)

From the studied L3 handover improvement solutions, Dual Active Protocol Stack (DAPS) has been specified in 3GPP NR Release-16. DAPS [ERI-MOB] is the most prominent L3 solution in terms of reduced interruption time. DAPS applies a make-before-break mechanism, where the connection to the target gNB is first established before releasing the source gNB. This allows DAPS to reach near 0 ms interruption time.

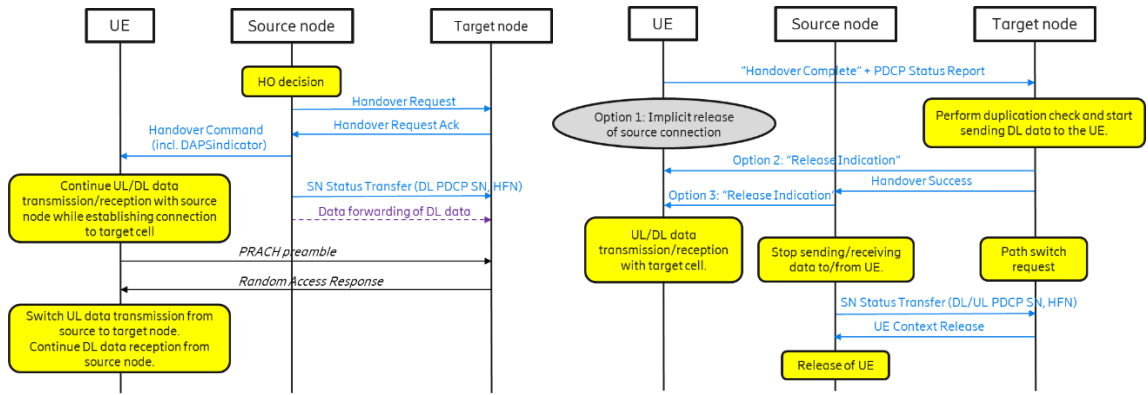

Figure 5  DAPS handover procedure, beginning on left, ending on right side.

While offering the lowest interruption times, DAPS has high complexity and a high implementation cost.

Conditional handover (CHO) [ERI-KMOB] is the other L3 solution that was studied. The aim of conditional handover is to improve the reliability of the handover rather than reducing the interruption time. Thus, the interruption time of conditional handover is the same as for NR Release-15 L3 handover. However, improved handover robustness is useful for various URLLC scenarios, especially when the UE experiences poor radio conditions as failed transmission of the handover command is a common reason for unsuccessful handovers.

The reliability of the conditional handover is increased by preparing the UE in advance for handover by sending the RRC message with handover command earlier, possibly including conditions for when the UE should execute the handover and configuring it with potential target gNBs. Also, the potential target gNBs may be prepared in advance, exchanging the configuration information between the source and target gNBs. Thus, sending the handover command message earlier in better radio conditions can increase the robustness of the handover procedure. Additionally, as the UE doesn't need to send the measurement report in order to trigger the handover command but rather determines handover triggering by its preconfigured conditions, the measurement report can't be lost, further increasing the robustness of the conditional handover.

### 3.2.2    L1/L2 Mobility (multi-TRP)

In local or confined areas, a less complex alternative to DAPS with equally low interruption time is L1/L2 intra-cell mobility, known also as the multi-TRP solution, which offers close to 0 ms interruption time in good channel quality. With L1/L2 intra-cell mobility, there is no need to perform reconfiguration or random access on the UE every time when the UE changes its serving TRP, which greatly improves the performance during mobility. With larger cells however, the UE still may need to be eventually reconfigured with the list of CSI-RS to be measured. This reconfiguration can be done in parallel to the user plane transmissions however, and doesn't discontinue or affect the URLLC features, thus it doesn't have negative impact on the URLLC QoS.

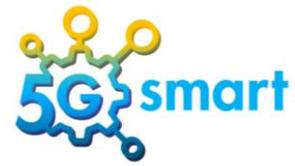

One limitation of L1/L2 intra-cell mobility solution is that when the serving area is larger than what can be covered with a single cell, a cell change mechanism becomes a necessity. One possibility to address this issue is the L1/L2 inter-cell mobility. However, L1/L2 inter-cell mobility is still in early phases and planned to be standardized in 3GPP Release-18, thus it has not yet been defined or specified in detail. The other option for inter-cell mobility is to apply one of the L3 handover solutions discussed above.

# 4 Performance trade-off analysis for various industrial URLLC deployments and services

In smart manufacturing environments, simplicity and flexibility for any reconfiguration of the factory is dramatically increased by replacing cables with dependable wireless connections. Sensors and machines, like robots, may then be operated by the central controller in a server room of the factory. Dynamic response to changes in the environment, and therefore rely on interaction between the wirelessly connected entities with short round-trip times, which for the use cases listed in D1.1, lie in the tens of millisecond to one millisecond range. It is important to reach these latencies with a very high probability (reliability) for the systems to operate properly (see Figure 6 below).

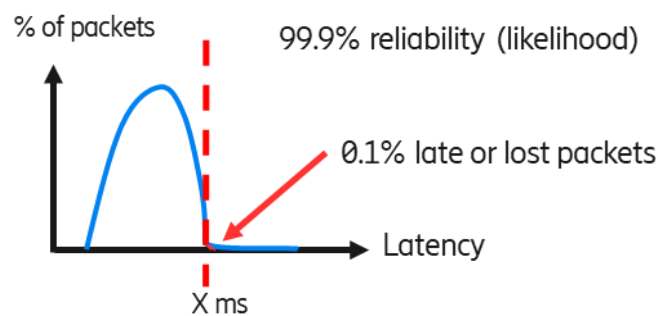

Figure 6 Guaranteeing latency at defined reliability.

A simple way to define reliability (see 3GPP TS 22.261) in the context of network layer packet transmissions, is the percentage value of the amount of sent network layer packets successfully delivered to a given system entity within the time constraint required by the targeted service, divided by the total number of sent network layer packets.

In wireless communication, one of the main reliability bottlenecks is traditionally the radio interface due to stochastic nature of this communication type. Over-the-air transmission reliability is limited by the available bandwidth and signal-to-noise ratio at the receiver, which is impacted by signal pathloss and link stability of the wireless transmission. Allocation of ample spectrum to make the transmission of the same information wider in frequency can keep the latency bound lower to meet the transmission reliability requirements. Alternatively, you could make the transmission longer in time or repeat the same information when needed, as is achieved with hybrid-automated repeat request (HARQ) retransmissions, which would, however, increase the latency. In cellular systems, spectrum resources are scarce and are shared among all connected devices in the cell. Devices require more or less time and spectrum resources to keep the connection up, depending on application requirements and on their current radio conditions. So, to improve reliability, the system needs to be organized so that the right amount of spectrum is used by the right device at the right time. Hence, we study the tradeoff analysis to understand and map the latency, capacity, throughput, spectral efficiency and reliability requirements to application requirements of different industrial services.

## 4.1 Latency, reliability, and throughput trade-off analysis

The planning and deployment of URLLC networks depends mainly on the use case and its required KPIs. If these KPIs are very stringent, then the deployment will require more robust configurations and denser deployments, more radio resources or configurations with more complex antenna type

solutions[5GS20-D14]. The wide range of use cases and the requirements of the industrial sector make it possible to relax some KPIs. For example, in today's market, AGVs are able to operate with latencies higher than 1 ms and/or lower data rates, which are less demanding KPIs than those defined by 3GPP. In this sense, there is a need to evaluate different use cases in the industrial sector, relaxing the KPIs defined by 3GPP in order to find a more realistic network configuration that allows, on the one hand, to optimize radio resources and, on the other hand, to optimize the operator's network deployment to meet real user requirements in terms of reliability, latency and throughput

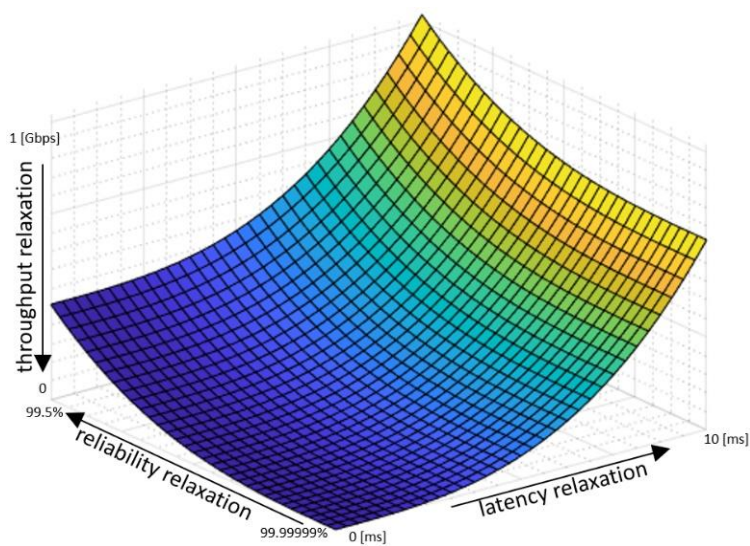

Figure 7 Trade-off representation between latency, reliability, and throughput.

To better understand the impact of relaxing some KPIs in different use cases, several simulations have been performed at system level simulation (SLS). The evaluations have taken into account different types of radio network deployment options. The question we want to answer in this study is whether a relationship can be established between the impact of relaxing the KPIs in terms of users satisfying the requirements and the impact on spectral efficiency.

### 4.1.1    Scenario

The SLS have been conducted in an area of 120 m by 50 m, following TR 38.901 [TR38.901], which consists of a factory network, as shown in Figure 8**Error! Reference source not found.**. By default, the factory network contains twelve ceiling- mounted omnidirectional gNBs that provide URLLC service to the UEs. The gNBs are 8 m height and the inter-site distance is equal to 20 m. All UEs are randomly distributed within the service area with a height of 1.5 meters. The clutter distribution corresponds to an InF-SH [TR38.901], environment with a density of 0.60 and with an average height of 6 meters and an average clutter size of 2 meters. For the simulations, no interference from surrounding macro-cells is considered because it is assumed that the attenuation of the factory walls is high enough to make the interference negligible. To simplify, the results are presented in terms of service areas according to the 5G-ACIA recommendation described in [TS22.104].

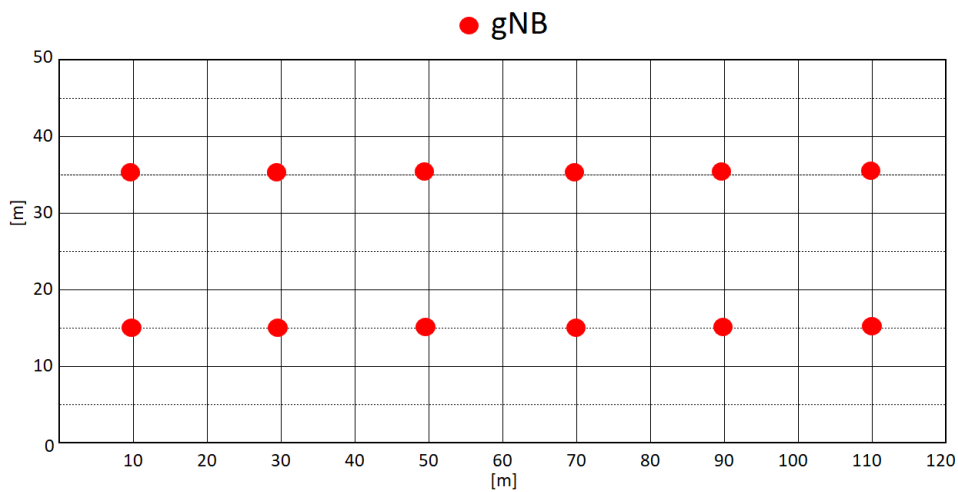

Figure 8. Indoor Factory Scenario Layout with 12 gNBs.

The use cases evaluated with these scenarios are described in Section 2. Mobile robot communications are time-critical, so the network configuration is focused on low latency and high reliability communications.. Additionally, in some cases a high data rate is also required because the robot needs to interact and navigate with cameras. With these criteria in mind, the following subsections describe the simulation methodology.

### 4.1.2   Simulation Parameters

The main SLS parameters are listed in Table 2. Sweeps of some of the parameters listed in the table have been performed to obtain the trade-off between the KPIs. Regarding the spectrum options, two operating bands have been considered. Mid Band: 4 GHz and High-band: 30 GHz with different bandwidths. Up to 100 users per service area have been positioned, which are randomly distributed. In terms of mobility, for the sake of simplicity of the simulation, a linear movement of the UE with a constant speed of 3 km/h has been chosen. The URLLC traffic consists of packets with a variable size. Furthermore, the performance of three TDD patterns, DDDU[2], DUDU and UUUD is also evaluated.

Table 2 Main system-level simulation parameters

| Parameters | Values | Comments |
|---|---|---|
| Factory hall size | 120x50m | TS 38.901 |
| Room height | 10m | |
| BS/TRP antenna height | 8m | InF-DH [TS38.901] |
| TX Antenna type | Omnidirectional | |
| Layout – BS/TRP deployment | Depending on the number of TRPs | See layout options |
| Channel model | InF-DH | TS 38.901 |
| Carrier frequency and simulation bandwidth | TDD<br>4.0 GHz: (20 -100)MHz<br>30 GHz: (100-400) MHz | a sweep of Bandwidth and TDD patterns is performed |

---

[2] In Time Division Duplexing pattern notation "D" stands for downlink NR slot, "U" is correspondingly uplink NR slot, "S" is a special slot used for transition from downlink to uplink transmission direction or vice versa.

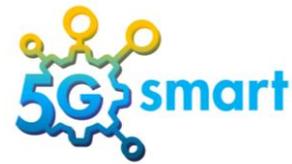

| TDD DL-UL configuration | DDDU, DUDU and UUUD | |
| Number of UEs per service area | Up to 100 per service area | |
| UE distribution | All UEs randomly distributed within the respective service area. | |
| Message size | 48,100,250, 500 and 1000 bytes | |
| DL traffic model | DL traffic arrival with option 1-3 | 5G-ACIA assumption [TS22.104]. |
| UL traffic model | UL traffic is symmetric with DL, and DL-UL traffic arrival time relationship with options 1-2 | 5G-ACIA assumption [TS22.104]. |
| Reliability | 99.99%, 99.9%, 99% | Simulation targets |
| Latency requirements | 1,2,3,5,10 ms | |
| Data rate | Different data rates | |

### 4.1.3    Performance evaluations: machine control for mobile robots

#### 4.1.3.1    Analysis of the impact of the TDD pattern and subcarrier spacing on latency and the percentage of UE satisfying requirements

One of the critical parameters affecting latency is the TDD pattern used. Moreover, as the TDD pattern defines the amount of resources available for the UL and DL link, and limits the data-rate offered to users. Furthermore, there is a direct relationship between latency and reliability. For example, if the latency obtained with the network configuration is lower than the target latency, more retransmissions can be made or more robust encodings can be used, which makes the communication more reliable.

In Figure 9 to Figure 12, the impact of the TDD pattern on the number of users satisfying the requirements for a given reliability is shown for a high- and mid-bands, using different subcarrier spacing (SCS) and a packet size of 48 bytes. It can be seen that for use cases where symmetric latencies for UL and DL are needed, the DUDU pattern will be the best candidate. In addition, it can be observed that if the latency is relaxed maintaining reliability, two options are available: to support more users that satisfy the requirements or to increase the reliability by making more retransmissions. In this sense, Table 3 to Table 6 show the worst case-latencies considering up to 5 retransmissions for 10, 20 and 30 UE/Serving Area (SA) for different TDD patterns and for both mid-band and high-band.

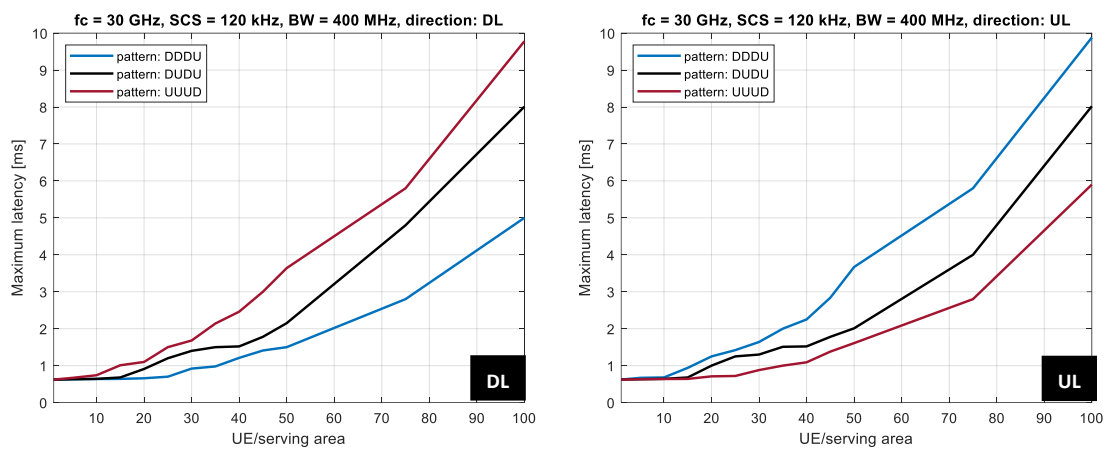

Figure 9 Impact of the TDD pattern on worst-case latency and the percentage of UE satisfying requirements.
**SCS = 120 kHz** (high-band)

Table 3 Downlink and uplink worst-case latencies assuming different TDD patterns, 1-5 transmission attempts, and a **SCS= 120 kHz.**

| | Tx attempts | DDDU | | DUDU | | UUUD | |
|---|---|---|---|---|---|---|---|
| | | DL [ms] | UL [ms] | DL [ms] | UL [ms] | DL [ms] | UL [ms] |
| **10 UE/SA** | 1st | 0.64 | 0.84 | 0.67 | 0.67 | 0.84 | 0.62 |
| | 2nd | 1.38 | 1.88 | 1.42 | 1.46 | 1.91 | 1.39 |
| | 3rd | 2.17 | 2.96 | 2.20 | 2.23 | 3.08 | 2.26 |
| | 4th | 2.95 | 4.05 | 2.99 | 3.00 | 4.10 | 2.99 |
| | 5th | 3.86 | 5.26 | 3.89 | 3.88 | 5.31 | 3.89 |
| **20 UE/SA** | 1st | 0.75 | 1.21 | 0.95 | 0.99 | 1.14 | 0.77 |
| | 2nd | 1.53 | 2.29 | 1.73 | 1.75 | 2.14 | 1.47 |
| | 3rd | 2.42 | 3.49 | 2.63 | 2.57 | 3.35 | 2.38 |
| | 4th | 3.14 | 4.50 | 3.34 | 3.37 | 4.45 | 3.18 |
| | 5th | 4.09 | 5.75 | 4.29 | 4.28 | 5.72 | 4.15 |
| **30 UE/SA** | 1st | 0.86 | 1.62 | 1.38 | 1.31 | 1.80 | 0.90 |
| | 2nd | 1.73 | 2.79 | 2.25 | 2.19 | 2.95 | 1.74 |
| | 3rd | 2.56 | 3.91 | 3.07 | 3.05 | 4.07 | 2.56 |
| | 4th | 3.57 | 5.22 | 4.08 | 3.97 | 5.29 | 3.48 |
| | 5th | 4.56 | 6.52 | 5.08 | 4.99 | 6.69 | 4.59 |

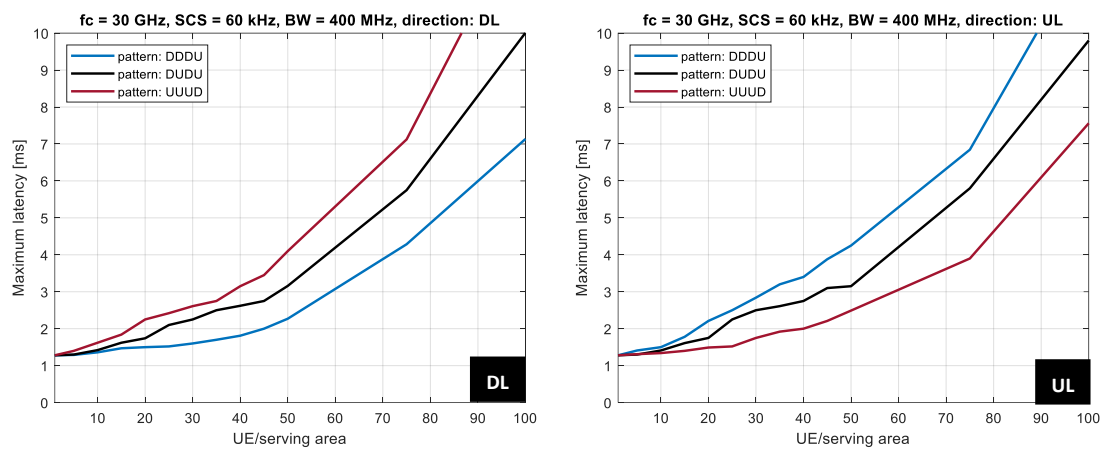

Figure 10. Impact of the TDD pattern on worst-case latency and the percentage of UE satisfying requirements. **SCS = 60 kHz** (high-band).

Table 4 Downlink and uplink worst-case latencies assuming different TDD patterns, 1-5 transmission attempts, and a **SCS: 60 kHz** (high-band).

| | Tx attempts | DDDU | | DUDU | | UUUD | |
|---|---|---|---|---|---|---|---|
| | | DL [ms] | UL [ms] | DL [ms] | UL [ms] | DL [ms] | UL [ms] |
| **10 UE/SA** | 1st | 1.27 | 1.47 | 1.38 | 1.35 | 1.64 | 1.22 |
| | 2nd | 2.25 | 2.73 | 2.35 | 2.32 | 3.01 | 2.29 |
| | 3rd | 3.21 | 4.05 | 3.36 | 3.33 | 4.27 | 3.26 |
| | 4th | 4.26 | 5.37 | 4.39 | 4.36 | 5.59 | 4.27 |
| | 5th | 5.29 | 6.69 | 5.41 | 5.38 | 6.86 | 5.25 |
| **20 UE/SA** | 1st | 1.47 | 2.21 | 1.84 | 1.80 | 2.15 | 1.46 |
| | 2nd | 2.56 | 3.61 | 2.94 | 2.86 | 3.56 | 2.56 |
| | 3rd | 3.60 | 4.94 | 3.97 | 3.91 | 4.88 | 3.58 |
| | 4th | 4.73 | 6.38 | 5.11 | 5.03 | 6.23 | 4.64 |
| | 5th | 5.89 | 7.84 | 6.27 | 6.11 | 7.61 | 5.72 |
| **30 UE/SA** | 1st | 1.64 | 2.91 | 2.31 | 2.51 | 2.62 | 1.74 |
| | 2nd | 2.78 | 4.35 | 3.45 | 3.64 | 4.09 | 2.91 |
| | 3rd | 4.08 | 5.95 | 4.75 | 4.84 | 5.61 | 4.12 |
| | 4th | 5.30 | 7.47 | 5.97 | 5.98 | 7.10 | 5.31 |
| | 5th | 6.53 | 9.00 | 7.20 | 7.15 | 8.67 | 6.59 |

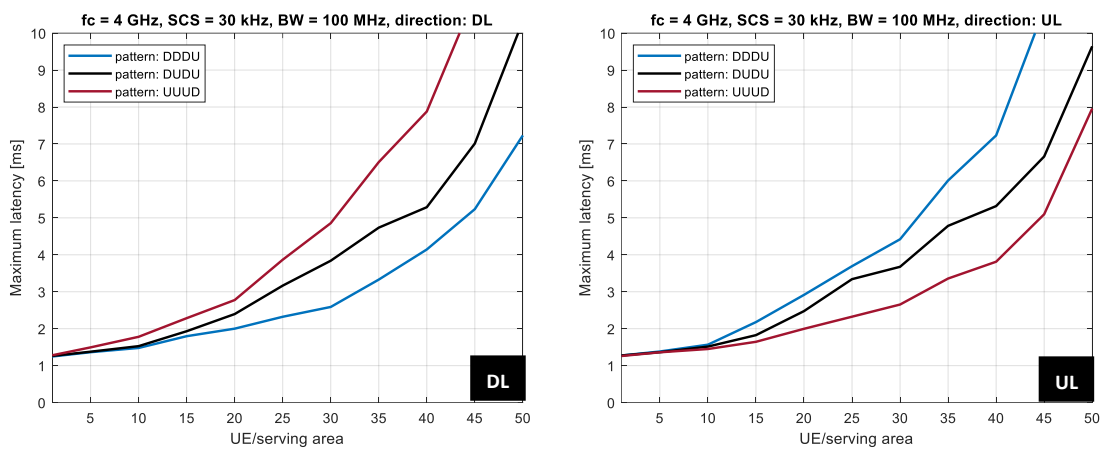

Figure 11 Impact of the TDD pattern on the worst-case latency and the percentage of UE satisfying requirements. **SCS = 30 KHz** (mid-band).

Table 5 Downlink and uplink worst-case latencies assuming different TDD patterns, 1-5 transmission attempts, and **SCS: 30 kHz** (mid-band).

| | Tx attempts | DDDU | | DUDU | | UUUD | |
|---|---|---|---|---|---|---|---|
| | | DL [ms] | UL [ms] | DL [ms] | UL [ms] | DL [ms] | UL [ms] |
| **10 UE/SA** | 1st | 1.48 | 1.57 | 1.53 | 1.52 | 1.79 | 1.45 |
| | 2nd | 2.81 | 3.37 | 2.78 | 2.77 | 3.57 | 2.89 |
| | 3rd | 3.84 | 4.93 | 3.90 | 3.89 | 5.16 | 3.88 |
| | 4th | 5.06 | 6.62 | 5.13 | 5.12 | 6.93 | 5.11 |
| | 5th | 6.30 | 8.29 | 6.36 | 6.34 | 8.71 | 6.32 |
| **20 UE/SA** | 1st | 2.00 | 2.91 | 2.40 | 2.46 | 2.78 | 2.00 |
| | 2nd | 3.42 | 4.85 | 3.82 | 3.99 | 4.74 | 3.40 |
| | 3rd | 4.69 | 6.43 | 5.09 | 5.11 | 6.50 | 4.67 |
| | 4th | 5.93 | 8.24 | 6.33 | 6.42 | 8.45 | 6.05 |
| | 5th | 7.12 | 10.00 | 7.51 | 7.69 | 10.23 | 7.34 |
| **30 UE/SA** | 1st | 2.59 | 4.42 | 3.84 | 3.68 | 4.86 | 2.65 |
| | 2nd | 4.00 | 6.24 | 5.25 | 5.01 | 6.66 | 3.94 |
| | 3rd | 5.42 | 8.23 | 6.67 | 6.42 | 8.40 | 5.26 |
| | 4th | 6.83 | 10.10 | 8.08 | 7.82 | 10.42 | 6.76 |
| | 5th | 8.15 | 11.88 | 9.40 | 9.19 | 12.39 | 8.23 |

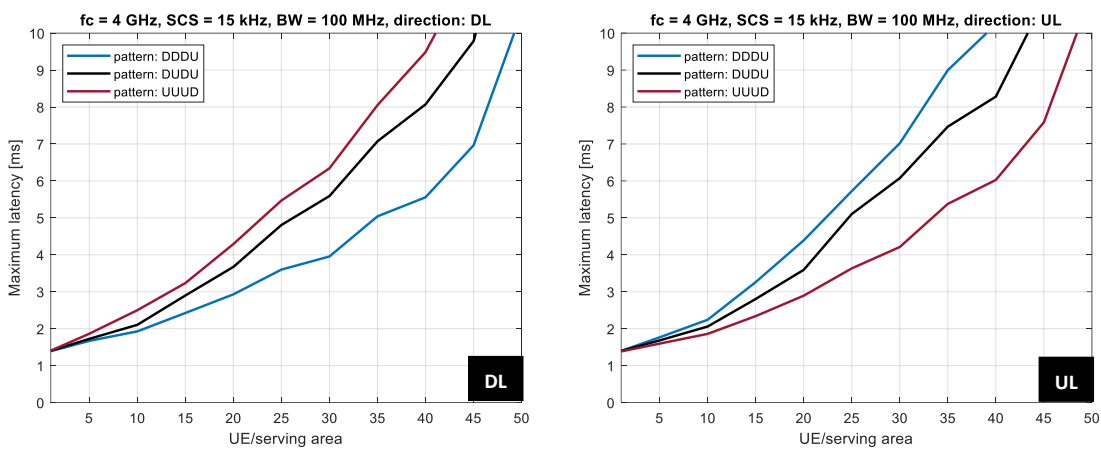

Figure 12 Impact of the TDD pattern on the worst-case latency and the percentage of UE satisfying requirements. **SCS = 15 kHz** (mid-band)

Table 6 Downlink and uplink worst-case latencies assuming different TDD patterns, 1-5 transmission attempts, and **a SCS: 15 kHz** (mid-band).

|   | Tx attempts | DDDU | | DUDU | | UUUD | |
|---|---|---|---|---|---|---|---|
|   |   | DL [ms] | UL [ms] | DL [ms] | UL [ms] | DL [ms] | UL [ms] |
| **10 UE/SA** | 1st | 1.93 | 2.24 | 2.10 | 2.06 | 2.50 | 1.86 |
|   | 2nd | 3.39 | 4.34 | 3.67 | 3.62 | 4.72 | 3.34 |
|   | 3rd | 5.14 | 6.55 | 5.32 | 5.27 | 6.96 | 5.07 |
|   | 4th | 6.91 | 8.92 | 7.09 | 7.04 | 9.39 | 6.82 |
|   | 5th | 8.68 | 11.30 | 8.85 | 8.81 | 11.64 | 8.52 |
| **20 UE/SA** | 1st | 2.919 | 4.39 | 3.672 | 3.592 | 4.29 | 2.89 |
|   | 2nd | 4.36 | 6.34 | 5.41 | 5.47 | 6.66 | 4.59 |
|   | 3rd | 6.13 | 8.67 | 6.98 | 6.84 | 8.99 | 6.16 |
|   | 4th | 7.84 | 10.96 | 8.89 | 8.91 | 11.46 | 7.99 |
|   | 5th | 9.59 | 13.47 | 10.94 | 10.71 | 13.90 | 9.88 |
| **30 UE/SA** | 1st | 3.954 | 7.018 | 5.59 | 6.07 | 6.34 | 4.21 |
|   | 2nd | 5.69 | 9.48 | 7.42 | 8.00 | 8.81 | 6.07 |
|   | 3rd | 7.61 | 12.06 | 9.25 | 9.85 | 11.46 | 8.04 |
|   | 4th | 9.52 | 14.76 | 11.16 | 11.83 | 14.20 | 9.99 |
|   | 5th | 11.49 | 17.47 | 13.13 | 13.84 | 16.90 | 12.05 |

As stated above, these results are useful to see the impact of relaxing latency as a function of the number of users. For example, if 10 UE are considered and the latency boundary is relaxed to 5ms, up to 3 transmission attempts can be made for high-band with SCS = 120 kHz and two transmission attempts with SCS = 60 kHz. While, using mid-bands two transmission attempts are supported using DUDU pattern for SCS= 30 and 15 kHz.

### 4.1.3.2  Supported number of users and achievable latency, reliability characteristics

To evaluate the impact of the TDD pattern and the relaxation of latency bounds (up to 10 ms) on reliability, a simulation campaign has been performed in both the mid and high-bands with different SCS for different numbers of users per service area. The results for high-band and mid-band are shown in Table 7 and Table 8 respectively. Three levels of reliability have been defined, A = 99.999%, B=99.99% and C=99.9%.

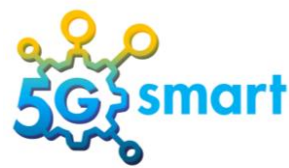

Table 7 Downlink and uplink reliability for different latency levels assuming different TDD patterns, and a sub-carrier spacing equal to 120 kHz (high-band), where A=99.999% (green), B = 99.99% (blue) and C = 99.9% (yellow).

| Frequency | SCS | UE/SA | Latency target [ms] | DDDU [Reliability] | | DUDU [Reliability] | | UUUD [Reliability] | |
|---|---|---|---|---|---|---|---|---|---|
| | | | | DL | UL | DL | UL | DL | UL |
| 30 GHz | 120 kHz | 10 | 1 | A | A | A | A | A | A |
| | | | 2 | A | A | A | A | A | A |
| | | | 5 | A | A | A | A | A | A |
| | | | 10 | A | A | A | A | A | A |
| | | 20 | 1 | A | C | B | B | C | A |
| | | | 2 | A | B | A | A | B | A |
| | | | 5 | A | A | A | A | A | A |
| | | | 10 | A | A | A | A | A | A |
| | | 30 | 1 | A | C | C | C | C | A |
| | | | 2 | A | B | B | B | B | A |
| | | | 5 | A | A | A | A | A | A |
| | | | 10 | A | A | A | A | A | A |
| | 60 kHz | 10 | 1 | B | B | B | B | B | B |
| | | | 2 | A | B | A | A | B | A |
| | | | 5 | A | A | A | A | A | A |
| | | | 10 | A | A | A | A | A | A |
| | | 20 | 1 | B | C | B | B | C | B |
| | | | 2 | B | B | B | B | B | B |
| | | | 5 | A | A | A | A | A | A |
| | | | 10 | A | A | A | A | A | A |
| | | 30 | 1 | B | C | C | C | C | B |
| | | | 2 | B | C | B | B | C | B |
| | | | 5 | A | A | A | A | A | A |
| | | | 10 | A | A | A | A | A | A |

Reliability levels correspond when users meet the latency requirements for UL and DL. For instance, it can be observed that for the high-band, if the latency is relaxed to 5 ms it is possible to support up to 30 users per service area with a reliability of 99.999% for both SCS= 120 and 60 kHz. It is further observed that if the reliability is relaxed to 99.99%, it is possible to achieve latencies of 2 ms for both numerologies using the DUDU pattern.

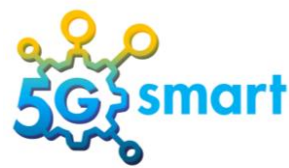

Table 8 Downlink and uplink reliability for different latency levels assuming different TDD patterns, and a sub-carrier spacing equal to 15 kHz (mid-band), where A=99.999%, B = 99.99% and C = 99.9%

| Frequency | SCS | UE/SA | Latency target [ms] | DDDU [Reliability] | | DUDU [Reliability] | | UUUD [Reliability] | |
|---|---|---|---|---|---|---|---|---|---|
| | | | | DL | UL | DL | UL | DL | UL |
| 4 GHz | 30 kHz | 10 | 1 | C | C | C | C | C | C |
| | | | 2 | A | B | A | A | B | A |
| | | | 5 | A | A | A | A | A | A |
| | | | 10 | A | A | A | A | A | A |
| | | 20 | 1 | C | C | C | C | C | C |
| | | | 2 | B | C | B | B | C | B |
| | | | 5 | A | A | A | A | A | A |
| | | | 10 | A | A | A | A | A | A |
| | | 30 | 1 | C | C | C | C | C | C |
| | | | 2 | B | C | C | C | C | B |
| | | | 5 | A | A | A | A | A | A |
| | | | 10 | A | A | A | A | A | A |
| | 15 kHz | 10 | 1 | C | C | C | C | C | C |
| | | | 2 | A | B | A | A | B | A |
| | | | 5 | A | A | A | A | A | A |
| | | | 10 | A | A | A | A | A | A |
| | | 20 | 1 | C | C | C | C | C | C |
| | | | 2 | C | C | C | C | C | C |
| | | | 5 | A | A | A | A | A | A |
| | | | 10 | A | A | A | A | A | A |
| | | 30 | 1 | C | C | C | C | C | C |
| | | | 2 | C | C | C | C | C | C |
| | | | 5 | A | B | A | A | B | A |
| | | | 10 | A | A | A | A | A | A |

It can be observed that for the mid-band, if the latency is relaxed to 5 ms, up to 30 users per service area can be supported with a reliability of 99.999% using the DUDU pattern. For the case of SCS = 30 kHz, if the reliability is relaxed to 99.99%, latencies of up to 2 ms can be achieved for 20 users in both UL and DL.

### 4.1.3.3   Impact of the available bandwidth on system latency

As the bandwidth is a limited and valuable resource, it is important to assess the impact it has on system KPIs in order to optimize spectrum usage planning.  For this purpose, a simulation campaign has been performed in the high-band considering two bandwidths: 100 and 400 MHz, TDD pattern: DUDU and packet sizes of 48 bytes. Figure 12 shows the worst latencies obtained for a number of users satisfying the requirements per serving area (99.999% of reliability). In the results, we can observe an expected impact when we reduce the bandwidth (less bandwidth, less users and vice versa). To broaden the understanding of this study, the figures show the latency CDFs for 10 and 30 users for both UL and DL.

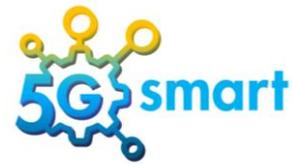

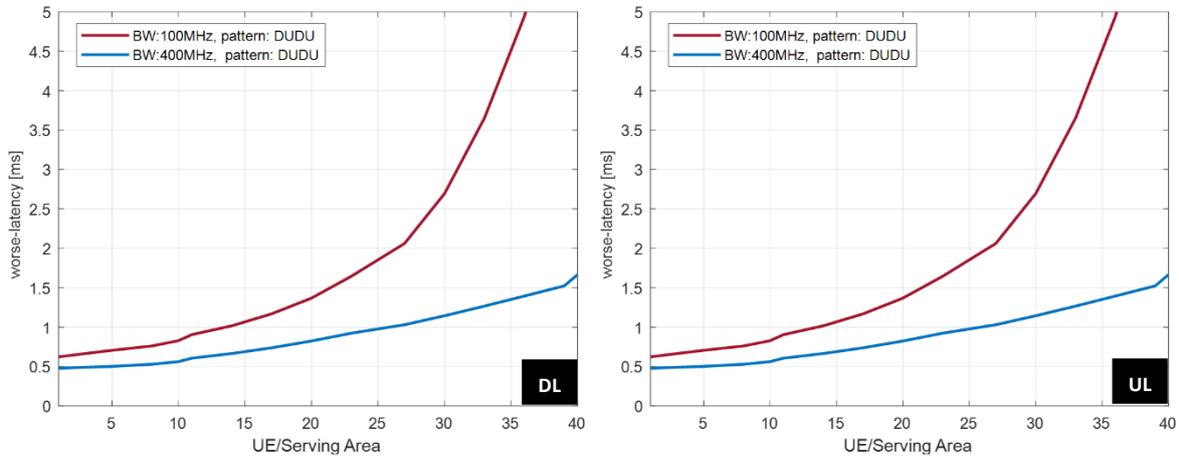

Figure 13 Impact of available bandwidth on system worst-case latency, high-band, SCS = 120 kHz, BW 100 and 400 MHz.

In that sense, it is necessary to evaluate the use of bandwidth by relaxing the latency requirement for the targeted reliability. For better understanding, Table 14 shows the latency CDFs for 10 and 30 users for both UL and DL. The difference is that if we relax the latency requirement, more retransmissions can be made if we use 400 MHz bandwidth and thus increase the reliability of the system, as studied in the previous section.

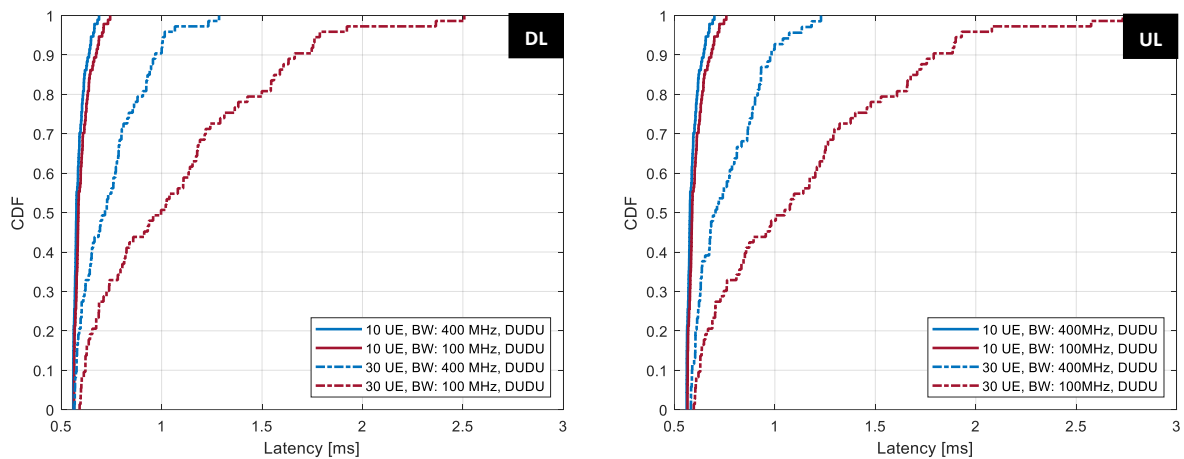

Figure 14 Latency CDF for 10 and 30 UE, BW 100 and 400 MHz.

#### 4.1.3.4   Impact of a packet size on system throughput

This section analyzes the impact of packet size on performance in terms of throughput. For this purpose, simulations have been performed in the mid- and high-bands, using different SCS and varying the packet size. Table 9 shows the different network configurations, where the throughput for DL and UL is evaluated. Note that for this evaluation, the latency target is 5 ms and the reliability target is 99.999%.

Table 9 Simulated configurations

| Configuration number | 1 | 2 | 3 | 4 | 5 | 6 | 7 | 8 |
|---|---|---|---|---|---|---|---|---|
| Carrier Frequency [GHz] | 4.0 | | 30.0 | | 4.0 | | 30.0 | |
| Bandwidth [MHz] | 100 | | 400 | | 100 | | 400 | |
| TDD pattern | DUDU | | | | | | | |
| SCS | 15 | 30 | 60 | 120 | 15 | 30 | 60 | 120 |
| Packet Size [Bytes] | 250 | | | | 500 | | | |
| Packet arrival interval [s] | 1/4000 | | | | 1/2000 | | | |
| Number of users | 10 | | | | | | | |

It is observed that the fact that since the TDD pattern is symmetrical, the capacity values obtained in the 50th percentile for UL are almost the same as for DL. On the other hand, it is observed that by increasing the packet size from 250 to 500 bytes, the average throughput is maintained. This can be explained by the fact that the network is not overloaded (50% load). Therefore, it can be concluded that if the traffic volume is the same and the network is not overloaded, similar throughput values will be obtained.

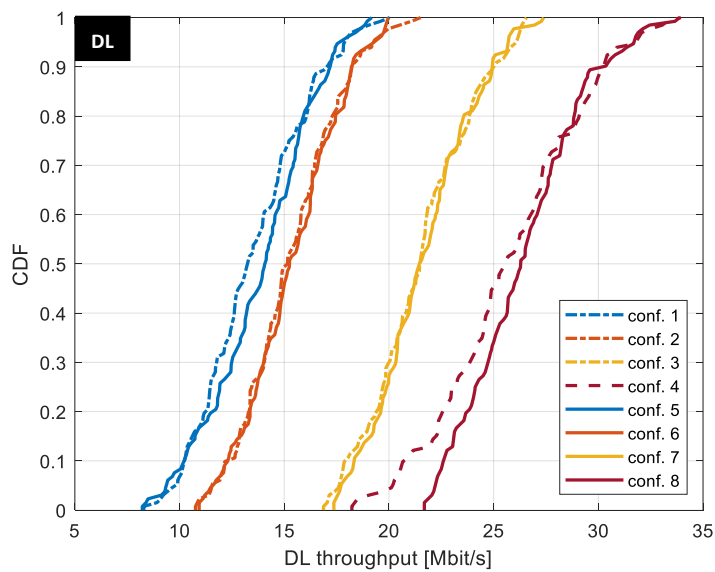

Figure 15 CDF of throughput for different configurations to satisfy UE requirements (downlink). Configurations given in Table 9.

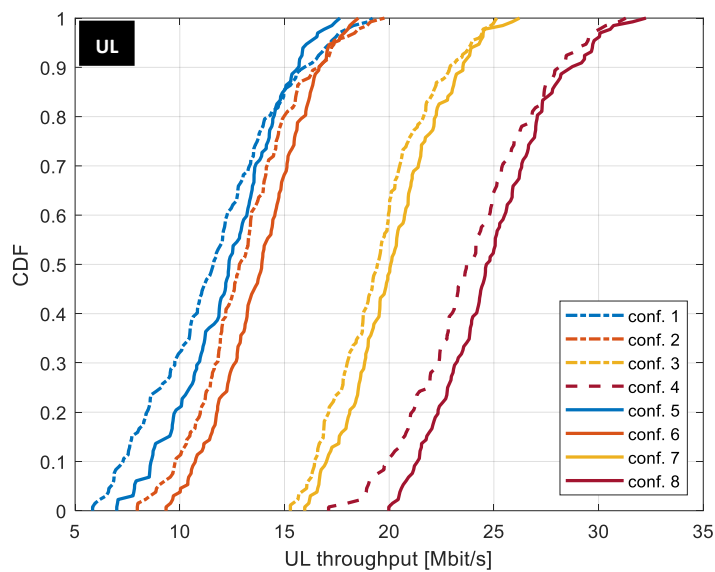

Figure 16 CDF of throughput for different configurations to satisfy UE requirements (uplink). Configurations given in Table 9.

#### 4.1.3.5   Impact of packet size on system latency

As we discussed in the previous sections, there is a predictable effect, the higher the numerology, the lower the latency. This section extends the analyse regarding with the impact of packet size on latency. For this purpose, simulations have been performed in the high-band for 10 randomly distributed UEs. Simulations have been performed with SCS =120 kHz, for both DL and UL considering a TDD: DUDU pattern, and packet sizes of 48, 100, 500 and 1000 bytes have been used.

By reducing the packet size and keeping the volume of traffic, users can be expected to have more resources and always have available slots in both uplink and downlink and achieve latencies close to 1 ms. In fact, as shown in Figure 16 50% of users achieve latencies of around 1 ms and close to 100% achieve latencies of less than 5 ms. On the other hand, when the packet size decreases, due to spatial diversity in frequency and channel encoders, there is a direct cost in latency performance. For example, it can be observed that for packets of 1000 bytes only 10% of users achieve latencies greater than 1 ms and that this percentage increases with smaller packets, up to 50% for packets of 48 bytes. This behavior can be explained by the spatial diversity in the frequency domain and the diversity that occurs in the channel encoder (LDCP). This means that the more resources to allocate, the higher the probability of failure, which results in higher latency with smaller packets. Therefore, it should be noted that this behavior is to be expected only when the network is not overloaded.

Therefore, we can conclude that, in order to develop and set the gNB parameters, it is essential to take into account the network load, number of UE per serving cell and to define the packet size to achieve the desired latencies.

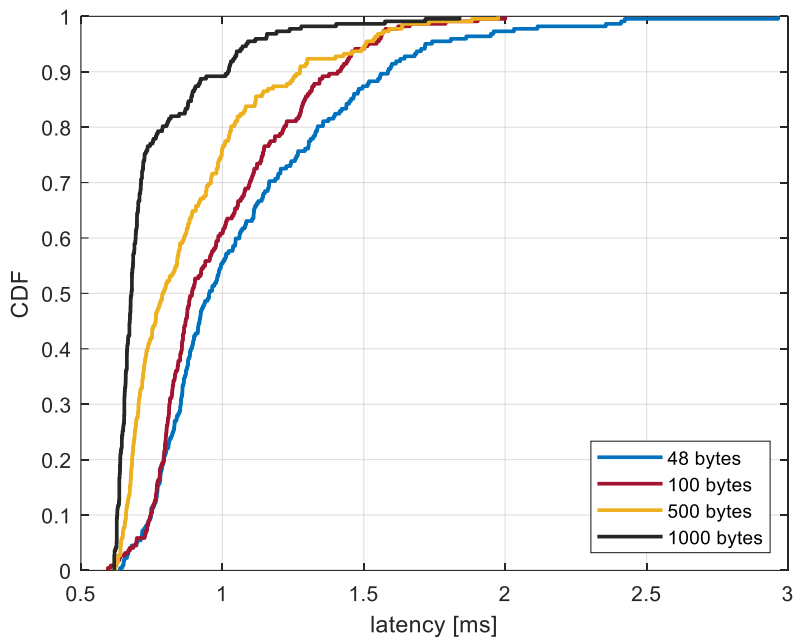

Figure 17 The impact of packet size on system latency in downlink.

## 4.2    Capacity of 5G IIoT deployments

### 4.2.1    Overview

This section continues the analysis of an independent standalone factory network presented in Section 5 of [5GS20-D14], considering an isolated scenario, without impact of external radio interference. The use of the same evaluation methodology makes it possible to understand the relative differences between results shown in [5GS20-D14] and those presented in this report. System level 5G NR performance evaluations are extended along several dimensions. Instead of focusing on a combination of the extremely low latency bound and high reliability target, in this deliverable a wider set of industrial applications is considered, where latency bounds range from 5 to 50 ms and network reliability targets are within 99% – 99.99%. This includes industrial applications presented in Section 2. It is important to note that applications with packet sizes of 250, 500 and 1000 bytes are considered, and this kind of analysis was not performed in [5GS20-D14]. Also, investigations on FDD band at 2.1 GHz carrier frequency is introduced in comparison with the other TDD spectrum bands. Additionally, a more elaborate evaluation of benefits for different antenna types and impact of the number of gNBs is presented.

### 4.2.2    Spectrum options

The configurations of the spectrum bands as listed in Table 10 are employed in the analysis presented in this section. Within NR slot there could be certain number of scheduling opportunities, which enable data transmissions within several OFDM symbols in a row. Length of this period is denoted as Transmission Time Interval (TTI).

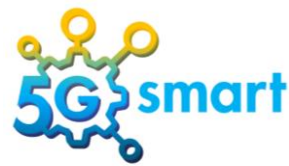

Table 10 Spectrum bands for capacity evaluation.

| Duplexing mode | TDD pattern | Carrier frequency (GHz) | Bandwidth (MHz) | Subcarrier spacing (kHz) | TTI length (OFDM symbols) |
|---|---|---|---|---|---|
| FDD | N/A | 2.1 | 20x2 | 30 | 14 |
| TDD | DUDU and DDDSU | 3.8 | 100 | 30 | 7 and 14 |
| TDD | DUDU and DDDSU | 26 | 400 | 120 | 14 |

Portions of spectrum in FDD bands could be provided by the Mobile Network Operators (MNOs) and therefore it is beneficial to understand how much capacity this NR system configuration could provide for industrial services. TDD mid-band spectrum can be provided by the MNOs, and in some countries access to TDD mid-band spectrum can be obtained by industrial players through local licensing, e.g., in Germany 3.7 − 3.8 GHz and in France 2570 − 2620 MHz frequency bands. TDD frequency bands in millimeter wave spectrum provide much larger system bandwidth; these bands can be accessed through local licensing in countries where this is supported or via MNOs.

### 4.2.3   Simulation tool and industrial scenario

The capacity evaluations of 5G NR industrial networks presented in this section are performed using the same methodology and simulation platform as described in Section 5 of [5GS20-D14]. It is important to note that in simulations used to generate results presented in this section downlink control channel and shared channel have realistic models. Other channels use simplistic modeling. We do not take into account overhead introduced by broadcast information (SSB blocks), RRC (re)configurations, random access messages, channel measurements and reporting (CSI-RS/IM, PUCCH). Ideal channel estimation is used. Additionally, we are using a simplified version of scheduler. Results with different implementations of scheduler may differ significantly.

The considered deployment scenario is according to 3GPP factory layout of 120 × 50 × 10 m with 3 and 12 gNBs, as shown in Figure 18. The propagation model is based on the 3GPP model for Indoor Factory with Dense clutter and High base station height (InF-DH) [3GPP19-38901]. Configurations of the industrial 5G NR network deployments with omnidirectional, distributed antenna system (DAS) with 12 antennas and advanced antenna system (AAS) with panel antenna array with beamforming are shown in Table 11, Table 12, Table 13.

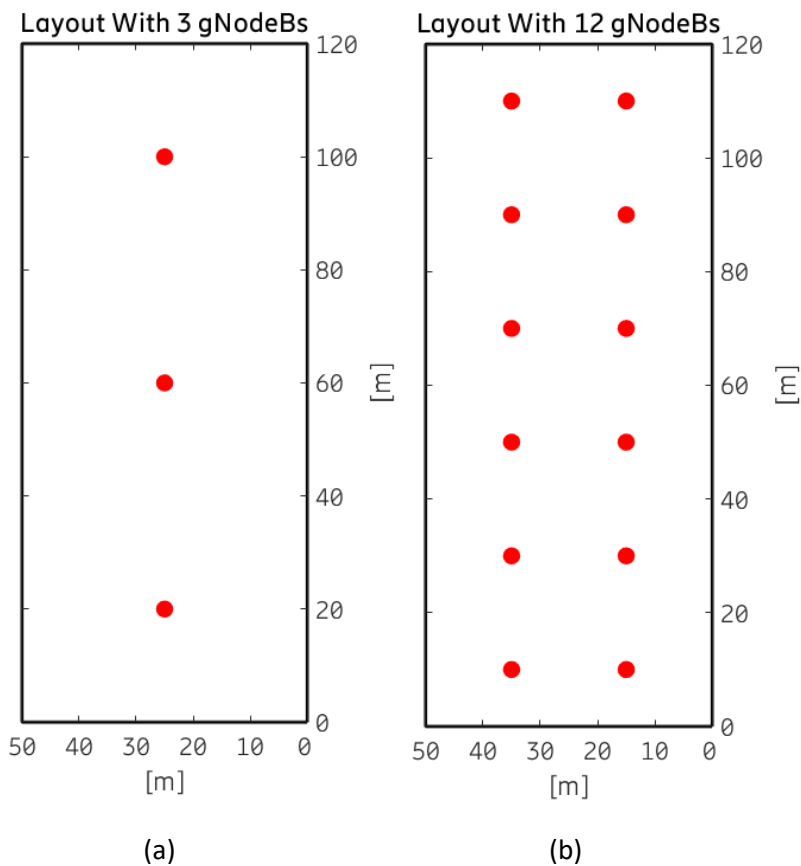

Figure 18: Simulation scenario of the indoor factory with 3 (a) and 12 (b) gNBs.

Table 11 5G NR system configuration with omni-directional antenna.

| Parameter | Mid-band | Mid-band | High-band |
|---|---|---|---|
| Frequency [GHz] | 2.1 | 3.6 | 26 |
| Channel bandwidth [MHz] | 20 | 100 | 400 |
| gNodeB transmission power [dBm] | 30 | 30 | 30 |
| UE transmission power [dBm] | 23 | 23 | 23 |
| gNodeB antenna gain (omni) [dBi] | 2 | 2 | 2 |
| UE antenna gain [dBi] | 0 | 0 | 9 |
| gNodeB receiver noise figure [dB] | 7 | 5 | 7 |
| UE receiver noise figure [dB] | 6 | 9 | 10 |
| Uplink power control setup | SNR target = 10 dB $\alpha$ = 0.8 | SNR target = 10 dB $\alpha$ = 0.8 | SNR target = 10 dB $\alpha$ = 0.8 |

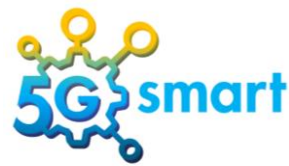

Table 12 5G NR system configuration with distributed antenna system (12 antennas).

| Parameter | Mid-band | Mid-Band | High-Band |
|---|---|---|---|
| Frequency [GHz] | 2.1 | 3.6 | 26 |
| Channel bandwidth [MHz] | 20 | 100 | 400 |
| gNodeB transmission power [dBm] | 30 | 30 | 30 |
| UE transmission power [dBm] | 23 | 23 | 23 |
| UE antenna gain [dBi] | 0 | 0 | 9 |
| gNodeB receiver noise figure [dB] | 19 | 19 | 19 |
| UE receiver noise figure [dB] | 6 | 9 | 10 |
| Uplink power control setup | SNR target = 10 dB $\alpha$ = 0.8 | SNR target = 10 dB $\alpha$ = 0.8 | SNR target = 10 dB $\alpha$ = 0.8 |

Table 13 5G NR system configuration with AAS and beamforming antenna.

| Parameter | Mid-band | Mid-Band | High-Band* |
|---|---|---|---|
| Frequency [GHz] | 2.1 | 3.6 | 26 |
| Channel bandwidth [MHz] | 20 | 100 | 400 |
| gNodeB transmission power [dBm] | 30 | 30 | 30 |
| UE transmission power [dBm] | 23 | 23 | 23 |
| gNodeB antenna element gain [dBi] | 5 | 5 | 5 |
| gNodeB antenna panel | 4x4x2 | 4x4x2 | 4x4x2/8x8x2 |
| UE antenna gain [dBi] | 0 | 0 | 9 |
| gNodeB receiver noise figure [dB] | 7 | 5 | 7 |
| UE receiver noise figure [dB] | 6 | 9 | 10 |
| Uplink power control setup | SNR target = 10 dB $\alpha$ = 0.8 | SNR target = 10 dB $\alpha$ = 0.8 | SNR target = 10 dB $\alpha$ = 0.8 |

*note that full digital beamforming is assumed for high-band, i.e. when the radio can form sufficient number of beams. Practical deployments may apply analog or hybrid beamforming, which would result in lower capacity values.

### 4.2.4    URLLC Service based System Capacity analysis

We have analyzed the performance of a 5G NR wireless network in several industrial-indoor use cases from 5G-SMART and those commonly used by 3GPP and 5G-ACIA in system level evaluations. The capacity in terms of maximum number of served UEs is calculated using the bitrate of a single user and a maximum bitrate achievable with a specific network deployment, given that 100% coverage can be provided. The most essential aspects of the analysis are related to the comparison of spectral efficiency and the maximum number of served users in different frequency bands, network setups, transmission directions and gNB deployment densities. To illustrate the performance for such a large variety of options we focus in particular on a single use case in this section. We selected the 5G-SMART UC1 robotics motion planning use case [5GS20-D11], which is characterized by 500 Byte packets being periodically transmitted in DL and UL at 0.8 Mbps bitrate per user. A latency bound of 5 ms needs to be met at 99.99% network reliability. Figure 19 and Figure 20 show the DL and UL spectral efficiency in terms of bits/second/Hz/cell, and Figure 21 and Figure 22 depict the system capacity for this use case in terms of the maximum number of served users. This section is dedicated to the analysis of the results presented in these figures as well as verification of these results by looking into radio performance characteristics. Beyond the 5G-SMART UC1, five more use cases have been investigated. Section 2 provides an overview of the achievable capacity for these use cases for a limited number of deployment options. Furthermore, in this section we provide an in-depth analysis for 5G-SMART UC1, but many of the general observations hold also for the other use cases. The results on capacity and spectral efficiency results for all use cases and deployment options can be found in the Appendix of this report.

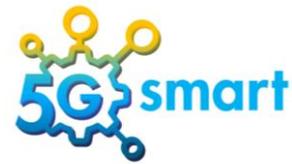

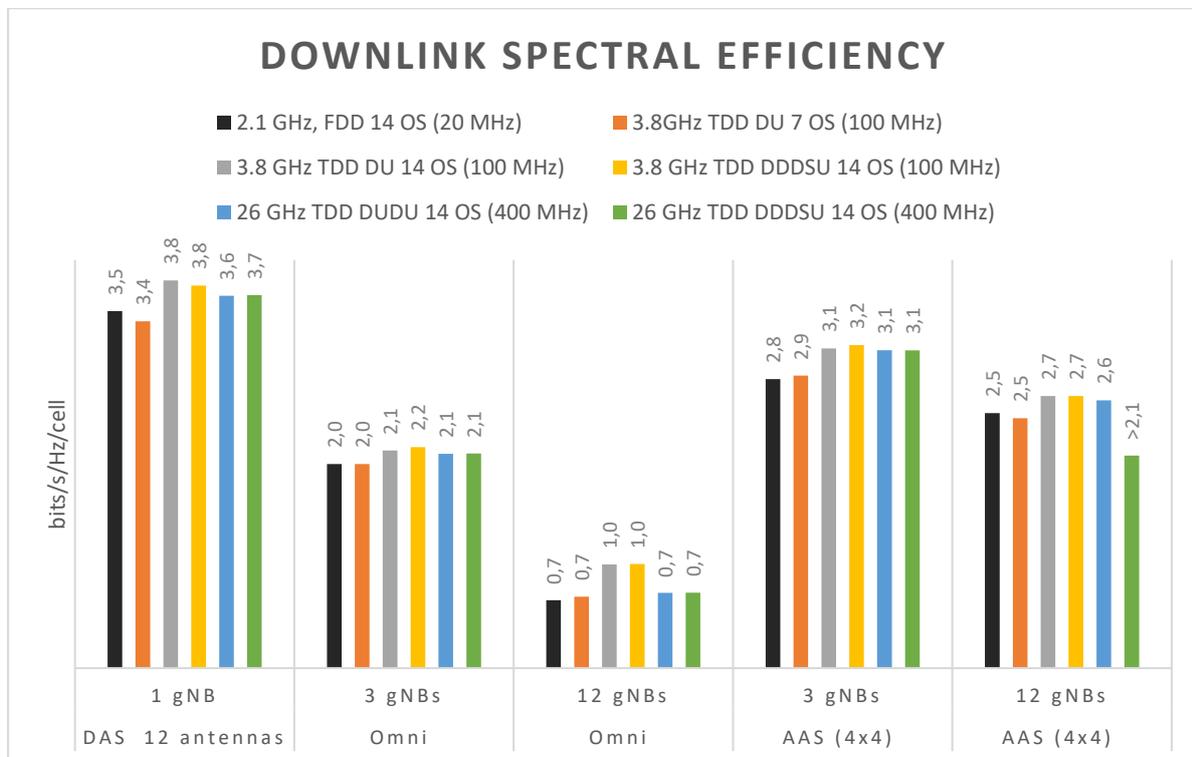

Figure 19: Downlink spectral efficiency for 5G-SMART UC1 for different deployment options of a 5G NR network in a factory.

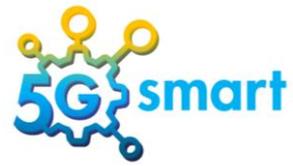

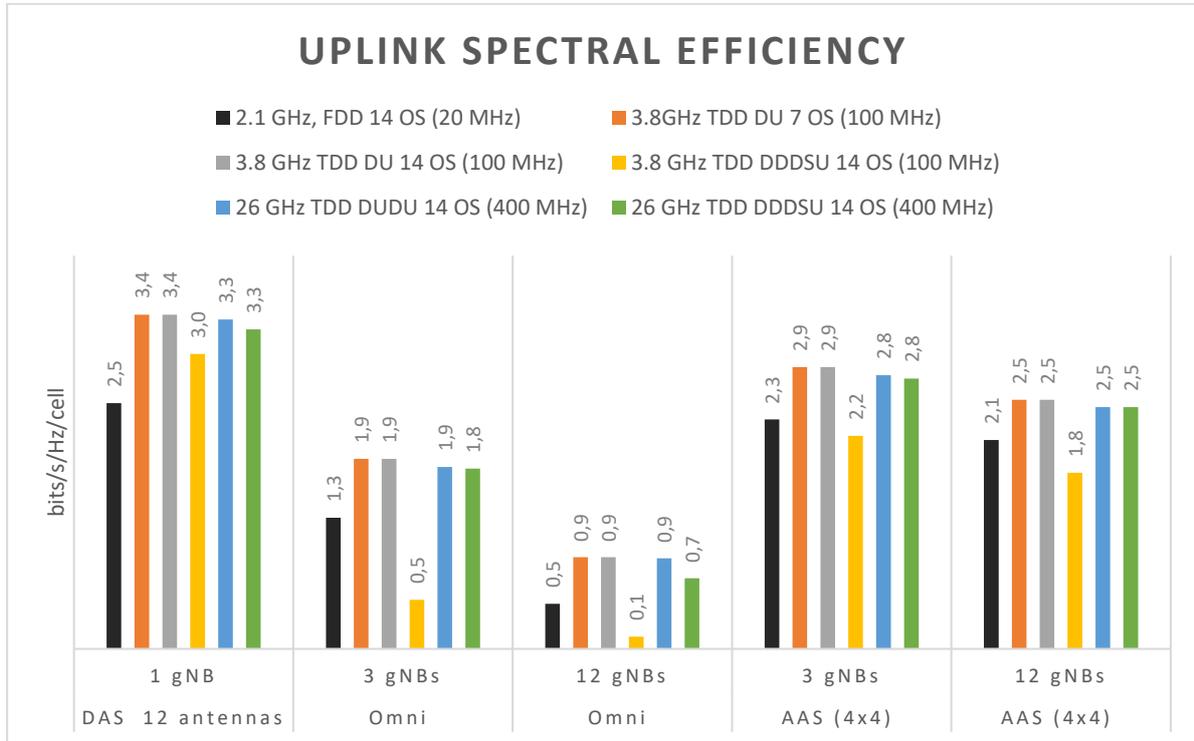

Figure 20: Uplink spectral efficiency for 5G-SMART UC1 for different deployment options of a 5G NR network in a factory.

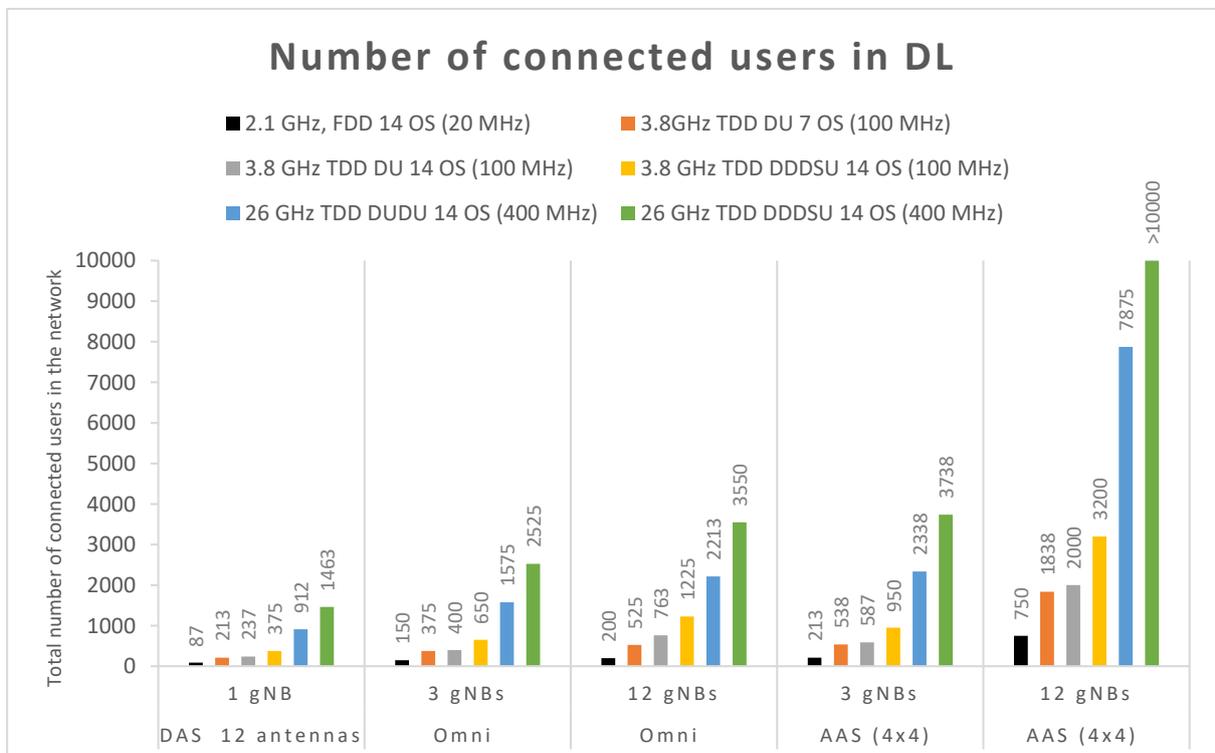

Figure 21: Maximum number of DL user connections for 5G-SMART UC1 in 5G NR network in a factory.

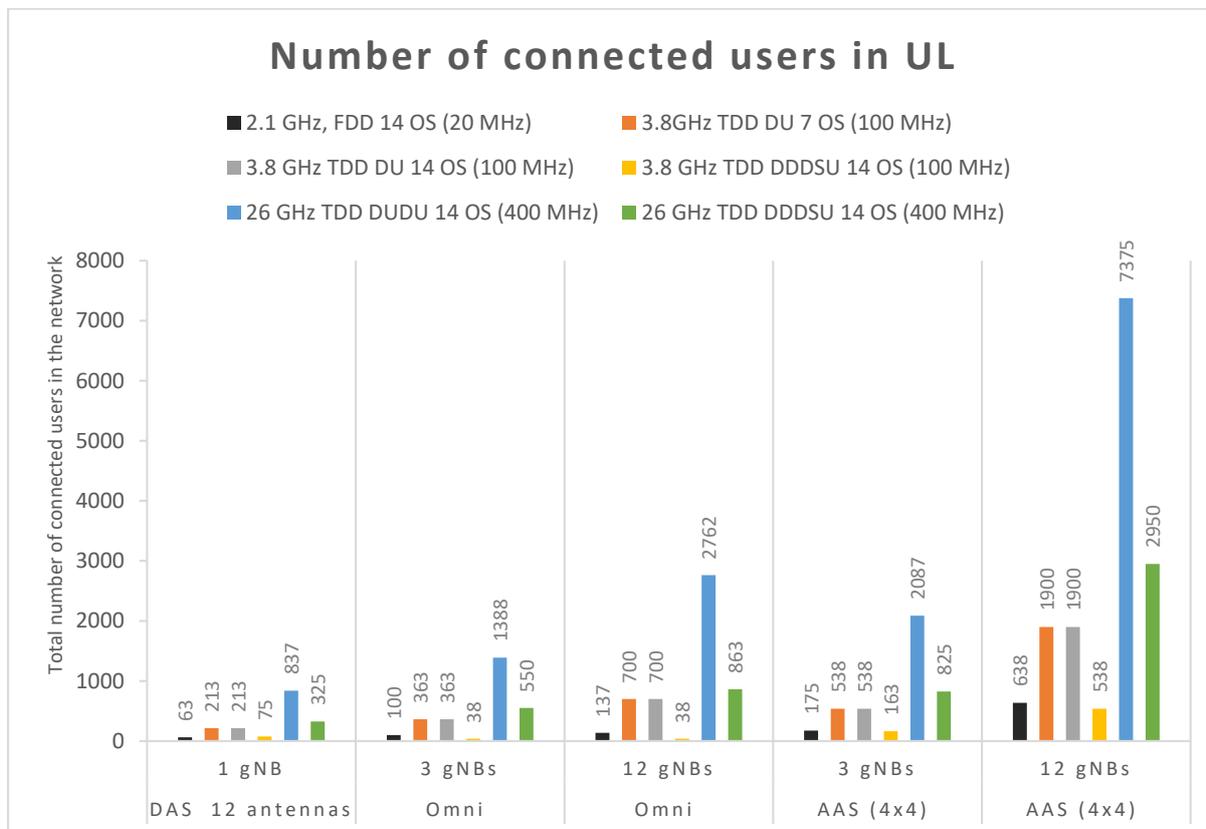

Figure 22: Maximum number of UL user connections for 5G-SMART UC1 in 5G NR network in a factory.

### 4.2.4.1    Frequency band, duplexing mode and TDD pattern

In a first step, we look at the capacity of TDD vs. FDD spectrum bands. In DL, spectral efficiency in FDD with 14 OS TTIs is on par with TDD DUDU at 3.8 GHz with 7 OS TTIs and TDD 26 GHz with 14 OS. The UL spectral efficiency of FDD is lower because control channel overhead compared to the amount of data resource blocks is slightly larger than in TDD.

The major difference in capacity of FDD, mid-band and high-band TDD is originating from the large differences in the amount of available bandwidth. FDD has only 2x 20 MHz for downlink and uplink compared to 100 MHz in 3.8 GHz band and 400 MHz in 26 GHz band, for uplink and downlink together.

Another important aspect is numerology used in mid-band and high-band, this influences the number of achievable transmission attempts, and as a result gives access to higher spectral efficiency even at lower latencies. For instance, in high-band it is possible to have more transmission attempts due to shorter slot duration of 125 μs compared to 0.5 ms in mid-band. The latter point is visible especially well from comparison of TDD pattern DDDSU uplink performance, shown with the yellow and green bars in Figure 20 and Figure 22. We can clearly see a drop for UL spectral efficiency and capacity for DDDSU in 3.8 GHz band where only 1 transmission can be performed within 5 ms latency bound, while at 26 GHz there's a possibility to do 4-5 transmissions depending on configuration. As a result, it is possible to use less robust and more spectrally efficient MCS and still guarantee high reliability.

On the other hand, a DDDSU pattern is more beneficial for downlink performance, compared to latency-oriented DUDU pattern suitable for balanced uplink-downlink traffic. This is well seen when comparing the grey and yellow, as well as blue and green bars respectively in Figure 21. In case of mid-band DL and e.g., AAS 4x4 antenna with 3 gNBs in the network, the number of connected UEs with DUDU TDD pattern is 2338, while for DDDSU it is 3738, which is about 60% more. This advantage of 60% higher DL capacity for DDDSU compared to DUDU holds for all compared configurations and deployment options. One exception is for AAS 4x4 with 12 gNBs, but this is only because the maximum system capacity for the DDDSU configuration has not been within the range of simulated load points, so it is simply not possible to calculate the exact percentage of the difference, see the right-most green bar in Figure 21. In the end, the traffic profile and service requirements define the traffic requirements of the service and thereby the capacity. All the services listed in chapter 2 have a symmetric traffic pattern and service requirements in UL and DL directions. The achievable capacity thus needs to be supported in both UL and DL directions. As a consequence, the lower of the UL and DL capacity values in Figure 21 and Figure 22 determine the achievable capacity. For example, for mid-band TDD with AAS and 3 gNBs, a TDD configuration of DDDSU can support 950 users of UC1 in DL and 163 users in UL. In consequence, only 163 users can be supported, as the UL constitutes the capacity bottleneck. For a TDD configuration of DUDU with 14 OFDM symbols, the corresponding DL and UL capacity figures are 587 users and 538, which means that 538 users can really be supported in both DL and UL. Even if the DL capacity of DUDU seems almost half as low for DL than for DDDSU, the overall capacity that can be supported with DUDU is more than 3 times higher than with DDDSU.

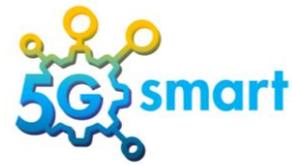

When it comes to the selection of TDD pattern it is interesting to compare the cost of more or less latency oriented TDD pattern, such as e.g. DUDU with 7 OS and 14 OS sub-slot scheduling opportunity. As it is shown in Figure 21, advantage of the configuration with 14 OS TTI length over 7 OS TTI in terms of DL capacity is on average 16% and it can vary between 6.6%-45% depending on the antenna type and number of gNBs. This difference comes from the necessity to send Downlink Control Information (DCI) more frequently when the TTI length is shorter. On the other hand, there is no UL capacity difference for 7 and 14 OS TTI configurations as it is shown in Figure 22. Thus, it is beneficial to choose a TDD pattern considering latency requirements of the industrial service, and not to opt by default for the most latency-oriented configuration.

### 4.2.4.2   Antenna type and deployment density

Let us first consider the impact of the antenna type, in terms of spectral efficiency in bits/s/Hz/cell. The downlink DAS scheme is demonstrating the highest spectral efficiency, regardless of the frequency band and TDD pattern as it can be seen from Figure 19. This is achieved mainly since there is only one cell in configuration with DAS, and as a result there is a lack of intercell interference and hence very high SINR values, as it is shown in Figure 23. Therefore, higher order MCSs are used, and this gives higher spectral efficiency per cell compared to the other deployment and antenna options with 3 or 12 cells. In uplink, a similar situation is observed mainly because SINR in case of DAS setup does not go below 10 dB, as it can be seen from Figure 24, the capacity of a single cell with DAS is limited by the total amount of time-frequency resources, i.e. the total number of schedulable resource blocks (RBs), rather than low SINR, as it is the case with the omni-directional or AAS antennas. It is important to note that SINR CDFs shown in Figure 23 and Figure 24, and SINR heatmaps in Figure 25, Figure 26, Figure 27 and Figure 28 for each system correspond to the highest capacity this system can fulfil guaranteeing 5 ms latency at 99.99% reliability for 100% of locations within factory network.

Thus, in terms of spectral efficiency a single cell with DAS is the most beneficial choice. However, when it comes to the system capacity, the situation is very different. Looking at the maximum number of connected UEs, DAS demonstrates the lowest capacity, then comes a network with omni-directional antenna and the highest capacity is achieved when AAS system is used.

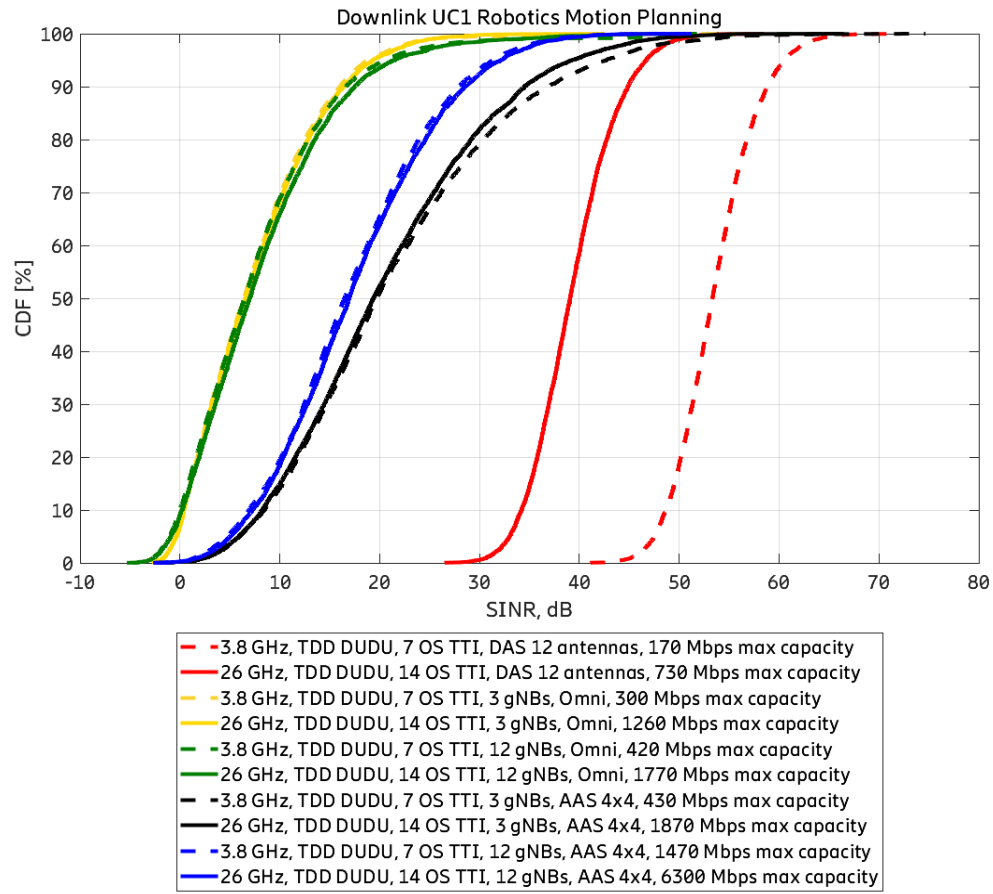

Figure 23. Downlink SINR of TDD 3.8 GHz, DUDU, 7 OS TTI and TDD 26 GHz, DUDU, 14 OS for Use Case 1: Robotics Motion Planning (latency bound = 5 ms, reliability target = 99.99%)

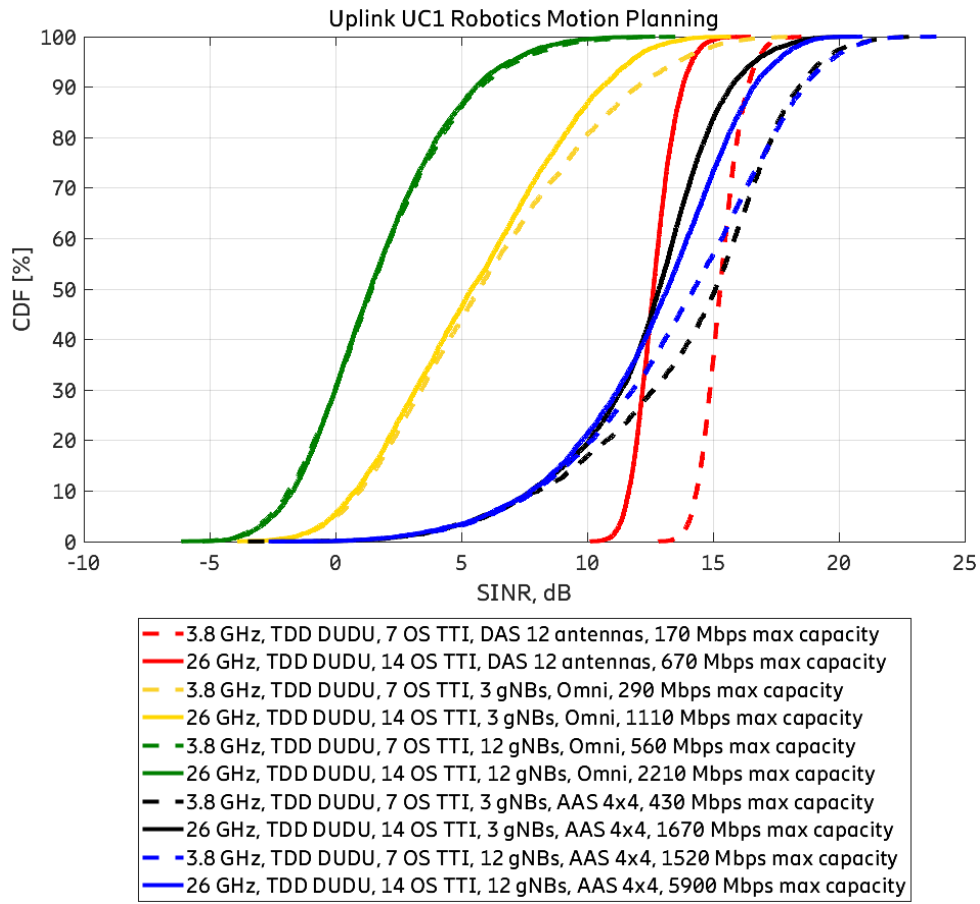

Figure 24. Uplink SINR of TDD 3.8 GHz, DUDU, 7 OS TTI and TDD 26 GHz, DUDU, 14 OS for Use Case 1: Robotics Motion Planning (latency bound = 5 ms, reliability target = 99.99%)

Densification helps to increase system capacity. As discussed above, DAS demonstrates the lowest capacity both in DL and UL and can be seen as a benchmark. An increase from 3 to 12 omni-directional antennas helps to boost downlink capacity by on average 55% and by about 90% for DDDSU configuration in 3.8 GHz TDD. In uplink adding more omni-antennas helps to improve capacity from 37% to 100%, but UL performance with DDDSU TDD configuration is very poor and cannot be improved with a denser deployment.

Network deployments with AAS in general demonstrate the best performance compared to other antennas both in absolute values and in terms of positive gain reached through densification. The reason is that with AAS, the 5G RAN is capable to much better suppress interference, resulting in a much better SINR contribution compared to omni-directional antennas, as can be seen in Figure 25 and Figure 26. The average growth of DL and UL capacity is 230% and 250% respectively when the number of cells is increased from 3 to 12.

It is worth mentioning that mid-band deployment with 12 omni-directional antennas is either on par or marginally better in DL than a network with 3 AAS antennas. And in high-band 3 AAS cells already have higher DL capacity than 12 cells with omni-directional antennas. In uplink 12 cells with omni are better than 3 cells with AAS in all setups, except for 3.8 GHz TDD with DDDSU, where uplink is nearly non-functional for omni-antenna deployments.

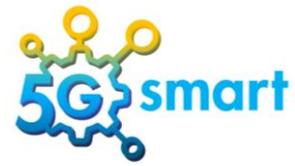

Below one can find DL and UL SINR heatmaps as an illustration of the coverage inside the factory with different antenna options in for 3.8 GHz DUDU 7 OS (Figure 25 and Figure 26) and 26 GHz DUDU 14 OS (Figure 27 and Figure 28). We can clearly see the differences between different antenna options in terms of SINR levels and potential with the deployments of certain types, e.g. deployment with omni-directional antennas is showing poorer coverage, and this matches well results discussed earlier in this section. These coverage maps are presented here to demonstrate not only statistical, but also spatial distribution of the SINR – as a key characteristic, which affects URLLC system performance and capacity.

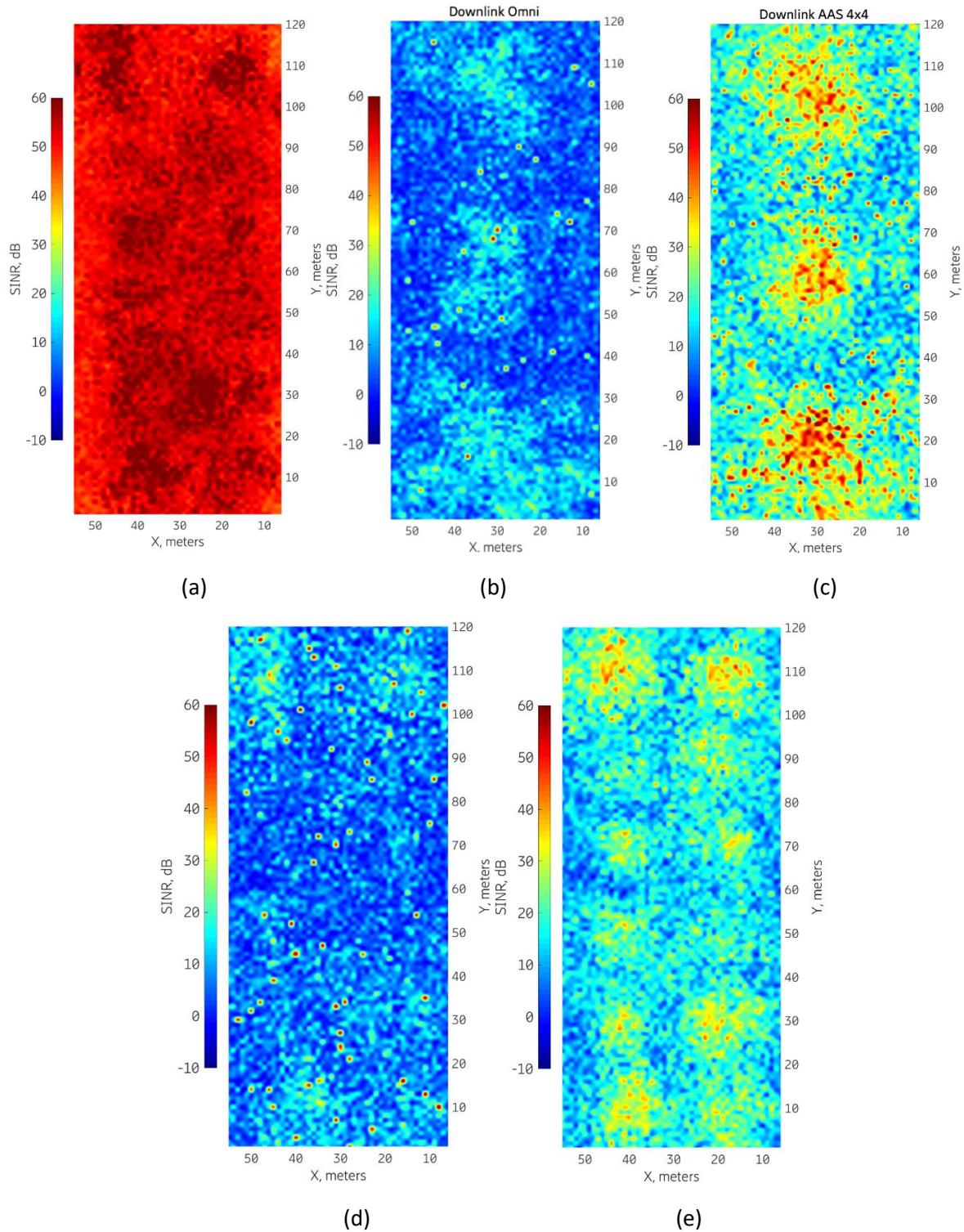

Figure 25. DL SINR in 3.8 GHz frequency band, TDD DUDU, 7 OS TTI (a) DAS 12 antennas, (b) Omni 3 gNBs, (c) AAS 4x4 3 gNBs (d) Omni 12 gNBs, (e) AAS 4x4 12 gNBs.

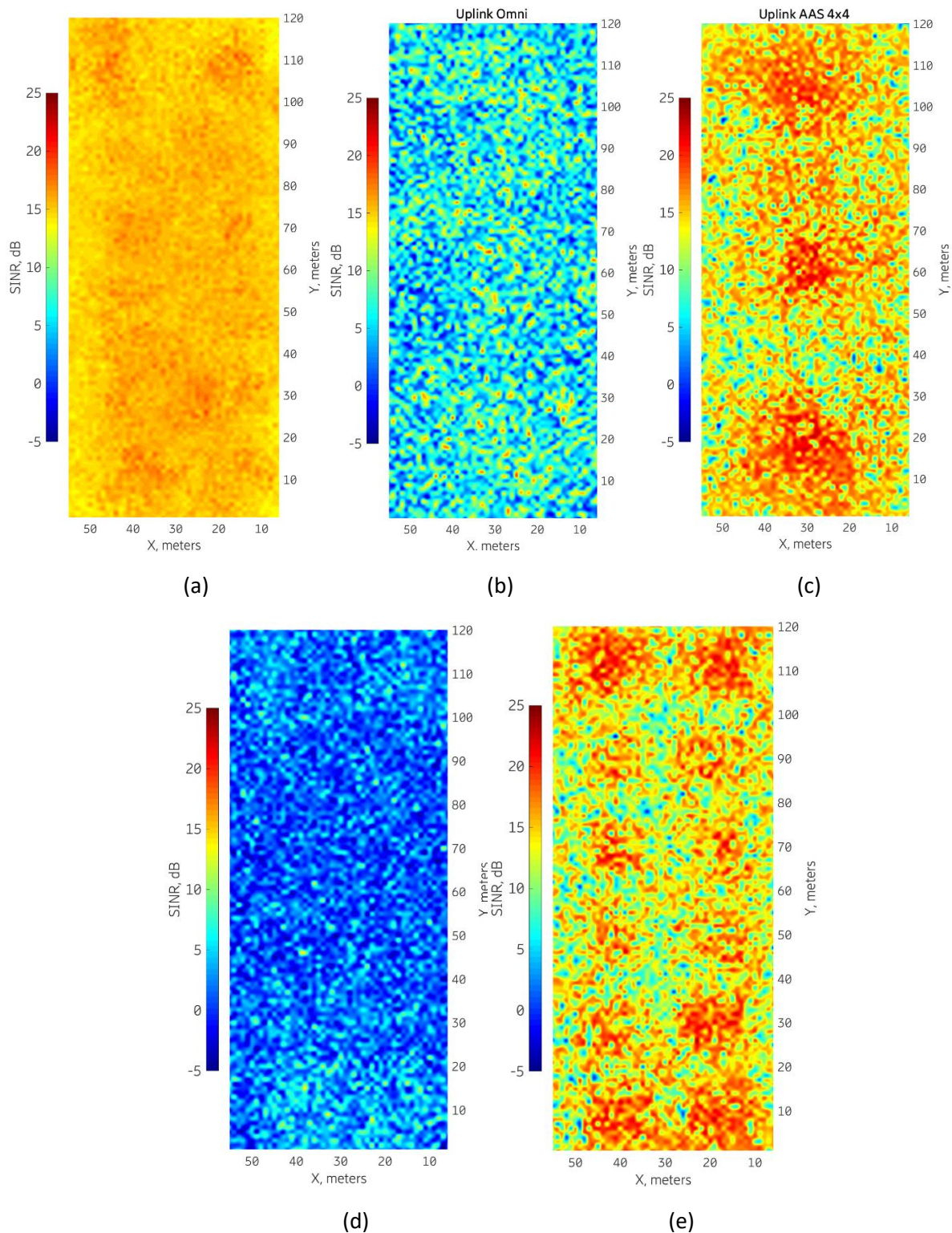

Figure 26. UL SINR in 3.8 GHz frequency band, TDD DUDU, 7 OS TTI (a) DAS 12 antennas, (b) Omni 3 gNBs, (c) AAS 4x4 3 gNBs (d) Omni 12 gNBs, (e) AAS 4x4 12 gNBs.

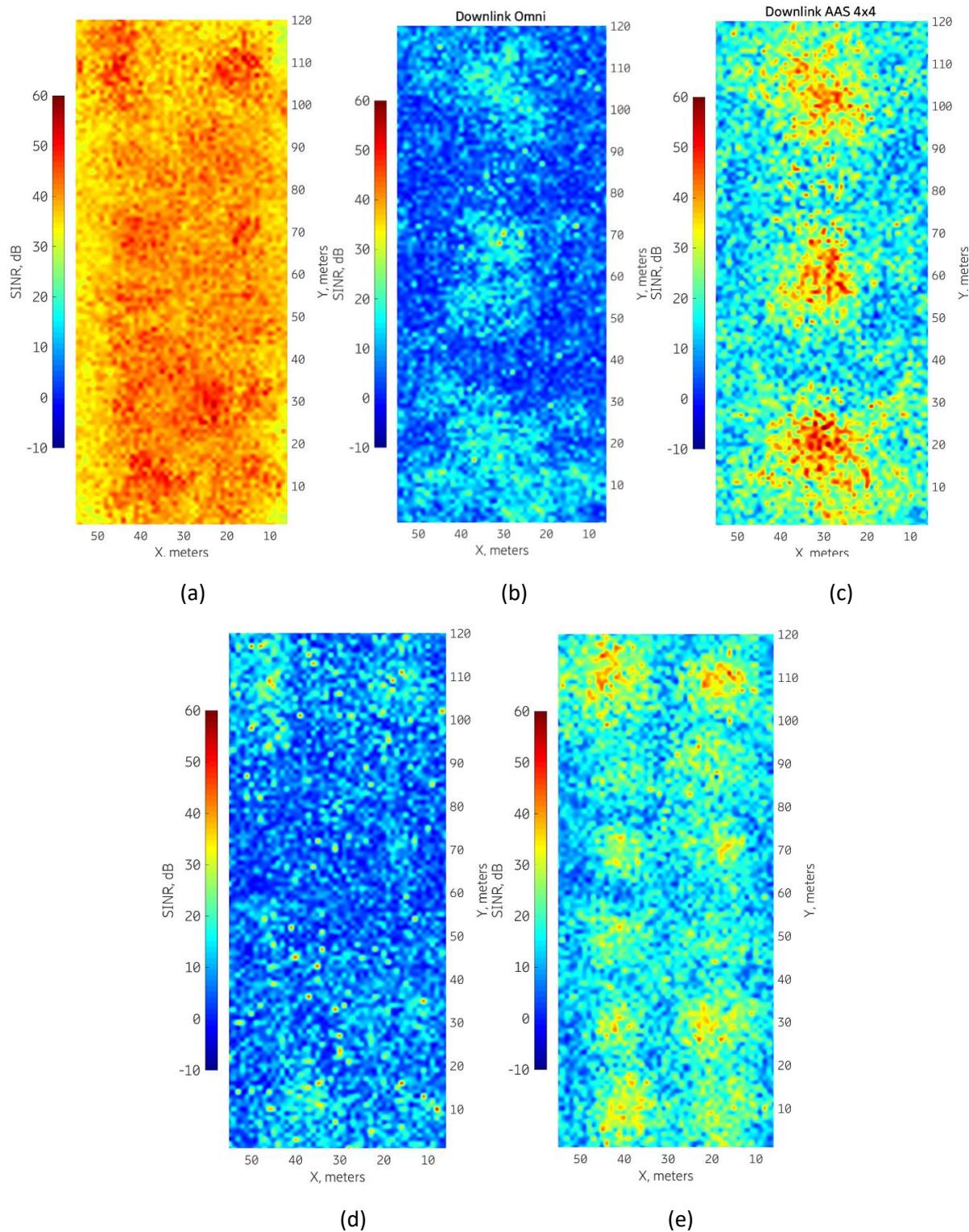

Figure 27. DL SINR in 26 GHz frequency band, TDD DUDU, 14 OS TTI (a) DAS 12 antennas, (b) Omni 3 gNBs, (c) AAS 4x4 3 gNBs (d) Omni 12 gNBs, (e) AAS 4x4 12 gNBs.

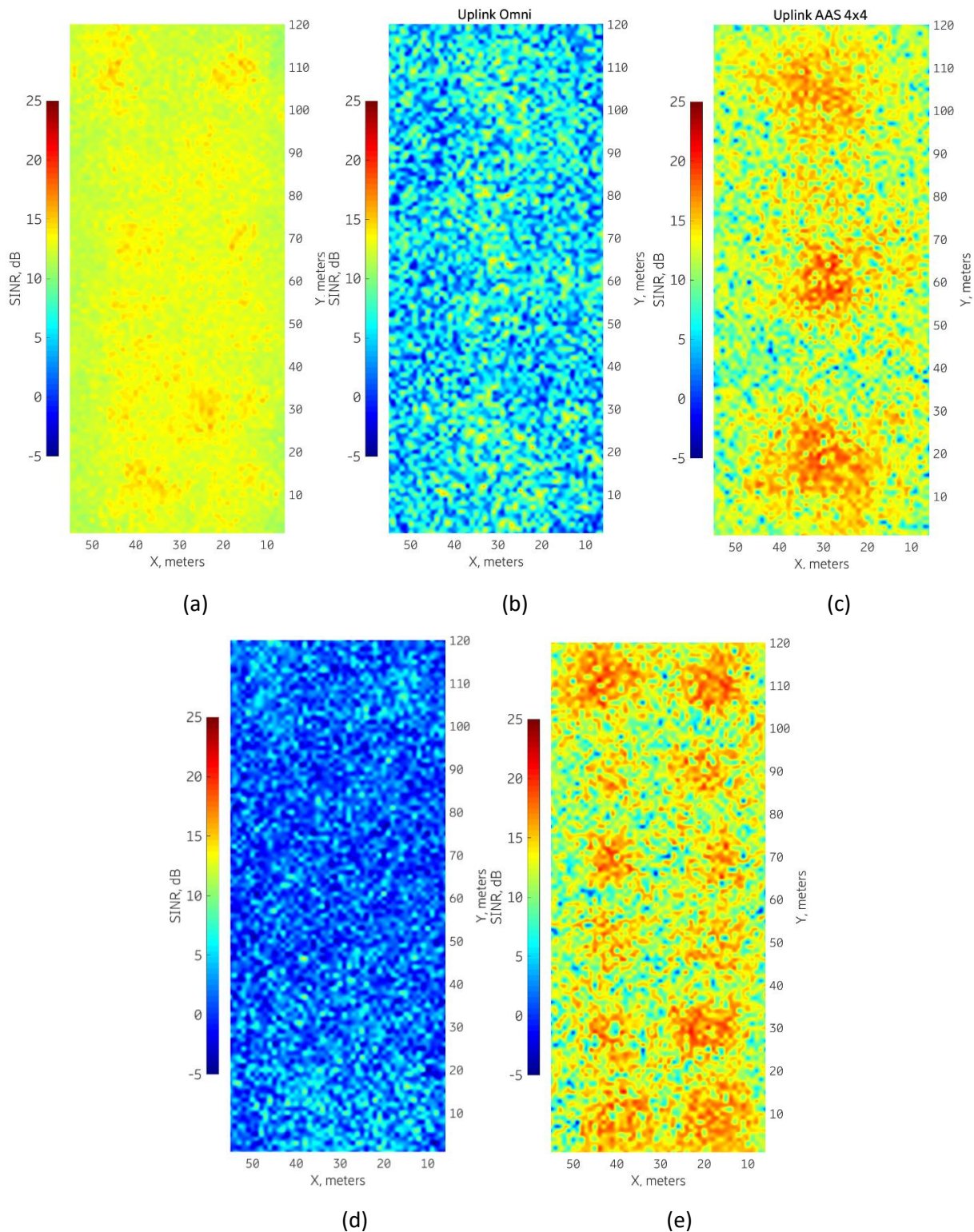

Figure 28. UL SINR in 26 GHz frequency band, TDD DUDU, 14 OS TTI (a) DAS 12 antennas, (b) Omni 3 gNBs, (c) AAS 4x4 3 gNBs (d) Omni 12 gNBs, (e) AAS 4x4 12 gNBs

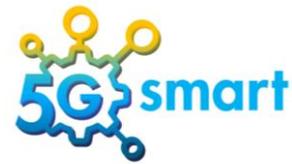

### 4.2.5    Summary

Earlier in this section we have shown the spectral efficiency and capacity numbers for different industrial service from those discussed in Chapter 2 considering the impact of deployment option, network setup, e.g. antenna type, number of cells and TDD configuration. Next, in Section 4.2.4.2 a more detailed comparison of statistical and spatial characteristics of radio quality for different network configuration options is presented. Below one can find a discussion about capacity numbers and network configuration impact.

#### 4.2.5.1    Capacity for different industrial services

The number of users that can be supported in a 5G deployment depends on several factors, as will be explained at the achievable capacity in Table 14. Different use cases from Section 2 are investigated. The use cases differ in their traffic characteristics, expressed as the message size and the cycle time. In general, it can be seen that the capacity in number of served users of a use case category depends on the data rate of the traffic: the higher the data rate required for a service, the lower the number of users that can be served.

The capacity also depends on the spectrum band and the deployment. In Table 14 we only look at a deployment of twelve 5G gNBs in the shopfloor (other options are included in Appendix). As expected, the capacity increases significantly if adaptive antenna systems (AAS) with beamforming are used, instead of omnidirectional antennas. In mid-band capacity in the FDD band with 2x 20 MHz of spectrum, is almost the same as for the 100 MHz of mid-band TDD spectrum with a DDDSU TDD pattern. The reason is that the service has a symmetric data rate requirement for UL and DL, but the TDD pattern has most resources allocated to downlink. The number of served users is thus limited by the UL capacity of the TDD band, which is similar to the FDD band. If a balanced DUDU TDD pattern is used, the capacity of the TDD band more than doubles, as the amount of radio resources that are provided to UL increase by approximately a factor of 2.5. In the high-band spectrum, a significant amount of spectrum is available which can serve many more users than in the mid-band allocations. This high capacity in high-band is assuming that digital beamforming is applied; for practical high-band deployments analog or hybrid beamforming may be used which would result in lower capacity values.

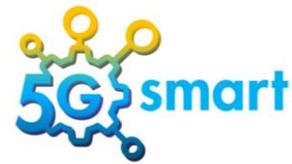

Table 14 5G system capacity in number of served uses for the different use cases, depending on the spectrum band and RAN deployment option.

| Use cases | | | | UC7 (controller-to-controller) | 22.104 mobile robots (1) - machine control | UC4 (process monitoring) | 22.104 (controller-to-controller) (1) | 22.104 (controller-to-controller) (2) | UC1 (robotics motion planning) |
|---|---|---|---|---|---|---|---|---|---|
| Message size   [byte] | | | | 500 | 250 | 1024 | 1000 | 1000 | 500 |
| cycle time [ms] | | | | 10 | 10 | 5 | 10 | 50 | 5 |
| Data rate [Mb/s] | | | | 0.4 | 0.2 | 1.6384 | 0.8 | 0.16 | 0.8 |
| Latency bound [ms] | | | | 10 | 10 | 10 | 10 | 50 | 5 |
| Communication Service Availability [# of 9's] | | | | 5 | 6 | 5 | 6 to 8 | 6 to 8 | 4 |
| Survival time   [cycle times] | | | | 2 | 1 | 1 | 1 | 1 | 0 |
| Network reliability (# of nines) | | | | 2 | 3 | 3 | 4 | 4 | 4 |
| Spectrum band | TDD pattern or FDD | Antenna type | Number of gNBs | Maximum number of served users (in UL and DL)[3] | | | | | |
| FDD 2.1 GHz (2x 20 MHz BW, SCS 30 kHz) | – | Omni | 12 | 550 | 1000 | 24 | 21 | 106 | 137 |
| | | AAS (4x4) | 12 | 1300 | 2550 | 317 | 288 | 1438 | 638 |
| TDD 3.7-3.8 GHz (100 MHz BW, SCS 30 Khz) | DU14 | Omni | 12 | 1750 | 3000 | 391 | 587 | 2937 | 700 |
| | DU14 | AAS (4x4) | 12 | 3850 | 7600 | 940 | 1900 | 9500 | 1900 |
| | DDDSU | Omni | 12 | 700 | 1200 | 24 | 19 | 94 | 38 |
| | DDDSU | AAS (4x4) | 12 | 1525 | 3000 | 372 | 737 | 3687 | 538 |
| TDD 26 GHz (400 MHz BW, SCS 120 kHz) | DU14 | AAS (4x4) | 12 | 15000 | 13200 | 3632 | 7375 | 36875 | 7375 |

#### 4.2.5.2   Discussion on the network configuration impact

This chapter capacity analysis for three relevant spectrum bands: an FDD at 2.1 GHz with 2x 20 MHz bandwidth, a TDD band at 3.8 GHz with 100 MHz bandwidth, and a TDD band at 26 GHz with 400 MHz bandwidth. This includes evaluation of 3 different TDD configurations for the TDD frequency bands. The capacity has been conducted for three different deployments in the factory: a single gNB with a distributed antenna system, three gNBs and 12 gNBs. The latter two deployments have been investigated with two different antenna configurations: with omnidirectional antennas and with beamformed transmission based on adaptive antenna systems.

Results show that adaptive antenna systems can improve the capacity significantly over omnidirectional antennas by providing better link performance and reducing the inter-cell interference in the network. Network densification brings further capacity gains. It is demonstrated that the TDD pattern may limit the capacity if it does not match the symmetric traffic load of the industrial use cases. For example, it is shown that the FDD band with 2x 20 MHz can provide a similar capacity as a 100 MHz mid-band TDD system that uses the typical DDDSU TDD pattern.

---

[3] Maximum number of served users may also be affected by specific scheduler implementation.

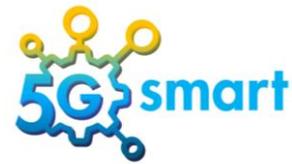

Also, it has been shown that it is beneficial to choose the TDD pattern considering latency requirements and capacity needs of the industrial service, and not to opt by default for the most latency-oriented configuration, especially if service allows for a larger latency bound.

In networks with omni-directional antennas uplink capacity for 3.8 GHz with TDD DDDSU pattern is extremely poor. This can be fixed with the use of denser deployment with AAS cells. Network deployments with AAS in general demonstrate the best performance compared to other antennas both in absolute values and in terms of positive gain reached through densification. Average growth of DL and UL capacity is 230% and 250% respectively when the number of cells is increased from 3 to 12.

In general, results presented in this chapter show that all use cases can be supported with sufficient capacity.

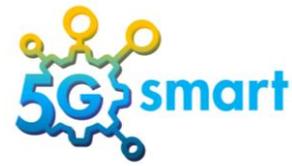

# 5    Coexistence Solutions

5G radio transmission is based on electromagnetic waves that are created at the transmitter and are then propagating through the transmission medium like e.g. air. It is desirable that a strong signal is received at receiver to achieve reliable communication at high date rates. However, the signal also propagates to other locations where other equipment may be located and may there create interference. Co-existence describes the impact of such interference on another system, or the mutual interference among multiple systems on each other. In this chapter we investigate two different co-existence scenarios.  It builds on our earlier work in [5GS20-D14].

If a 5G non-public network is deployed locally in a factory, another wide-area 5G network in the same channel may be used outside the factory for e.g. public network services. These two separate 5G networks may create mutual interference on each other. We have evaluated the impact that a wide-area 5G network has on industrial 5G services in the local non-public network, as presented in section 5.1.

In a factory, some equipment may be used for e.g. high-accuracy measurements, and this operation may be potentially interfered by emitted radio signals of a 5G network. The impact of the radio signal on the equipment operation is also denoted as electromagnetic compatibility (EMC). In section 5.2 we investigate if it is generally possible, to shape the 5G transmission by means of beamforming, in order to ensure that the received signal at a predefined location (micro-exclusion zone) have very low energy. The investigation is made for the 5G downlink transmission, where the isolation from an operating network within a micro-exclusion zone will be achieved via radio network deployment planning and a proper design of the codebook of precoders.

## 5.1    Empirical coexistence measurements between a local and a wide-are 5G network

In this section, we describe how the performance of an indoor non-public network in a factory is impacted due to a coexisting outdoor (public) network operating in the same spectrum. In this section, we only focus on the cochannel scenarios where coexisting networks use the same spectrum. Adjacent channel coexistence aspects are not covered in this section. We describe our measurement methodology and present the key findings of our detailed empirical study on the cochannel coexistence behavior.

### 5.1.1    Coexistence Scenarios

Non-public networks (NPNs) deployed locally for indoor industrial applications may receive interference from other networks operating on the same spectrum in the neighborhood. There are different coexistence scenarios for a local industrial 5G network, as e.g., described and investigated in [5GS20-D14]. In this evaluation we focus on coexistence between a local industrial 5G network and a neighboring outdoor network, for which we provide an evaluation based on measurements in a corresponding trial setup. Furthermore, we focus on co-channel coexistence, which means that the local network and the interfering outdoor network use the same frequency band. Practical coexistence situations depend on the way networks are deployed, which in turn depends on how access to the spectrum is regulated for local deployments.

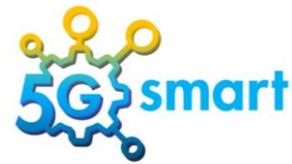

The scenarios that are addressed with our measurement-based coexistence study are depicted in Figure 29. A mobile network operator (MNO) may have a national or regional 5G license and use this license to provide public 5G services in the wide-area; in addition, the MNO may provide non-public network services to industries locally. This scenario is depicted in the option (A) of Figure 29, indicating the potential interference between the public network and local NPN by a red arrow. In several countries, certain frequency bands have been reserved for non-public 5G deployments via local licensing. In this case, there are no public wide-area 5G networks using the same channel and creating interference to the local NPN that has the dedicated local license. However, other local NPNs may exist in vicinity to the NPN with their own separate local licenses, all using the same spectrum. This scenario is depicted in option (B) in Figure 29. In contrast to the scenario (A) on the left of Figure 29, the likelihood of an interfering neighboring network is significantly lower. For example, 184 local spectrum licenses at 3.7-3.8 GHz have been assigned in Germany between 2019 and October 2021[4], which cover only a marginal fraction of the approximate 357.000 square kilometer area of Germany. Hence, when deploying a local 5G network, the likelihood for co-channel interference may appear to be small. However, it can be expected that local spectrum licenses are clustered, around e.g., industrial areas, which may still lead to neighboring local NPNs with the corresponding coexistence scenarios.

Some countries combine the allocation of a spectrum band to an MNO for public use via a national spectrum license with the possibility for local industries to obtain spectrum usage rights for building their own local networks. This can be done by an obligation to the national spectrum license to locally lease spectrum to industries who want to deploy a local NPN (e.g., in Denmark). In other countries, the regulator provides a secondary local license for a local area to the industry, in alignment with the primary licensee (e.g., in UK). Such cases correspond to option (A) in Figure 29.

However, coexistence scenarios can also occur internally within a local campus network, as depicted in option (C) in Figure 29. For example, a local spectrum license may include multiple factory buildings but also an outdoor section on the industry premises. The outdoor network requires close coordination with other outdoor networks on the same or adjacent channels, which means it needs to apply a TDD configuration that is aligned with the public 5G networks in the same frequency bands on the same or adjacent channels. However, for the indoor deployment in the factory another 5G network configuration may be desirable that is optimized to the indoor use cases. An example of such a local campus network with indoor and outdoor deployments is the 5G Industry Campus Europe in Aachen[5], which includes the 5G-SMART trial network where our coexistence measurements were performed.

Our coexistence measurements are made in the TDD spectrum band at 3.7-3.8 GHz, which is part of the European harmonized 5G band at 3.4-3.8 GHz. In other countries also other bands in the TDD mid-band range from 2 to 6 GHz are in focus for local spectrum allocations [Eri21]. Our coexistence studies should be generally relevant for local 5G usage in the TDD mid-band spectrum.

---

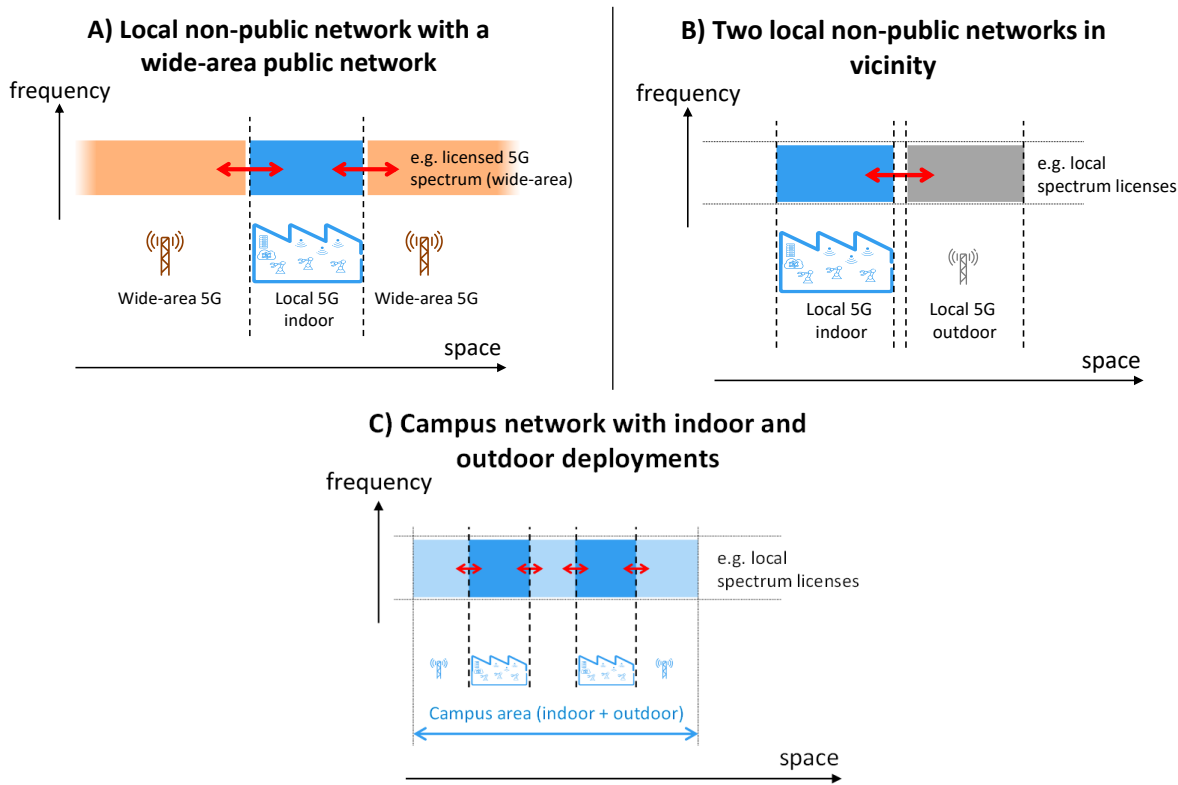

Figure 29: Coexistence scenarios for local non-public network with other networks operating in the same spectrum. The red arrow indicates the investigated interference situations.

### 5.1.2    Interference Situations

Figure 30 shows the various interference situations that arise due to the spectrum coexistence issue in TDD spectrum. The indoor non-public network base station may receive interference from the base station as well as the UEs of the outdoor networks. The indoor non-public network UEs may also receive interference from the outdoor base station as well as the UEs of the outdoor non-public network. The coexistence problem is also bidirectional; the indoor non-public network UEs and base station may also inflict interference to outdoor network UEs and potentially to the outdoor network base station. Figure 30 illustrates the different interference situations with red arrows. When different TDD patterns, i.e., downlink (DL) and uplink (UL) transmission slots, are configured in the indoor non-public network and the outdoor network, this can create additional mutual interference. This is shown by the dotted red arrows in Figure 30 and is known as cross-link interference. The case when DL slots in one network interfere with the DL slots in the other network or when the UL slots of two networks interfere, is known as near-far interference, represented by solid red arrows in Figure 30.

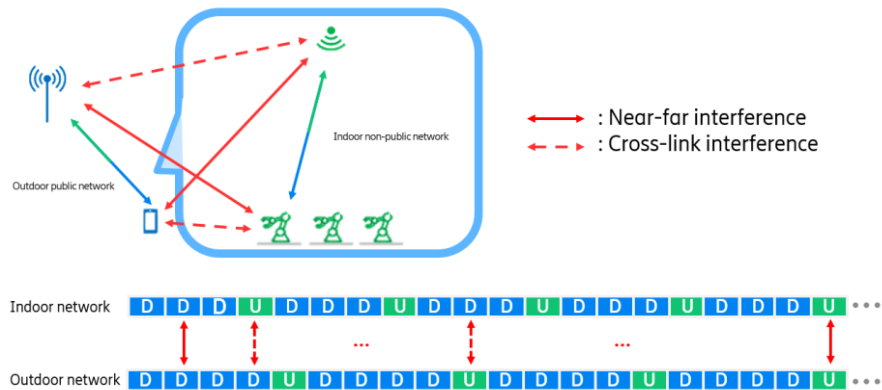

Figure 30: Interference situation among the base stations and UEs of the indoor non-public network and the outdoor network. When a different TDD pattern is used in the non-public network indoors, besides the near-far interference, cross link interference also exists.

### 5.1.3    5G Industry Campus Europe Deployment

The 5G-SMART empirical cochannel coexistence study has been carried out at the 5G Industry Campus Europe in Aachen where 100 MHz of system bandwidth has been allocated by German Spectrum Regulatory Authority for industrial use in the local $3.7 - 3.8$ GHz 5G N78 TDD band [AUC-ALLO]. The outdoor network deployment covers an area of approximately 1 sq. km and uses four outdoor sites shown by yellowish dots with an arrow in Figure 31. The arrow points in the direction of the macro radio antenna panel deployment. The indoor non-public network involved in the coexistence studies is the network deployed at the 5G-SMART trial site, inside the Fraunhofer IPT (Institute for Production Technology) shown as a green rectangle in Figure 31 with an area of approximately 2700 sq. m. All the outdoor sites are configured to use a transmit power level of 5 W, while the indoor network in the Fraunhofer IPT shopfloor uses a transmit power level of 900 mW.

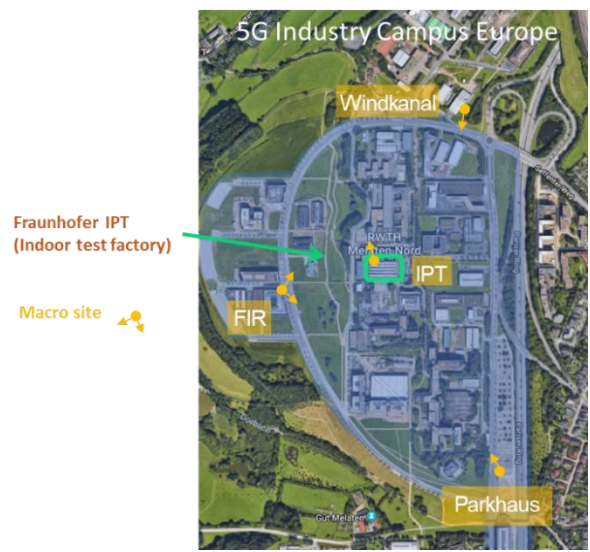

Figure 31: 5G Industry Campus Europe in Aachen where the coexistence measurements have been carried out. The network uses 100 MHz system bandwidth in the 5G N78 TDD band.

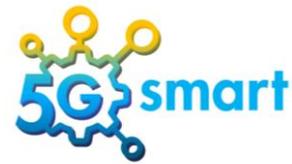

### 5.1.4    Measurement Methodology and Setup Details

In the empirical study, we analyzed the following:

- Effects due to the test UEs location for the indoor and outdoor networks
- The amount of traffic load in the indoor network in UL, DL and both UL/DL directions
- The level of interference coming to the indoor network from the outdoor network in UL, DL and both UL/DL
- The impact of different TDD patterns (DL/UL split) selected indoors. Please note that we did not change the TDD pattern (DDDDU)[6] in the outdoor network as it was kept harmonized according to the regulations with the rest of the MNOs in Germany.

The above combinations lead to several different study scenarios. For each scenario, we have first established a baseline indoor network performance when the outdoor network is made inactive. For each baseline test and coexistence performance test, we have obtained a large measurement sample trace. Our overall empirical data gathering amounts to a duration of multiple days. In these measurements, we have used large packet sizes of about 1400 bytes UDP payloads with high transmission rates compared to our other studies focusing on pure latency measurements from the industrial automation traffic perspective. As a result, we expect to see queuing and processing effects both in the device side and in the network leading to slightly increased latency values.

In the empirical study, we had the opportunity to use both the indoor and outdoor networks dedicatedly and exclusively for coexistence measurements, and we could configure the networks as desired. For the duration of the testing, we further ensured that no other devices were active in the network besides the devices used in the empirical study. The on-premise setup allowed us to use the local breakout connection. We used one test UE for the indoor network and one test UE for the outdoor network. The test UEs were based on the Qualcomm x55 modem chipset. We restricted handover of the UEs for the measurements so that "indoor" and "outdoor" UEs remained attached to their respective cells. The measurement setup is shown in Figure 32 and Figure 33.

---

[6] TDD configurations typically have a *special slot* (S) when the DL transmission changes to UL transmission, e.g. a common TDD pattern is DDDSU. However, since most OFDM symbols in the special slot considered here are assigned to DL transmission, we simplify this in our description as a downlink slot, in this case as DDDDU.

<table>
| (A) | (B) | (C) |
| --- | --- | --- |
| Outdoor UE next to window of the factory shopfloor | NPN UE on the shopfloor at the window | NPN UE on the shopfloor with some distance to the window |
</table>

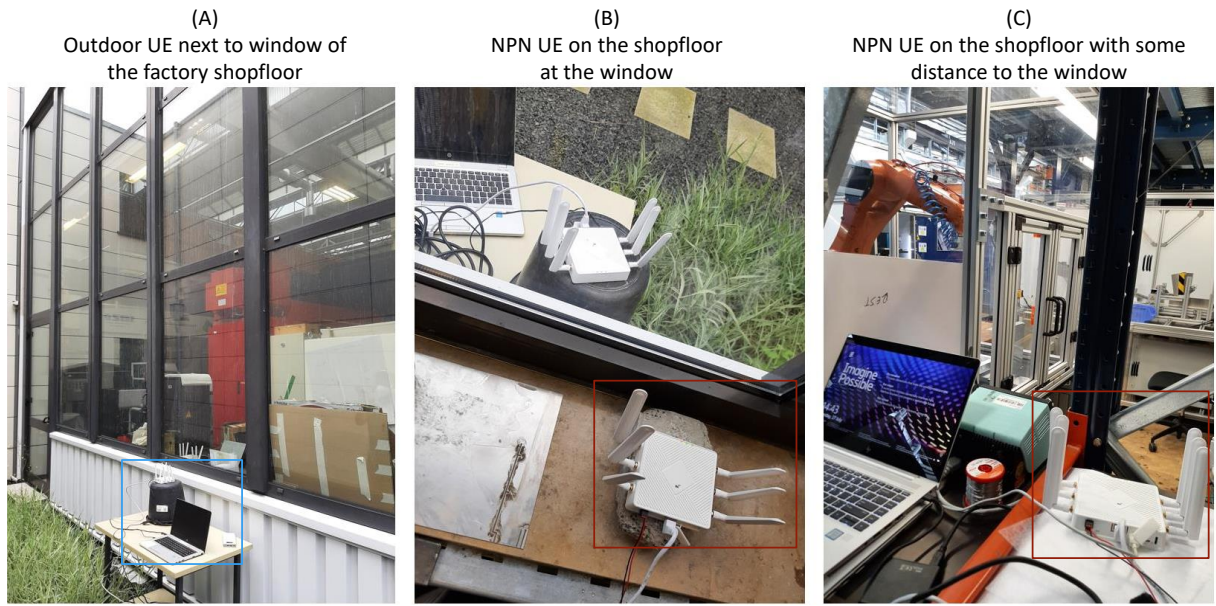

Figure 32:  UEs and their locations for coexistence measurements.

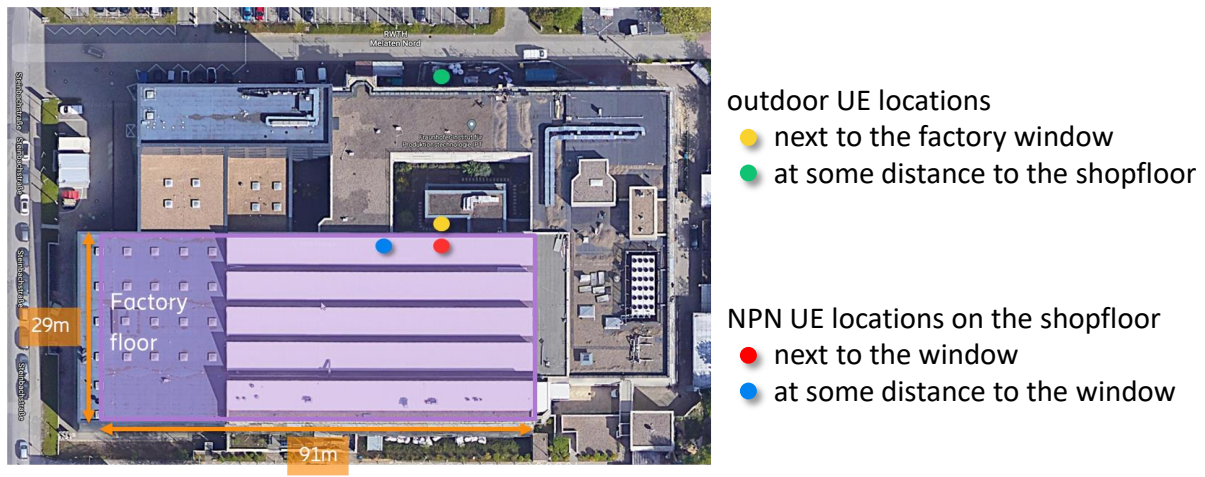

Figure 33:  Map of UE locations for coexistence measurements.

As per the 3GPP Technical Specifications 38.401, Section 9.1 [TR38.401], the indoor and outdoor networks have a common clock reference. We used a synchronized TDD pattern start for the indoor and outdoor networks in our measurements. The indoor and outdoor networks are configured to use the same subcarrier spacing of 30 kHz and have the same slot boundaries and duration.

In order to systematically carry out coexistence measurements, we have developed a tool for traffic generation and for carrying out accurate one-way (DL/UL) latency and throughput measurements. In the cochannel coexistence measurements, we studied the impact on the performance of the indoor network in the presence of the coexisting outdoor network using the same spectrum. In other words, our study quantifies the baseline performance of the indoor network and compare it with performance in the presence of the coexisting outdoor network.

An example of a measurement configuration is shown in Figure 34. In this scenario the two networks use the same TDD pattern DDDDU. The interference in the outdoor network is configured to be present only in the DL direction, i.e., only the DL slots are occupied in the outdoor network. The example scenario depicts a situation when the indoor network generates traffic load only in the DL direction. In this case, only near-far interference is present, and all DL slots of the indoor network can be interfered by the outdoor network. In another setup, it is possible to configure the outdoor network to create interference only in UL direction and measure the impact of the interference. Similarly, for indoor UL transmissions the setup can be configured in order to measure relative impacts from the different transmission directions of the outdoor network.

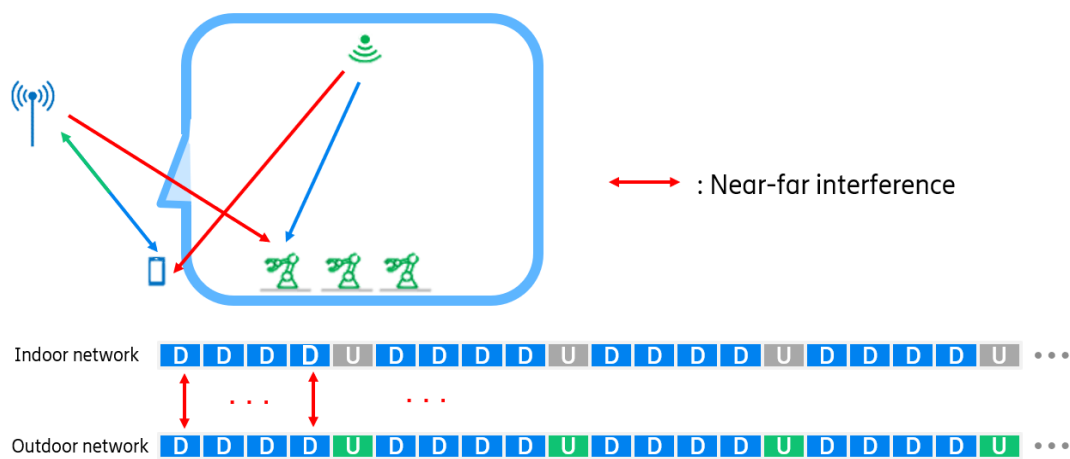

Figure 34: An example scenario where the same TDD pattern is used in the indoor and outdoor networks. In this measurement configuration, only DL transmission is carried out in the indoor network in the presence of only DL interference from the outdoor network.

### 5.1.5    Coexistence Study Results

In the following, we present our key results. In particular, we analyze the impact on the throughput and latency of the indoor network due to the coexisting cochannel outdoor network. Besides the mean value for latency, we consider the 99th percentile and the 99.9th percentile values because the typical industrial application scenarios considered in 5G-SMART have higher reliability targets for the given low latency requirements.

### 5.1.5.1    Scenario with Strong Interference

In order to study the case for an extremely strong interference from the outdoor public network, we placed the outdoor public network UE just outside the shopfloor while the indoor non-public network UE was placed inside the shopfloor such that the inter-UE distance was 1 m, as shown in the left two pictures of Figure 32, and the red and yellow dots in Figure 33. In order to create a worst-case scenario that maximizes the impact of interference on the non-public indoor network, the two UEs were merely separated by a normal double glass window inducing a signal loss of about 8 dB, as observed from Reference Signal Received Power (RSRP) measurements. Moreover, depending on the case, the outdoor network UE and/or outdoor network base station may generate as much as possible interfering transmissions in this scenario by configuring high traffic loads.

We observed that with high interference in DL and UL directions from the outdoor network, the throughput and latency are affected for the indoor non-public network UE. The throughput degradation in the heavily interfered scenario is not so pronounced, we observed a degradation in the baseline throughput of ca. 5% in medium traffic load conditions for both DL and UL. In contrast, the impact on the latency is visible compared to the baseline latency performance.

Figure 35 shows that when the outdoor network creates interference to the indoor network and uses the same TDD pattern, there is almost no impact on the median latency values for both UL and DL transmissions of the non-public indoor network. However, we observe an increase of the 99[th] percentile and the 99.9[th] percentile values for DL latency by 1 ms and 5.2 ms respectively in the presence of heavy DL interference. Having the two UEs in close proximity naturally causes DL-DL interference. We also observed an increase in the UL latency for the non-public indoor network by 5.3 ms and 7.6 ms for the 99[th] and the 99.9[th] percentile values in the presence of heavy UL interference. Part of the UL-UL interference is attributed to the poor beamforming capability at the UEs.[08]

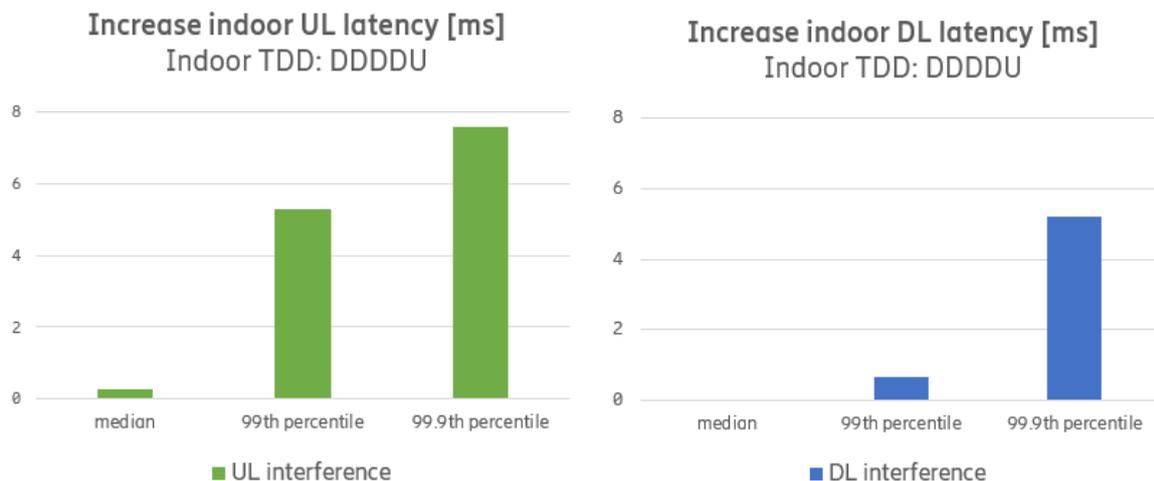

Figure 35: Increase in UL and DL latency of the indoor non-public network when adding heavily interfering transmissions in the outdoor public network at very close distance. Both the networks are using the same TDD pattern, DDDDU.

The increase in latency is because of some HARQ retransmissions caused by interference compared to the baseline case without any outdoor transmissions. In general, the impact of outdoor network interference on the indoor network latency performance is small but can lead to some latency increase in rare cases. We expect that by configuring Radio Access Network (RAN) with more robust link adaption parameter settings for the indoor network, the impact due to a coexisting outdoor network can be reduced. This aspect is left for further studies.

### 5.1.5.2   Effect of the TDD Patterns

Outdoor networks using TDD spectrum bands are generally required to use the same harmonized TDD pattern to avoid interference to/from other outdoor networks, including public 5G networks. In Germany, the fixed TDD pattern of DDDDU is used in outdoor 5G networks in the 3.4-3.8 GHz spectrum band, which is also used in the outdoor component of our trial network. For indoor non-public networks different TDD patterns are possible, and we configured the same and different TDD patterns as in the outdoor network. Using the same TDD pattern for the indoor non-public network inflicts only the near-far interference while restricting the indoor network configuration to the exact DL/UL split of the outdoor network. With a different TDD pattern used in the NPN, not only near-far interference but also cross-link interference is present (cf. Figure 30). The ratio and location of the DL slots versus the UL slots in the TDD patterns determine how much near-far or cross-link interference is present from the outdoor public network using the TDD pattern DDDDU. In our studies we have investigated two TDD patterns for the indoor non-public network: DDDDUDDDUU and DDDU.

When the TDD pattern in the outdoor network is DDDDU and the indoor TDD pattern for the non-public network is DDDDUDDDUU, for UL transmission in the indoor network theoretically up to 34% of the slots have cross-link interference while up to 66% of the symbols have near-far interference as can be seen in Figure 36 below. One specific characteristic of this indoor TDD pattern is that there is never an UL slot of the outdoor network during a DL slot of the indoor network. As a consequence, cross-link interference from the outdoor UE to the indoor UE can be avoided during indoor DL transmissions. Instead, the NPN is only experiencing cross link interference at the base station for the UL transmissions, in addition to the normal near-far interference, see Figure 36.

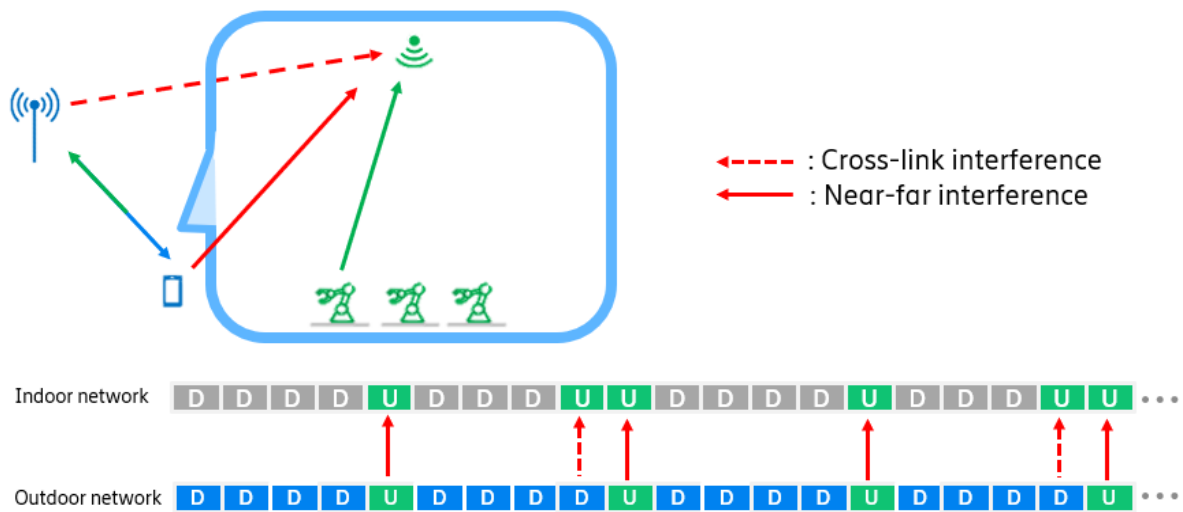

Figure 36: Scenario where different TDD pattern is used in the indoor and outdoor networks. Only UL transmission is carried out in the indoor network in the presence of DL and UL interference from the outdoor network.

Figure 37 shows how the latency in the indoor NPN is impacted by an outdoor UE with high traffic load. We assume the outdoor UE to be located directly at the window of the factory shopfloor (see options (A) and (B) in Figure 32, and the yellow point in Figure 33) and observe the performance of an NPN UE on the inside of the window at a distance of around 1 m from the outdoor UE. It can be seen that the median latency of the NPN UE is not affected in DL, but an increase of around 1.7-1.8 ms of the median UL latency can be seen. For the 99[th] percentile and 99.9[th] percentile, the UL latency increases by 4.6-5.4 ms and 5.7-7.3 ms respectively due to the close by interfering UE, depending on if the interference is caused by the outdoor DL or UL transmissions. For the DL performance of the NPN UE, the 99[th] and 99.9[th] percentiles of the DL latency increase by 3.6 ms and 5.9 ms compared to the baseline situation without outdoor network traffic respectively. Moreover, for the TDD pattern DDDDUDDDUU, one can see that the non-public indoor DL latency performance in case of UL interference does not suffer. This can be explained by that the outdoor UE is not creating any interference on the indoor DL transmission as explained above.

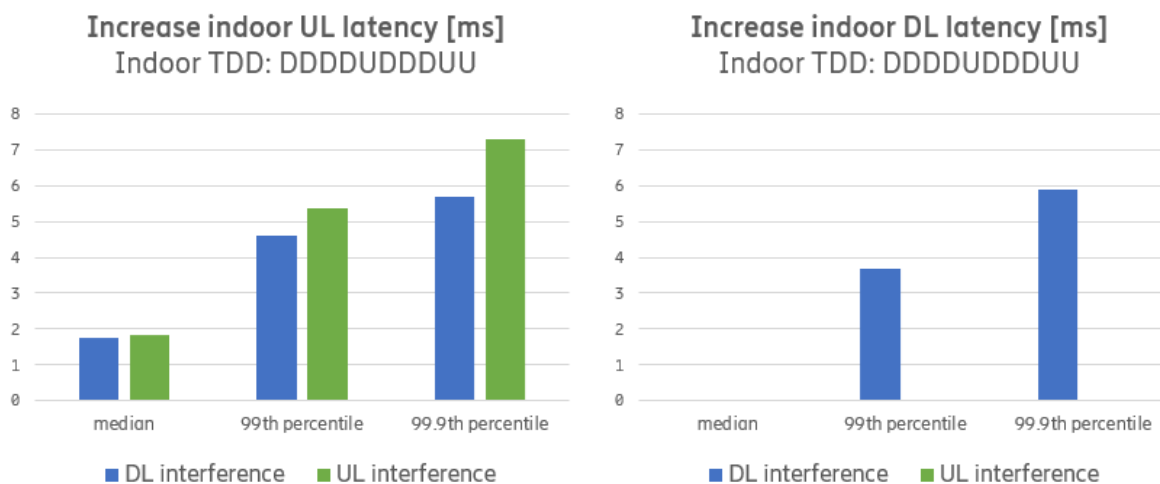

Figure 37: Increase in UL and DL latency of the indoor non-public network when adding heavily interfering transmissions in the outdoor public network at very close distance. The indoor non-public network is configured to use the TDD pattern DDDDUDDDUU while the outdoor public network uses the TDD pattern DDDDU.

When applying the TDD pattern DDDU to the indoor NPN with the outdoor TDD pattern DDDDU, the indoor NPN transmissions are arbitrarily exposed to all four constellations of the near-far and cross-link interference as depicted in Figure 30. While the interference constellations are similar for the two unsynchronized TDD patterns DDDU and DDDDUDDDUU, there is one significant difference. For the DDDU pattern, the outdoor UL transmission can interfere with the DL transmission of the indoor NPN. In Figure 30 we see that for indoor DL transmissions with DDDU, 80% of the DL slots have near-far interference (during outdoor downlink transmissions) and the remaining 20% have cross-link interference (during outdoor uplink transmissions). The result of this cross-link interference on the indoor DL can be seen from Figure 38. Even if the median value of the indoor DL transmission in the NPN is barely affected, very large latency increases of more than 300 ms can occur in rare events due to UL transmission in the outdoor network. A clear difference can be seen to the DDDDUDDDUU pattern in Figure 37, where no DL latency impact of outdoor UL transmissions is noticed. As a result, there is a clear benefit to select wisely an indoor TDD pattern that is aligned to the outdoor TDD pattern so that outdoor UL interference to the indoor device downlink transmissions can be avoided. In order to help mitigate such impact from UL interference, it is also useful to fine-tune the RAN parameters (for instance, link robustness) or use different already existing techniques to leverage the indoor DL slots suffering from near-far interference (DL interference).

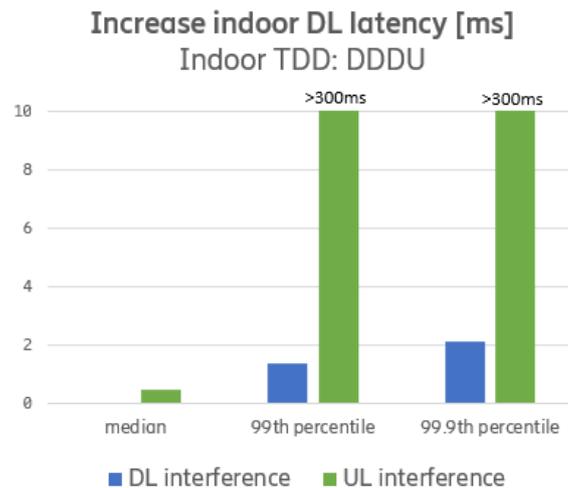

Figure 38: Increase in DL latency of the indoor non-public network UE when adding heavily interfering transmissions in the outdoor public network at very close distance. The indoor non-public network is configured to use the TDD pattern DDDU while the outdoor public network uses the TDD pattern DDDDU.

### 5.1.5.3  Effect of the Test UEs Deployment Locations

Our experiments so far have been focused on device locations where a) the outdoor UE is only ~50 cm away from the window of the factory shopfloor, and b) the interfered NPN UE is located at the inside wall of the factory right at the window. This setup creates an extreme case example of what interference the communication in the outdoor network can create on a 5G device in the shopfloor. However, these assumptions are less likely to exist in practical deployment scenarios. Locations outside the window of factories are typically not the hotspots for mobile broadband users. Furthermore, the area around a factory hall is typically only accessible for personnel. Also, the production equipment – robots, AGVs, machines, production lines – are typically not placed at the outer wall at the window. In this section we consider how the location of either the outdoor UE or the indoor NPN UE impact the interference between the networks. To this end, we have considered four different scenarios of placement of the UEs:

1. Worst case interference with both UEs at the window (less likely)
   - The interfering outdoor UE is located some ~50 cm away from the factory shopfloor, just outside the window. This corresponds to the yellow dot in Figure 33 and can be seen in options (A) and (B) in Figure 32.
   - The interfered indoor NPN UE is located at the edge of the shopfloor at the outer wall, some ~50 cm away from window and ~1 m away from the interfering outdoor UE. This corresponds to the red dot in Figure 33 and can be seen in option (B) in Figure 32.
   This scenario has been used in the measurements in Sections 5.1.5.1 and 5.1.5.2. It provides an unlikely worst case scenario of what impact the interference may have.

2. Indoor NPN UE on the shopfloor with outdoor UE at the window (i.e., worst case location)
   - The interfering outdoor UE is located some ~50 cm away from the factory shopfloor, just outside the window. This corresponds to the yellow dot in Figure 33 and can be seen in options (A) and (B) in Figure 32.

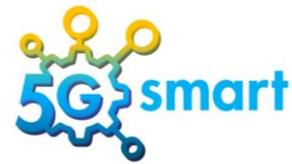

- The interfered indoor NPN UE is located at a robot cell on the shopfloor at some 10 m distance to the window where the outdoor interfering UE is located. This corresponds to the blue dot in Figure 33 and can be seen in option (C) in Figure 32.

3. Indoor NPN UE at the window (i.e., worst case location) and the outdoor UE is at a parking area.
   - The interfering outdoor UE is located at a parking area next to the factory building which is some 15 m away from the factory wall. This corresponds to the green dot in Figure 33.
   - The interfered indoor NPN UE is located at the edge of the shopfloor at the outer wall, some ~50 cm away from window. This corresponds to the red dot in Figure 33 and can be seen in option (B) in Figure 32.

4. Indoor NPN UE on the shopfloor and the outdoor UE is at a parking area.
   - The interfering outdoor UE is located at a parking area next to the factory building which is some 15 m away from the factory wall. This corresponds to the green dot in Figure 33.
   - The interfered indoor NPN UE is located at a robot cell located on the shopfloor at some 10 m distance to the window where the outdoor interfering UE is located. This corresponds to the blue dot in Figure 33 and can be seen in option (C) in Figure 32.

   This scenario corresponds to the most realistic interference scenario in a practical situation. The impact of the interference from the outdoor network on the indoor NPN communication is the lowest of all the four scenarios. We have not investigated this scenario explicitly, since already in scenario 3 the impact of the interference becomes negligible, as shown in Figure 39.

Figure 39 shows the increase in UL latency in the indoor NPN caused by the outdoor interference on the left. The strongest UL latency increase in Sections 5.1.5.1 and 5.1.5.2 occurred with the indoor TDD configurations DDDDU and DDDDDUDDDUU, where the former had a slightly stronger impact on the high percentile values. When the outdoor UE remains next to the factory window and the indoor UE is moved some meters into the shopfloor area, the outdoor UE still creates interference to the indoor NPN base station which increases the UL delay for the NPN UE. When the outdoor UE is moved away from the factory window to the parking area, the impact of interference on the UL transmissions on the shopfloor disappears.

The increase in DL latency for the indoor NPN transmissions due to the interference of the outdoor network can be seen in Figure 39 on the right. The largest latency increase in Sections 5.1.5.1 and 5.1.5.2 occurred for an indoor DDDU TDD configuration. The main source of interference in this setup is the UL transmission of the outdoor UE interfering the DL transmission at the indoor NPN UE. It can be seen that if either of the two UEs is moved away from the window (the indoor NPN UE to the robot cell, or the outdoor UE to the parking area), the impact on DL latency in the factory in principle disappears. When we moved the indoor NPN UE some 10 m away from the window to a robot cell, the interference power from the outdoor UE at the window decreased by approximately 5 dB.

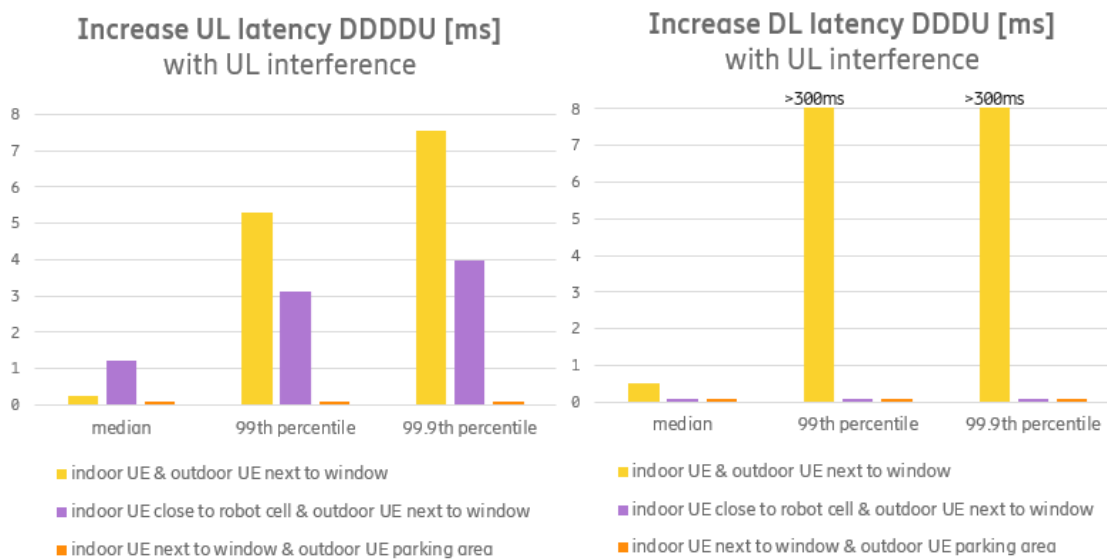

Figure 39: Increase of UL and DL latency compared to an interference-free baseline scenario, for the three different deployment options: (1. left bar) the indoor non-public network UE and the outdoor UE are placed next to the window; (2. middle bar) the indoor non-public network UE is placed in a robot cell and the outdoor UE is at the window; (3. right bar) the indoor non-public UE is located at the window and the outdoor UE is in a parking area. The performance is evaluated for the indoor TDD configurations which caused the largest latency increase for worst case deployment (1.), which is DDDU for UL and DDDDU for DL.

### 5.1.6    Conclusions

Our empirical results give quantitative insight into the performance impact of co-channel coexisting networks. We evaluated the performance of an indoor non-public network using different TDD patterns, different UE locations in the outdoor and non-public networks, different co-channel interference levels from the outdoor network in UL, DL and both directions. Our main conclusions are the following.

There can be scenarios when the transmission in an outdoor network in close proximity to an indoor NPN can create interference to the indoor network that increases the latency of the indoor NPN. Latency increases are very small for the median transmission latency but may impact the 99[th] and 99.9[th] percentile of the latency distribution.

Such interference can in particular appear in a practically uncommon situation: when an outdoor UE is located directly next to the factory, like just outside an unshielded factory window. For the NPN UL this creates increased transmission latencies for both the cases that the indoor TDD pattern is synchronized or unsynchronized to the outdoor network. As the outdoor UE moves away from the factory wall, the UL interference becomes negligible. For the NPN DL, the latency increase appears only when both the outdoor UE and the indoor UE are in close proximity, like the outside and the inside of a window in the factory wall. In case of a TDD-unsynchronized indoor NPN, the short distance to the interfering outdoor UE can create very large increase for the 99[th] and 99.9[th] percentile of the DL latency. However, this latency increase can be avoided, if the unsynchronized TDD pattern in the indoor network has only UL transmission slots when the outdoor network has UL slots, for instance a DDDDUDDDUU pattern indoors for an outdoor pattern of DDDDU. If either of the outdoor or indoor

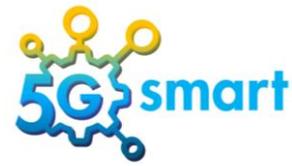

UEs move away from the window and the distance between them increases, the impact on UL latency in the NPN disappears.

## 5.2    Micro-exclusion zone analysis

### 5.2.1    Motivation

For the deployment of 5G networks in industry, in addition to ensuring full coverage, some constraints could arise depending on the characteristics of the scenario and of the manufacturing work undertaken. For example, if the production environment is sensitive to radiation, electromagnetic compatibility (EMC) levels must be assessed to ensure that they do not interfere with the production process. This is the case of the Bosch semiconductor factory in Reutlingen. In this factory, there is an area dedicated to the injection of tracks to the wafers being fabricated. This procedure requires adequate electromagnetic isolation, since if an electromagnetic signal with sufficient power reaches the area, it could dope the PN junctions and affect the final functionality of the product. Therefore, one option is to create a micro-exclusion zone within the factory, where power levels are maintained below a set threshold.

As it is well known, the radio channel affects any wireless transmission. In order to offer a solution to the aforementioned problem, the first action is to know the channel as precisely as possible in the scenario in consideration.

### 5.2.2    Channel modeling

#### 5.2.2.1    Deterministic Channel modeling

Deterministic models are based on a detailed description of the propagation environment and antenna configuration. The purpose of deterministic models is to recreate the geometry and electromagnetic parameters of propagation environments such as streets, buildings, trees, offices, desks, etc. It is for this reason that deterministic models are geometrically and computationally complex, however, they are the most accurate models. For example, ray-tracing is a pure deterministic model that considers the electromagnetic properties of materials, the geometry and positioning of all the elements of the scenario. It is an accurate model with detailed angular information and is suitable for the exhaustive estimation of channel variables for both large scale (LS) and small scale (SS) parameters.

#### 5.2.2.2    Ray-tracing based channel modeling

UPV's ray-tracing tool has been developed by iTEAM group to provide a highly accurate channel model. Its processing is based on both GPU and CPU units and optimizes threads to improve computational cost and processing time. This tool is calibrated for the sub 6 GHz and up to 72 GHz bands for both indoor and outdoor scenarios. One of the most important points of the tool is the capability to reproduce an accurate model of the scenario, which allows to calculate propagation phenomena more accurately and reliably. Due to the fact that in industrial scenarios there are a large number of obstacles, mostly metal and also moving objects such as AGVs or operators, calibration campaigns have been carried out and are described below.

#### 5.2.2.3    Diffuse-Scattering Calibration for metal surfaces

For the measurement campaign, the scenario shown in the Figure 43 has been implemented. TX and RX were equipped with a horn antenna, model 22240-20, operating on frequencies from 26 to 40 GHz, with a nominal gain of 20 dBi. The beamwidth at 3 dB is 18 degrees for both the vertical and horizontal planes. They are connected to ports 1 and 2 of the VNA model E8363A. 18 positions have been considered for the TX and 18 positions for the RX, from 0 to 90 degrees for the TX and from 90 to 180 degrees for the RX. Adjacent positions are separated by 5 degrees. No anechoic chamber is used and this could be a source of error. To minimize that error, potential sources of reflected rays were far from the material under test. The nearest surface was the floor, which was at 50 cm from the antennas. In addition, the directivity of the antennas and the use of millimiter-wave (mmW) minimizes the probability of having reflected rays with significant power coming from materials different to the material under test.

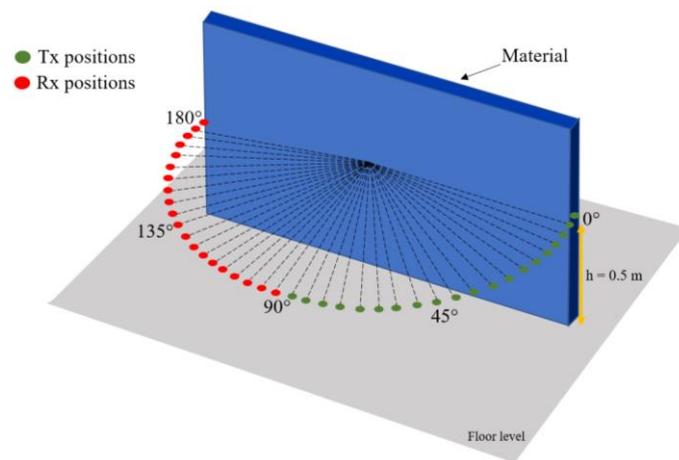

Figure 40 Setup and calibration procedure

For the measuring process, the TX is set at the first measuring point (0 degrees) pointing to the center of the material, at a height of 0.50 m. Then, a sweep is made with the RX in the positions of the semicircle, from 90 to 180 degrees. This procedure is repeated until all the combinations between TX-RX are obtained.

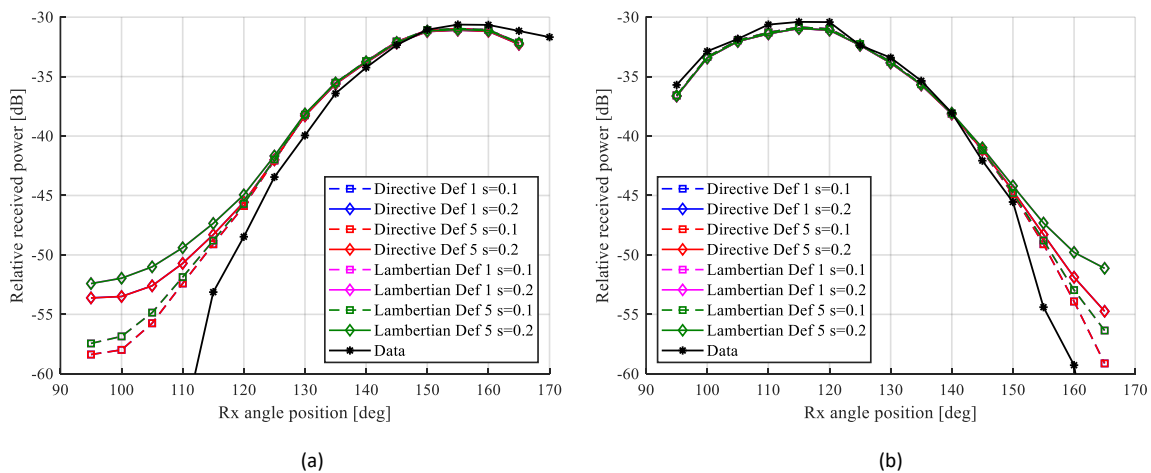

Figure 41 Calibration results for different diffuse scattering models. In (a) the TX is placed at 35º and in (b) TX is placed at 65º

The scattering models defined in [SPO-DIFF] have been implemented. After an exhaustive iteration, the scattering coefficient (S) that best fits the results has been obtained. It can be noted from the Figure 41 that for metal, specular components are the most dominant. In fact, since this material is very reflective, the value of the parameter S that best fits the measurements is between 0 and 0.1 with the directive Def. 5 model. Based on the results obtained, it can be indicated that values of relative received power from the measurement campaign are valid to be used in the industrial scenarios.

### 5.2.2.4  Body blockage calibration

The blockage effect at mmW band has been a reason of concern due to the difficulty to transmit effectively in NLoS conditions at these frequencies. One of the elements that blocks the mmW signals is the human body. In fact, in open-space indoor environments, typically with high human density, it can be the most important blocker. Indeed, it is important to model accurately each blocker but also to consider appropriately the presence of multiple blockers.

In this sense, UPV's ray-tracing tool considers the model proposed in [PRA-BODY]. This body blockage model is based in the 3GPP body blockage model but, additionally, it is complemented with a set of modifications to obtain more realistic results. The model is validated with real measurements.

### 5.2.3  Coverage optimization for micro-exclusion zone

The proposed solution aims to fixed maximum power levels in the exclusion zone, based on the design and optimization of the precoders. To do this, we first need to have a set of test points appropriately located in the exclusion zone. These points will permit us to know the power reaching the zone.  The optimization problem to be solved is a max min problem where the spectral efficiency of the user with lower spectral efficiency is maximized considering power constraints per AP and ensuring that the power received in the exclusion zone is below a certain threshold. Mathematically, it is expressed as follows:

$$SE_k = \log\left(1 + \frac{\sum_{m=1}^{M} \rho_{mk} h_{mk} h_{mk}^*}{\sum_{k' \neq k}^{K} \sum_{m=1}^{M} \rho_{mk'} h_{mk'} h_{mk'}^* + \frac{n}{P_T}}\right)$$

$$\max_{\rho_{mk}} \min_{k=1,\dots,K} SE_k$$

$$s.t. \sum_{k=1}^{K} \rho_{mk} \leq 1, m = 1, \dots, M$$

$$\rho_{mk} \geq 0, k = 1, \dots, K, m = 1, \dots, M$$

$$g_l^* \left(\sum_k Q_k\right) g_l \leq \gamma, \forall l$$

where $K$ is the number of users, $M$ is the number of APs, $SE_k$ is the spectral efficiency for the user $k$, $\rho_{mk}$ is the power coefficient, $h_{mk}$ is the he channel between the user $k^{th}$ and all the APs and $Q_k$ is the covariance channel matrix associated to user $k^{th}$. $L$ is the number of channel estimator located in the exclusion zone, $g_l$ is the channel between the user $l^{th}$ and $\gamma$ is the normalized maximum power allowed in the exclusion zone.

In order to obtain more realistic results in solving the optimization problem described above, a deterministic channel matrix has been used as input. The elements of the channel matrix obtained from a ray tracing - based simulator have the following form:

$$H_{k,m} = \sum_{\alpha=1}^{q} A_\alpha e^{-j} \omega_0 \tau_\alpha$$

where $A_\alpha$ is the amplitude of the $\alpha$th ray, $q$ is the total number of rays, $\tau$ is the time delay of the $\alpha$th ray, and $\omega_0$ is the carrier frequency.

### 5.2.3.1 Evaluation setup

To evaluate this scenario, the Bosch semiconductor factory in Reutlingen has been geometrically reproduced to make channel measurements using a ray-tracing tool considering various aspects that interfere with radio propagation as shown in Figure 42. The propagation environment consists of a semiconductor production plant with several corridors formed by metallic machinery. This type of environment is characterized by high dispersion, specular and diffuse components that enrich the radio channel.

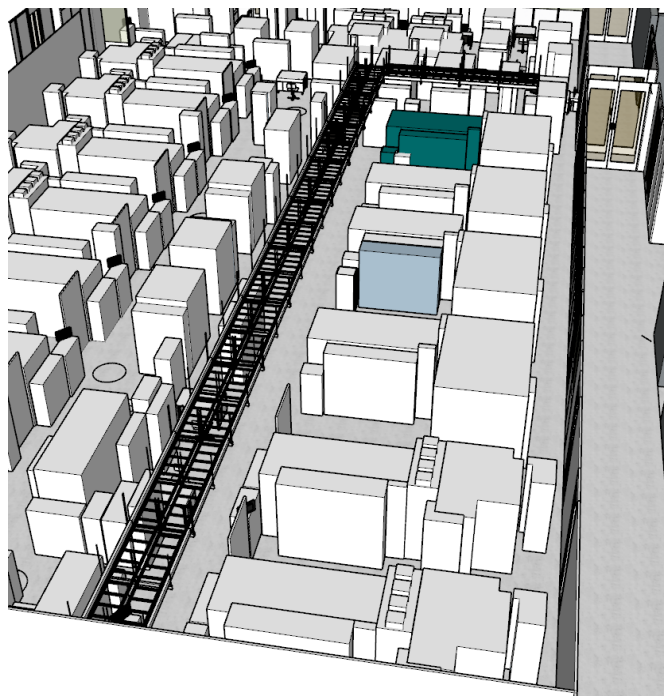

Figure 42 Bosch semiconductor factory in Reutlingen.

For this study, a section of the scenario has been selected and within this section it is required to create an electromagnetic exclusion zone. The dimensions of the selected section are 15 m × 30 m. The scenario height is 5m. The exclusion zone is delimited by a volume of 1 and the top of the volume is at 1.5 m of height. For more details, see Figure 43.

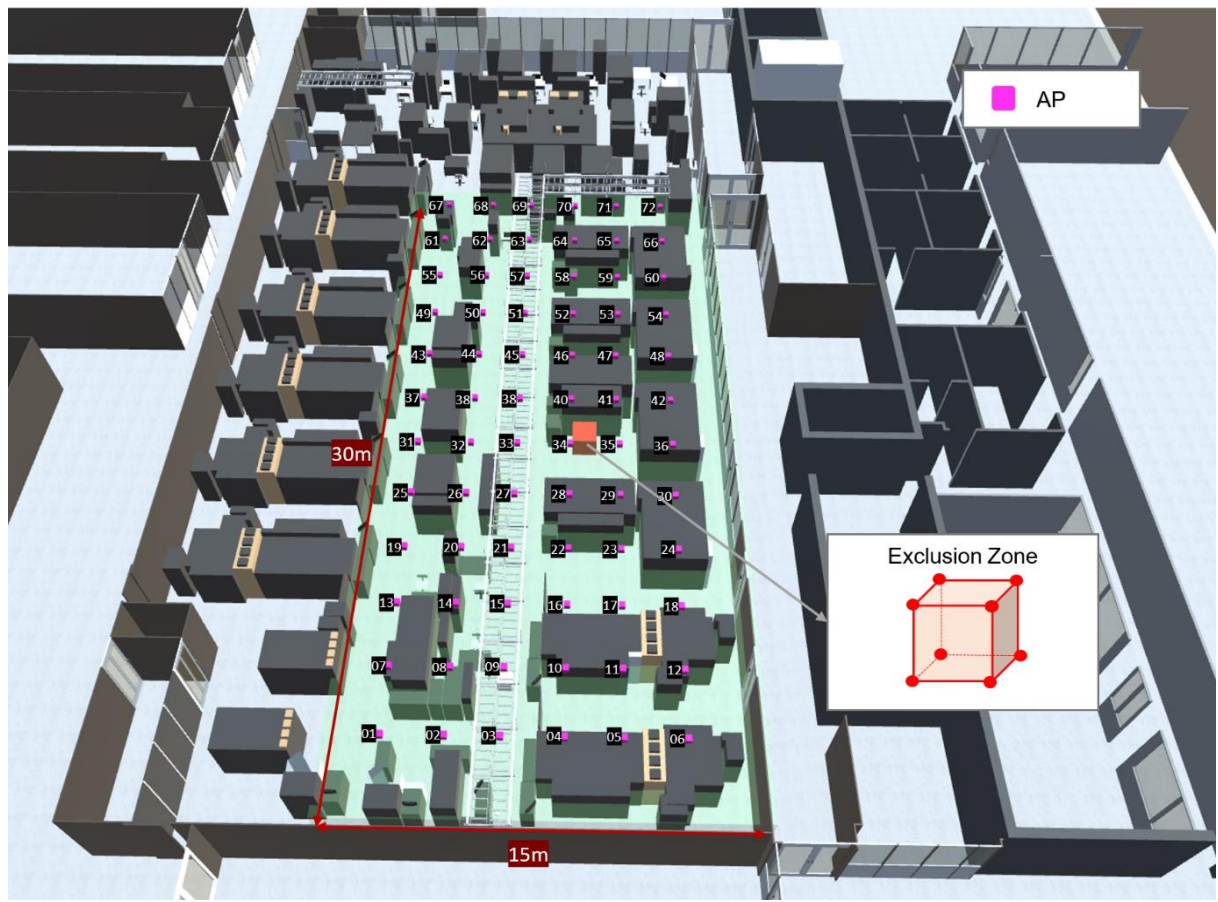

Figure 43 Scenario dimensions.

The deployment consists of 72 single-antenna APs mounted on the ceiling, uniformly distributed on the scenario with a spacing of 2.5 m and connected to a single processing unit, which allows the coordination of the APs. Additionally, there are 8 testing points, each one allocated in each vertex of the exclusion zone volume. Note that this deployment is for simulation purposes only and does not represent the actual factory deployment.

The selected values for the simulation parameters are described in Table 15.

Table 15. Simulation parameters.

| Parameter | Value |
|---|---|
| Number of APs ($M$) | 72 |
| Number of testing points ($L$) | 8 |
| Measurement points ($K$) | 112 |
| Inter-site distance | 2.5 m |
| Scenario dimensions | 30 m × 15 m |
| AP height | 4 m |
| Exclusion zone dimensions | 1 m × 1 m × 1.5 m |
| Frequency | 30 GHz |
| Transmitted power per AP ($P_T$) | 200 mW |
| Power threshold in the exclusion zone ($\gamma$) | -120 dBm |

### 5.2.3.2  Results and analysis

In order to know the starting point, the received power level at each testing point have been obtained considering all contributions (signal coming from the 72 APs) in azimuth and elevation.

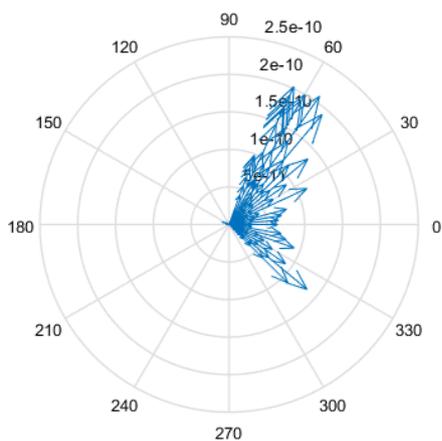

Figure 45 shows the numbering of the points, while Figure 45 show the received power results for testing point 6 as an example of the results of the ray tracing process.

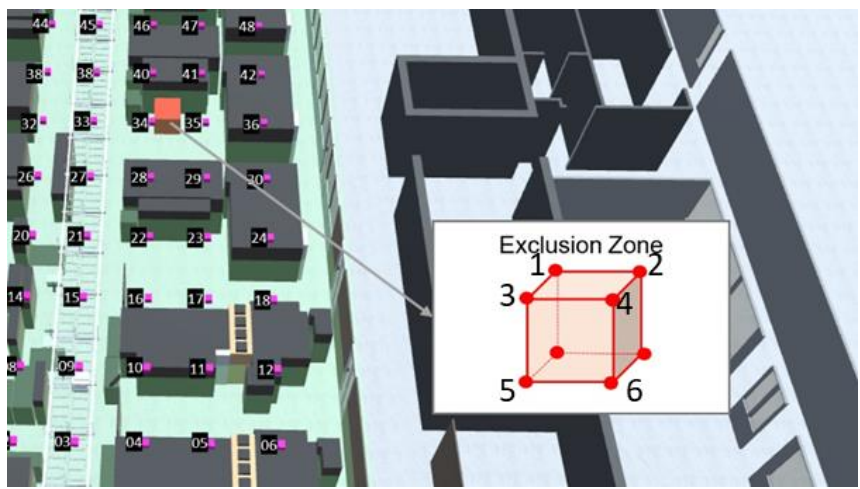

Figure 44 Restriction points - Exclusion Zone

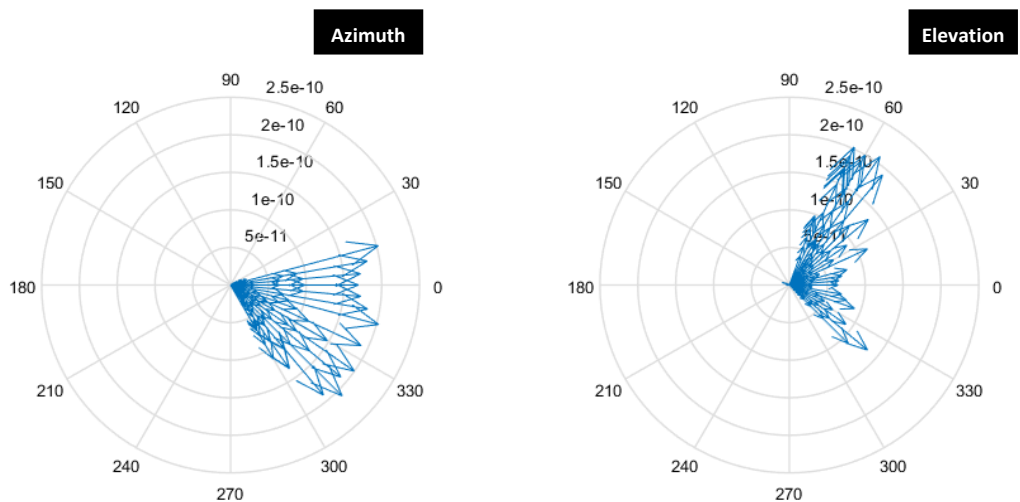

Figure 45 Received power for testing point 6.

The results in terms of SINR and power level at the testing points are shown in Figure 46 and Figure 47. In both figures, the curve in blue is obtained when all APs transmit with maximum power and considering the channel conditions; the curve in red is obtained when considering the optimization problem solution without taking into account the exclusion zone constraint; and finally, the curve in yellow is obtained from the solution of the optimization problem when the exclusion zone constraint is considered.

Figure 45 shows the power levels for the 8 testing points.  The graph shows not only the effect of the optimization but also of the power restriction in the exclusion zone. The power level decreases considerably, about 50 dB from the no optimized to the optimized solution and about 40 dB from the optimized solution to the optimized solution considering the exclusion zone restriction. In order to have a measure of the costs of maintaining certain power levels in the exclusion zone, Figure 47 shows the SINR values. As it can be seen the SINR obtained in each point of the scenario varies from -17 dB to 7 dB when no optimization is considered. This means that the quality of the service perceived by the users will depend on their position within the scenario. However, when the optimization problem is solved, the fairness among users can be guaranteed, i. e., all users will experiment similar levels of SINR. Additionally, when the power constraint for the exclusion zone is considered, the SINR decreases as part of the trade-off between the SINR levels and power levels in the exclusion zone.

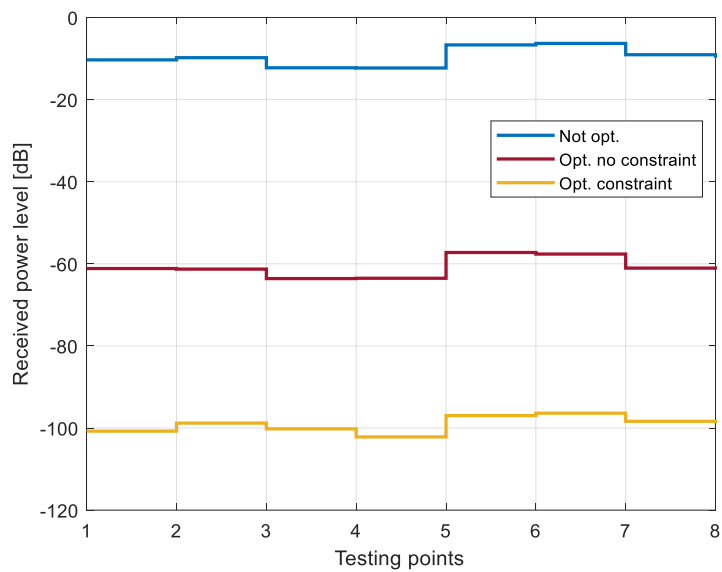

Figure 46 Exclusion zone interference level.

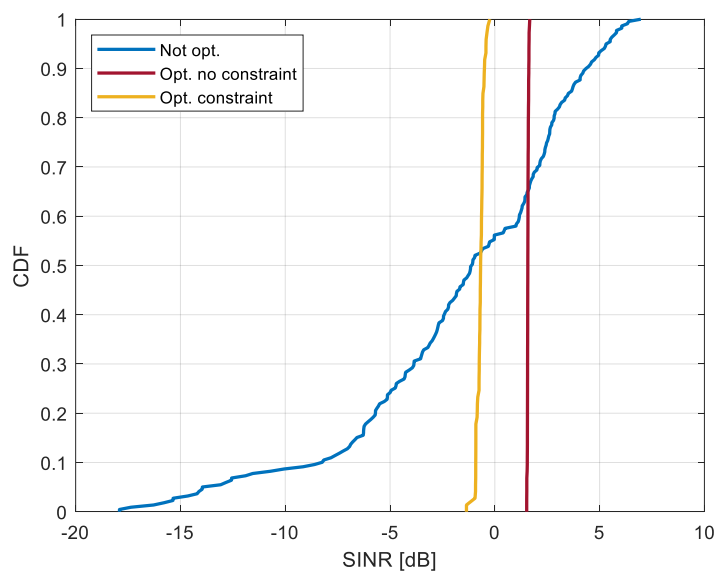

Figure 47 CDF of SINR.

In order to get a clearer understanding of the results before discussed, the heatmaps for the 3D scenario have been obtained. The heatmaps represent the power level in the highlighted region of the scenario. Figure 48 shows the heatmap when all APs transmit with maximum power while Figure 49 and Figure 50 show the heatmaps when the power coefficients are adjusted satisfying the optimization problem described in Section 6.3 without and with the exclusion zone constraint.

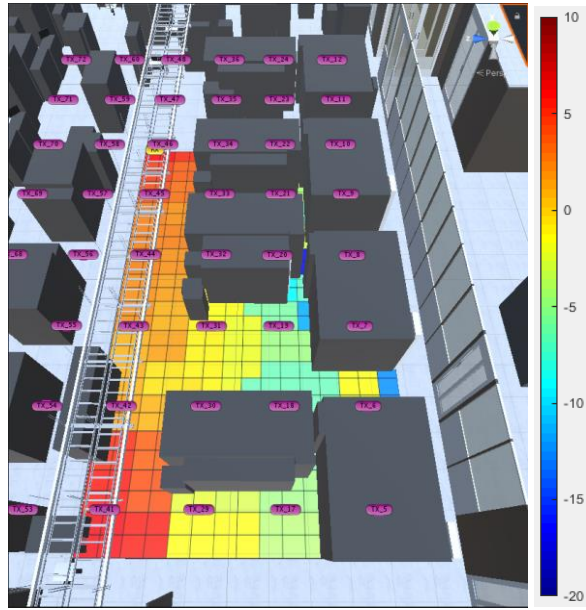

Figure 48 No power control, heatmap of SINR.

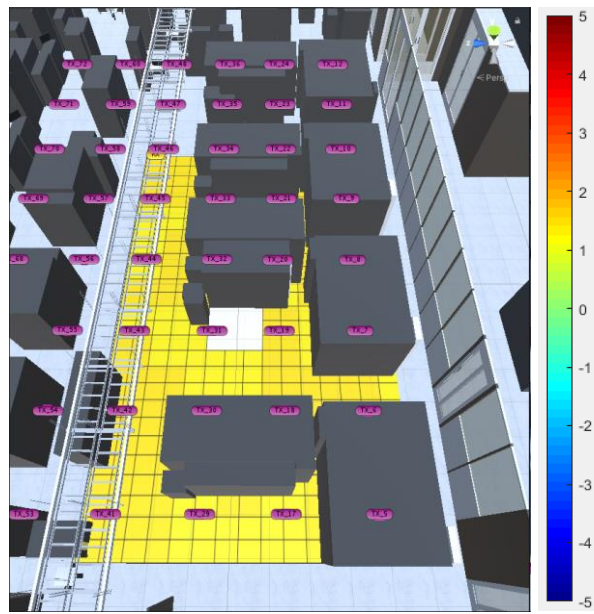

Figure 49 Max-min power control, no exclusion zone power constraint, heatmap of SINR.

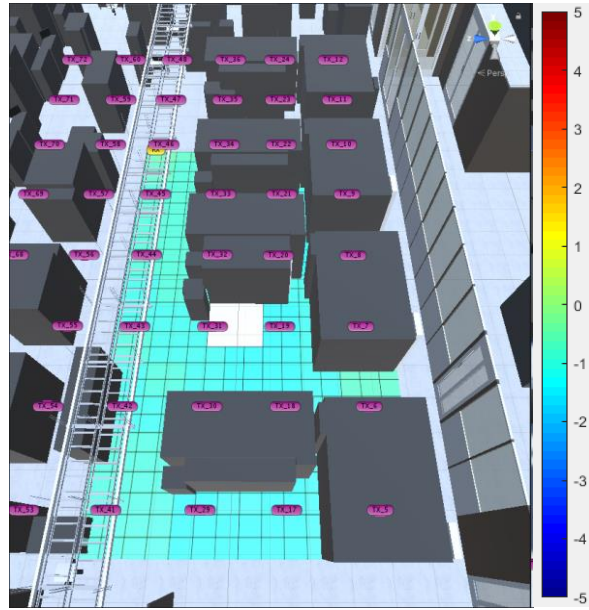

Figure 50 Max-min power control with exclusion zone power constraint, heatmap of SINR.

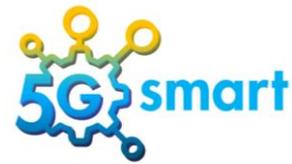

# 6    Conclusions

This deliverable presents evaluations of radio network deployment options of 5G NR in a factory to enable smart manufacturing use cases. Several aspects of the 5G radio access network are considered, like the capabilities of 5G NR regarding latency reduction and mobility enhancements, analysis of performance trade-offs for different KPIs of the industrial network, and coexistence aspects of locally deployed 5G non-public networks with surrounding networks.

In Section 3.1 an overview of 5G NR features for latency reduction is provided. Also, the achievable latencies for different 5G NR configurations are shown. Mainly, delay reduction is achieved using the UL configured grant, a decrease of UE/gNB processing times, as well as denser control signaling that is used together with sub-slot scheduling. The analysis of the air-interface delays for the 3GPP standard baseline, as well as the technology potential with latency optimization features, demonstrates that sub-millisecond delay can be achieved even when data for specific transmission direction arrives at the most inconvenient moment, i.e., with the worst-case alignment delay. The latency calculation model employed for air-interface latency calculation considers one user in the network and therefore does not take into account scheduling delays nor does it account for inter- and intra-cell interference. The impact of such effects would also need to be considered at higher network loads with many users in the system.

Some of the industrial use cases may require UE mobility to be combined with URLLC communication, which demands QoS requirements both in terms of latency and reliability. The analysis presented in Section 3.2 is performed to understand the commonly used mobility mechanisms in 5G networks and describe mobility solutions for URLLC services. First, it is identified that according to 3GPP Release-15 specifications, NR handover causes interruption time in the range of 40-60 milliseconds, for both FDD and TDD modes. Since for URLLC services such large delays are intolerable, several handover enhancements for reducing the interruption time were investigated. One of the enhancements specified in 3GPP is Layer 3 based DAPS (Dual Active Protocol Stack) handover, which is based on a make-before-break handover mechanism, allowing to reach down to zero ms handover interruption time with the cost of higher complexity. Another mechanism which can provide zero ms mobility interruption time is Layer 1/Layer 2 multi-TRP based intra-cell mobility, where the necessity to have several cells is eliminated by extension of cell boundaries for a single cell with the use of multiple transmission-reception points (TRPs). A further mechanism is Layer 3 based CHO (Conditional Handover), which is not aimed for reduction of handover interruption times but improves the robustness of the handover. Therefore, CHO is a useful mechanism in poor radio conditions when communication reliability requirements are demanding. With the use of handover enhancements discussed in section 3.2 chapter, 5G NR can support URLLC industrial services also during mobility.

With help of system level simulations, we have quantified the impact of different network configuration on latency, reliability, and throughput. Results are presented in Section 4.1. It has been observed that the TDD pattern has a direct impact on latency for both DL and UL and that to achieve latencies below 1ms, it is necessary to use the high-bands and SCS 120kHz and symmetric pattern to guarantee bi-directional low latencies. However, for the assumption that the latency target is not so strict, the impact of this latency relaxation up to 10ms has been evaluated on two parameters: the number of users satisfying the requirements per service area and the reliability achieved as a result of the relaxation. We have provided an evaluation of the trade-off between spectrum options, numerology, TDD pattern and number of users on latency and reliability. On the other hand, we have evaluated the impact of reducing the packet sizes while maintaining a constant traffic load in terms of system throughput and latency. It can be concluded that when the packet size decreases, there is a direct cost in latency performance. This behavior can be explained by the frequency diversity and the diversity that occurs in the channel encoder, which can be better exploited by larger packets. Therefore, it is essential to take into account the network load, number of UEs per serving cell and to define the packet size, playing for instance with a higher transmission periodicity, to be able to meet the desired latencies.

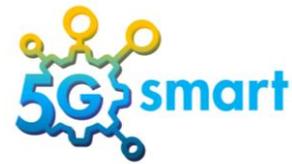

One essential question for a 5G deployment in a factory is the achievable capacity to serve industrial use cases in Section 4.2. We have investigated a number of different industrial use cases from those discussed in chapter 2 and determined the system capacity for a 5G deployment. The number of users that can be served by the network so that the service requirements are fulfilled for all users has been evaluated, and spectral efficiency values have been derived. This capacity analysis has been made for three relevant spectrum bands: an FDD at 2.1 GHz with 2x 20 MHz bandwidth, a TDD band at 3.8 GHz with 100 MHz bandwidth, and a TDD band at 26 GHz with 400 MHz bandwidth. For the TDD bands, different TDD configurations have been evaluated. The capacity has been conducted for three different deployments in the factory: a single gNB with a distributed antenna system, three gNBs and 12 gNBs. The latter two deployments have been investigated with two different antenna configurations: with omnidirectional antennas and with beamformed transmission based on adaptive antenna systems. In general, it is shown that all use cases can be supported with sufficient capacity. Adaptive antenna systems can improve the capacity significantly over omnidirectional antennas by providing better link performance and reducing the inter-cell interference in the network. Network densification brings further capacity gains. It is shown that the TDD pattern may restrict the capacity if it does not match the symmetric traffic load of the industrial use cases. For example, it is shown that the FDD band with 2x 20 MHz can provide a similar capacity as a 100 MHz mid-band TDD system that uses the typically DDDSU TDD pattern.

When a 5G network is locally deployed in a factory, there may exist other 5G networks in the neighboring areas using the same frequency. This leads to coexistence situations that may create interference in between the neighboring networks. We have performed coexistence measurements in the 5G-SMART trial system at 3.7-3.8 GHz at the 5G Industry Campus Europe, and are presented in section 5.1. We show under what conditions and to what level outdoor 5G transmissions in proximity to a factory can impact the 5G transmission within the factory and increase the transmission latencies. We have investigated different locations of indoor and outdoor devices and investigated different TDD patterns for the indoor network, where the outdoor network uses the harmonized outdoor TDD pattern DDDSU for that band in Germany. There can be scenarios when the outdoor transmission can create interference to the indoor network that increases the latency of the indoor NPN. This is the case when the outdoor UE is located at the window to the factory. This can cause a latency increases to the indoor UEs located on the other side near the window. While such an increase is relatively small for the median latency, 99th and 99.9th percentile of the latency are more highly affected in relative percentage. When the indoor network is not TDD-synchronized to the outdoor network, strong interference can occur if an outdoor UE transmits in UL in close proximity to an indoor UE received data in downlink. This situation can be avoided by a smart selection of the indoor TDD pattern, as shown at the example of DDDDUDDDUU. When the outdoor UE is at some distance to the factory, like a parking area outside the building, no latency increase due to interference can be noticed.

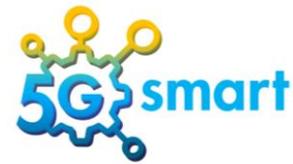

The problem of electromagnetic isolation of a particular zone of a factory network is discussed in Section 5.2 and an optimization solution is proposed. First, the scenario and the exclusion zone were modeled. Subsequently, a cell-free massive MIMO deployment was proposed for the industrial scenario in focus. Then, the deterministic channel estimation was performed by using a ray-tracing based simulator. As a final step, the SINR and the received power in the exclusion zone were calculated for three cases: maximum power transmission by the antenna radio, power coefficients of each transmitter are optimized to maximize the SINR of the scenario point with the lowest SINR (max-min fairness optimization problem) and finally, similar to the previous one but adding a power constraint for the exclusion zone in the optimization problem. It was shown that a good trade-off is possible between the achieved SINR and the power levels in the exclusion zone. In particular, for the configuration evaluated in Section 5.2.3, the power is reduced by around 100 dB while the SINR mean goes from -0.93 dB to -2.84 dB. As it can be intuited, the more restricted power level in the exclusion zone, the lower SINR level will be obtained. It is important to mention that this solution is not applicable to factory networks where guaranteed latency or reliability are required by the industrial service.

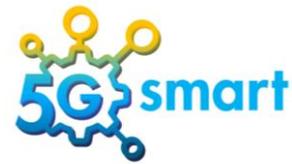

# References

| [5GA19-1] | 5G-ACIA, "5G for Connected Industries and Automation", February 2019, https://5g-acia.org/wp-content/uploads/5G-ACIA_WP_5G-for-Connected-Industries-and-Automation-Second-Edition_SinglePages.pdf |
|---|---|
| [5GA19-2] | 5G-ACIA, "5G for Automation in Industry", July 2019, https://5g-acia.org/wp-content/uploads/5G-ACIA_WP_5G-for-Automation-in-Industry_SinglePages.pdf |
| [5GA20] | 5G-ACIA, "Key 5G Use Cases and Requirements", May 2020, https://5g-acia.org/wp-content/uploads/5G-ACIA_WP_Key-5G-Use-Cases-and-Requirements_SinglePages.pdf |
| [5GS20-D11] | 5G-SMART, Deliverable 1.1, "Forward looking smart manufacturing use cases, requirements and KPIs", June 2020, https://5gsmart.eu/wp-content/uploads/5G-SMART-D1.1.pdf |
| [5GS20-D14] | 5G-SMART, Deliverable 1.4, "Radio Network Deployment Options for Smart Manufacturing", November 2020, https://5gsmart.eu/wp-content/uploads/5G-SMART-D1.4-v1.0.pdf |
| [5GS20-D21] | 5G-SMART, Deliverable 2.1, "Design of 5G-based Testbed for Industrial Robotics", May 2020, https://5gsmart.eu/wp-content/uploads/5G-SMART-D2.1.pdf |
| [5GS20-D32] | 5G-SMART, Deliverable 3.2, "Report on System Design Options for Monitoring of Workpieces and Machines", May 2020, https://5gsmart.eu/wp-content/uploads/5G-SMART-D3.2.pdf |
| [5GS21-D33] | 5G-SMART, Deliverable 3.3, "Report on Implementation of Options for Monitoring of Workpieces and Machines", May 2021, https://5gsmart.eu/wp-content/uploads/5G-SMART-D3.3-v1.0.pdf |
| [5GS21-D43] | 5G-SMART, Deliverable 4.3, "Report on Development of the 5G Use Cases", May 2021, https://5gsmart.eu/wp-content/uploads/5G-SMART-D4.3-v1.0.pdf |
| [5GS20-D52] | 5G-SMART, Deliverable 5.2, "5G network architecture options and assessments", November 2020, https://5gsmart.eu/wp-content/uploads/5G-SMART-D5.2-v1.0.pdf |
| [5GS20-Term] | 5G-SMART, "5G common terminology", June 2020, https://5gsmart.eu/wp-content/uploads/5G-SMART-common-terminology.pdf |
| [Eri2004] | O. Ohlsson, P. Wallentin, C.-G. Persson, "Reducing mobility interruption time in 5G networks", Ericsson blog post, April 2020, https://www.ericsson.com/en/blog/2020/4/reducing-mobility-interruption-time-5g-networks. |
| [Eri2005] | I. L. Da Silva, C. Eklöf, J. Muller, R. Zhohov, "This is the key to mobility robustness in 5G networks", Ericsson blog post, May 2020, https://www.ericsson.com/en/blog/2020/5/the-key-to-mobility-robustness-5g-networks. |
| [Eri21] | M. Norin, R. Högman, M. Buchmayer, G. Lemne, F. Pedersen, A. Zaidi, "5G spectrum for local industrial networks", Ericsson white paper, April 2021, https://www.ericsson.com/en/reports-and-papers/white-papers/5g-spectrum-for-local-industrial-networks. |

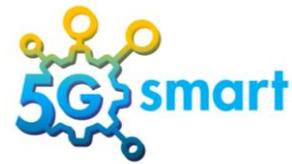

| [SKA18] | A. Shapin, K. Kittichokechai, N. Andgart, M. Sundberg and G. Wikström, "Physical Layer Performance for Low Latency and High Reliability in 5G," 2018 15th International Symposium on Wireless Communication Systems (ISWCS), 2018, pp. 1-6, doi: 10.1109/ISWCS.2018.8491084 |
| --- | --- |
| [PRA-BODY] | Prado-Alvarez, Danaisy, et al. "Millimeter-Wave Human Blockage Model Enhancements for Directional Antennas and Multiple Blockers." IEEE Communications Letters 25.9 (2021): 2776-2780. |
| [RP-210884] | 3GPP, "Reply LS on 3GPP NR Rel-16 URLLC and IIoT performance evaluation", RP-210884, March 2021, https://www.3gpp.org/ftp/TSG_RAN/TSG_RAN/TSGR_91e/Docs/RP-210884.zip |
| [SPO-DIFF] | Mani, Francesco, et al. "Parameterization of a polarimetric diffuse scattering model in indoor environments." IEEE transactions on antennas and propagation 62.8 (2014): 4361-4364. |
| [TR36.881] | 3GPP TR 36.881 version 14.0.0 Release 14, "Study on latency reduction techniques for LTE". |
| [TR38.901] | 3GPP TR 38.901 version 15.0.0 Release 15, "Study on channel model for frequencies from 0.5 to 100 GHz". |
| [TS22.104] | 3GPP TS 22.104 V16.1.0 (2019-03), "Service requirements for cyber-physical control applications in vertical domains; Stage 1 (Release 16)". |
| [TS22.261] | 3GPP TS 22.261 V17.8.1 (2021-09), "Service requirements for the 5G system; Stage 1 (Release 17)". |
| [TS38.331] | 3GPP TS 38.331 version 16.6.0 Release 16, "NR; Radio Resource Control (RRC); protocol specification". |

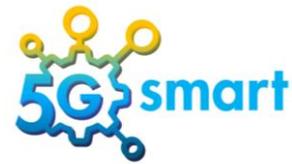

# Appendix

This appendix contains a complete set of capacity results discussed in Section 4.2. Table 16 shows the maximum number of connected users in downlink and uplink independently. Note that for symmetric traffic, the minimum value from DL/UL pair should be taken. Correspondingly, pretty much any DL/UL traffic ratio can be considered, by just applying this proportion to the values presented in the Table 16. Spectral efficiency in bits/second/Hz/cell is shown in Table 17. Color coding of cells in both tables is done from red, which corresponds to the lowest values, to green, which is used to represent the highest values.

Table 16 5G system capacity in number of served uses[3] for the different use cases, depending on the spectrum band and RAN deployment option.

| Band | DL/UL | TDD pattern | Antenna type | Number of gNBs | UC7 (controller-to-controller) [Packet size: 500 Bytes, Latency: 10 ms, Reliability 99%] 0.4 Mbps | 22.104 mobile robots (1) - machine control [Packet size: 250 Bytes, Latency: 10 ms, Reliability 99.9%] 0.2 Mbps | UC4 (process monitoring) [Packet size: 1024 Bytes, Latency: 10 ms, Reliability 99.9%] 1.6384 Mbps | 22.104 (controller-to-controller) (1) [Packet size: 1000 Bytes, Latency: 10 ms, Reliability 99.99%] 0.8 Mbps | 22.104 (controller-to-controller) (2) [Packet size: 1000 Bytes, Latency: 50 ms, Reliability 99.99%] 0.16 Mbps | UC1 (robotics motion planning) [Packet size: 500 Bytes, Latency: 5 ms, Reliability 99.99%] 0.8 Mbps |
|---|---|---|---|---|---|---|---|---|---|---|
| FDD 2.1 GHz (20 MHz BW, SCS 30 kHz) | DL | FDD | DAS 12 antennas | 1 | 175 | 350 | 43 | 87 | 437 | 87 |
| | | | Omni | 3 | 350 | 600 | 79 | 150 | 750 | 150 |
| | | | AAS (4x4) | 3 | 450 | 850 | 110 | 137 | 687 | 213 |
| | | | Omni | 12 | 725 | 1250 | 79 | 100 | 500 | 200 |
| | | | AAS (4x48) | 12 | 1525 | 2950 | 330 | 288 | 1438 | 750 |
| | UL | FDD | DAS 12 antennas | 1 | 138 | 275 | 34 | 69 | 344 | 69 |
| | | | Omni | 3 | 275 | 500 | 18 | 19 | 94 | 100 |
| | | | AAS (4x4) | 3 | 375 | 700 | 92 | 175 | 875 | 175 |
| | | | Omni | 12 | 550 | 1000 | 24 | 21 | 106 | 137 |
| | | | AAS (4x4) | 12 | 1300 | 2550 | 317 | 625 | 3125 | 638 |
| TDD 3.7-3.8 GHz (100 MHz BW, SCS 30 kHz) | DL | DU7 | DAS 12 antennas | 1 | 425 | 850 | 104 | 213 | 1063 | 213 |
| | | | Omni | 3 | 875 | 1550 | 201 | 387 | 1937 | 375 |
| | | | AAS (4x4) | 3 | 1100 | 2150 | 269 | 550 | 2750 | 538 |
| | | | Omni | 12 | 1825 | 3100 | 269 | 258 | 1250 | 525 |
| | | | AAS (4x4) | 12 | 3775 | 7300 | 916 | 1862 | 9312 | 1838 |
| | | DU14 | DAS 12 antennas | 1 | 475 | 900 | 116 | 237 | 1187 | 237 |
| | | | Omni | 3 | 950 | 1650 | 220 | 425 | 2125 | 400 |
| | | | AAS (4x4) | 3 | 1200 | 2350 | 293 | 600 | 3000 | 587 |
| | | | Omni | 12 | 2000 | 3400 | 391 | 587 | 2937 | 763 |
| | | | AAS (4x4) | 12 | 4100 | 7950 | 1001 | 2025 | 10125 | 2000 |
| | | DDDSU | DAS 12 antennas | 1 | 750 | 1500 | 183 | 375 | 1875 | 375 |
| | | | Omni | 3 | 1550 | 2700 | 360 | 675 | 3375 | 650 |
| | | | AAS (4x4) | 3 | 1925 | 3750 | 478 | 950 | 4750 | 958 |
| | | | Omni | 12 | 3200 | 5450 | 720 | 1263 | 6313 | 1225 |
| | | | AAS (4x4) | 12 | 6575 | 12750 | 1599 | 3225 | 16125 | 3200 |
| | UL | DU7 | DAS 12 antennas | 1 | 425 | 850 | 104 | 213 | 1063 | 213 |
| | | | Omni | 3 | 850 | 1500 | 195 | 338 | 1688 | 363 |
| | | | AAS (4x4) | 3 | 1100 | 2150 | 269 | 538 | 2688 | 538 |
| | | | Omni | 12 | 1750 | 3000 | 250 | 437 | 2187 | 700 |
| | | | AAS (4x4) | 12 | 3875 | 7550 | 934 | 1900 | 9500 | 1900 |
| | | DU14 | DAS 12 antennas | 1 | 425 | 850 | 104 | 213 | 1063 | 213 |
| | | | Omni | 3 | 850 | 1500 | 195 | 375 | 1875 | 363 |
| | | | AAS (4x4) | 3 | 1100 | 2150 | 269 | 538 | 2688 | 538 |
| | | | Omni | 12 | 1750 | 3000 | 397 | 688 | 3438 | 700 |
| | | | AAS (4x4) | 12 | 3850 | 7600 | 940 | 1900 | 9500 | 1900 |
| | | DDDSU | DAS 12 antennas | 1 | 150 | 300 | 37 | 75 | 406 | 81 |
| | | | Omni | 3 | 325 | 600 | 18 | 25 | 125 | 38 |
| | | | AAS (4x4) | 3 | 425 | 850 | 104 | 213 | 1063 | 163 |
| | | | Omni | 12 | 700 | 1200 | 24 | 19 | 94 | 50 |
| | | | AAS (4x4) | 12 | 1525 | 3000 | 372 | 737 | 3687 | 538 |
| TDD 26 GHz (400 MHz BW, SCS 120 kHz) | DL | DU14 | DAS 12 antennas | 1 | 1825 | 3600 | 452 | 925 | 4625 | 912 |
| | | | Omni | 3 | 3750 | 1700 | 867 | 1637 | 8187 | 1575 |
| | | | AAS (4x4) | 3 | 4750 | 5550 | 1160 | 2363 | 11813 | 2338 |
| | | | Omni | 12 | 7750 | 2900 | 1215 | 2300 | 11500 | 2213 |
| | | | AAS (4x4) | 12 | 16125 | 16500 | 3967 | 8000 | 40000 | 7875 |
| | | DDDSU | DAS 12 antennas | 1 | 2925 | 5000 | 720 | 1475 | 7375 | 1463 |
| | | | Omni | 3 | 6000 | 2700 | 1392 | 2625 | 13125 | 2525 |
| | | | AAS (4x4) | 3 | 7500 | 8000 | 1831 | 3750 | 18750 | 3738 |
| | | | Omni | 12 | 12625 | 5300 | 2000 | 3687 | 18437 | 3550 |
| | | | AAS (4x4) | 12 | 20000 | 26500 | 4883 | 10000 | 50000 | 10000 |
| | UL | DU14 | DAS 12 antennas | 1 | 1675 | 1750 | 409 | 837 | 4187 | 837 |
| | | | Omni | 3 | 3325 | 1400 | 763 | 1437 | 7187 | 1380 |
| | | | AAS (4x4) | 3 | 4325 | 3700 | 1044 | 2100 | 10500 | 2087 |
| | | | Omni | 12 | 6875 | 2650 | 1514 | 2350 | 11750 | 2762 |
| | | | AAS (4x4) | 12 | 15000 | 13200 | 3632 | 7375 | 36875 | 7375 |
| | | DDDSU | DAS 12 antennas | 1 | 650 | 700 | 159 | 325 | 1625 | 325 |
| | | | Omni | 3 | 1325 | 550 | 128 | 50 | 250 | 550 |
| | | | AAS (4x4) | 3 | 1725 | 1500 | 415 | 638 | 3188 | 825 |
| | | | Omni | 12 | 2750 | 1050 | 183 | 163 | 813 | 863 |
| | | | AAS (4x4) | 12 | 6025 | 5150 | 1459 | 2950 | 14750 | 2950 |

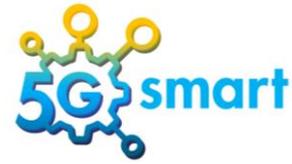

Table 17 5G system spectral efficiency for the different use cases, depending on the spectrum band and RAN deployment option.

| Band | DL/UL | TDD pattern | Antenna type | Number of gNBs | UC7 (controller-to-controller) [Packet size: 500 Bytes, Latency: 10 ms, Reliability 99%] 0.4 Mbps | 22.184 mobile robots (1) – machine control [Packet size: 250 Bytes, Latency: 10 ms, Reliability 99.9%] 0.2 Mbps | UC4 (process monitoring) [Packet size: 1024 Bytes, Latency: 10 ms, Reliability 99.9%] 1.6384 Mbps | 22.184 (controller-to-controller) (1) [Packet size: 1000 Bytes, Latency: 10 ms, Reliability 99.99%] 0.8 Mbps | 22.184 (controller-to-controller) (2) [Packet size: 1000 Bytes, Latency: 50 ms, Reliability 99.99%] 0.16 Mbps | UC1 (robotics motion planning) [Packet size: 500 Bytes, Latency: 5 ms, Reliability 99.99%] 0.8 Mbps |
|---|---|---|---|---|---|---|---|---|---|---|
| | | | | | Spectral efficiency [bits/s/Hz/cell] | | | | | |
| FDD 2.1 GHz (20 MHz BW, SCS 30 kHz) | DL | FDD | DAS 12 antennas | 1 | 3.50 | 3.50 | 3.50 | 3.50 | 3.50 | 3.50 |
| | | | Omni | 3 | 2.33 | 2.00 | 2.17 | 2.00 | 2.00 | 2.00 |
| | | | AAS (4x4) | 3 | 3.00 | 2.83 | 3.00 | 1.83 | 1.83 | 2.83 |
| | | | Omni | 12 | 1.21 | 1.04 | 0.54 | 0.33 | 0.33 | 0.67 |
| | | | AAS (4x4) | 12 | 2.54 | 2.46 | 2.25 | 0.96 | 0.96 | 2.50 |
| | UL | FDD | DAS 12 antennas | 1 | 2.75 | 2.75 | 2.75 | 2.75 | 2.75 | 2.75 |
| | | | Omni | 3 | 1.83 | 1.67 | 0.50 | 0.25 | 0.25 | 1.33 |
| | | | AAS (4x4) | 3 | 2.50 | 2.33 | 2.33 | 2.33 | 2.33 | 2.33 |
| | | | Omni | 12 | 0.92 | 0.83 | 0.17 | 0.07 | 0.07 | 0.42 |
| | | | AAS (4x4) | 12 | 2.17 | 2.13 | 2.17 | 2.08 | 2.08 | 2.13 |
| TDD 3.7-3.8 GHz (100 MHz BW, SCS 30 kHz) | DL | DU7 | DAS 12 antennas | 1 | 3.40 | 1.70 | 1.70 | 1.70 | 1.70 | 3.40 |
| | | | Omni | 3 | 2.33 | 1.03 | 1.10 | 1.03 | 1.03 | 2.00 |
| | | | AAS (4x4) | 3 | 2.93 | 1.43 | 1.47 | 1.43 | 1.43 | 2.87 |
| | | | Omni | 12 | 1.22 | 0.52 | 0.57 | 0.17 | 0.17 | 0.70 |
| | | | AAS (4x4) | 12 | 2.52 | 1.22 | 1.25 | 1.24 | 1.24 | 2.45 |
| | | DU14 | DAS 12 antennas | 1 | 3.80 | 1.90 | 1.90 | 1.90 | 1.90 | 3.80 |
| | | | Omni | 3 | 2.53 | 1.10 | 1.20 | 1.13 | 1.13 | 2.13 |
| | | | AAS (4x4) | 3 | 3.20 | 1.57 | 1.60 | 1.60 | 1.60 | 3.13 |
| | | | Omni | 12 | 1.33 | 0.57 | 0.53 | 0.39 | 0.39 | 1.02 |
| | | | AAS (4x4) | 12 | 2.73 | 1.33 | 1.37 | 1.35 | 1.35 | 2.67 |
| | | DDDSU | DAS 12 antennas | 1 | 3.75 | 3.00 | 3.00 | 3.00 | 3.00 | 3.75 |
| | | | Omni | 3 | 2.58 | 1.80 | 1.97 | 1.80 | 1.80 | 2.17 |
| | | | AAS (4x4) | 3 | 3.21 | 2.50 | 2.57 | 2.53 | 2.53 | 3.17 |
| | | | Omni | 12 | 1.33 | 0.91 | 0.88 | 0.84 | 0.84 | 1.02 |
| | | | AAS (4x4) | 12 | 2.74 | 2.13 | 2.18 | 2.15 | 2.15 | 2.67 |
| | UL | DU7 | DAS 12 antennas | 1 | 3.40 | 1.70 | 1.70 | 1.70 | 1.70 | 3.40 |
| | | | Omni | 3 | 2.27 | 1.00 | 1.07 | 0.90 | 0.90 | 1.93 |
| | | | AAS (4x4) | 3 | 2.93 | 1.43 | 1.47 | 1.43 | 1.43 | 2.87 |
| | | | Omni | 12 | 1.17 | 0.50 | 0.54 | 0.29 | 0.29 | 0.93 |
| | | | AAS (4x4) | 12 | 2.58 | 1.26 | 1.28 | 1.27 | 1.27 | 2.53 |
| | | DU14 | DAS 12 antennas | 1 | 3.40 | 1.70 | 1.70 | 1.70 | 1.70 | 3.40 |
| | | | Omni | 3 | 2.27 | 1.00 | 1.07 | 1.00 | 1.00 | 1.93 |
| | | | AAS (4x4) | 3 | 2.93 | 1.43 | 1.47 | 1.43 | 1.43 | 2.87 |
| | | | Omni | 12 | 1.17 | 0.50 | 0.54 | 0.46 | 0.46 | 0.93 |
| | | | AAS (4x4) | 12 | 2.57 | 1.27 | 1.28 | 1.27 | 1.27 | 2.53 |
| | | DDDSU | DAS 12 antennas | 1 | 3.00 | 0.60 | 0.60 | 0.60 | 0.65 | 3.25 |
| | | | Omni | 3 | 2.17 | 0.40 | 0.10 | 0.08 | 0.07 | 0.50 |
| | | | AAS (4x4) | 3 | 2.83 | 0.57 | 0.57 | 0.57 | 0.57 | 2.17 |
| | | | Omni | 12 | 1.17 | 0.20 | 0.03 | 0.01 | 0.01 | 0.13 |
| | | | AAS (4x4) | 12 | 2.54 | 0.50 | 0.51 | 0.49 | 0.49 | 1.79 |
| TDD 26 GHz (400 MHz BW, SCS 120 kHz) | DL | DU14 | DAS 12 antennas | 1 | 3.65 | 1.85 | 1.85 | 1.85 | 1.85 | 3.65 |
| | | | Omni | 3 | 2.50 | 0.28 | 1.18 | 1.09 | 1.09 | 2.10 |
| | | | AAS (4x4) | 3 | 3.17 | 0.93 | 1.58 | 1.58 | 1.58 | 3.12 |
| | | | Omni | 12 | 1.29 | 0.12 | 0.41 | 0.38 | 0.38 | 0.74 |
| | | | AAS (4x4) | 12 | 2.69 | 0.69 | 1.35 | 1.33 | 1.33 | 2.63 |
| | | DDDSU | DAS 12 antennas | 1 | 3.66 | 2.90 | 2.95 | 2.95 | 2.95 | 3.66 |
| | | | Omni | 3 | 2.50 | 0.45 | 1.90 | 1.75 | 1.75 | 2.10 |
| | | | AAS (4x4) | 3 | 3.13 | 1.47 | 2.50 | 2.50 | 2.50 | 3.11 |
| | | | Omni | 12 | 1.52 | 0.22 | 0.96 | 0.61 | 0.61 | 0.74 |
| | | | AAS (4x4) | 12 | 2.08 | 1.10 | 1.67 | 1.67 | 1.67 | 2.08 |
| | UL | DU14 | DAS 12 antennas | 1 | 3.35 | 0.87 | 1.67 | 1.67 | 1.67 | 3.35 |
| | | | Omni | 3 | 2.22 | 0.25 | 1.04 | 0.96 | 0.96 | 1.85 |
| | | | AAS (4x4) | 3 | 2.88 | 0.62 | 1.43 | 1.40 | 1.40 | 2.78 |
| | | | Omni | 12 | 1.15 | 0.11 | 0.52 | 0.59 | 0.59 | 0.92 |
| | | | AAS (4x4) | 12 | 2.50 | 0.55 | 1.24 | 1.23 | 1.23 | 2.46 |
| | | DDDSU | DAS 12 antennas | 1 | 3.25 | 0.35 | 0.65 | 0.65 | 0.65 | 3.25 |
| | | | Omni | 3 | 2.21 | 0.09 | 0.18 | 0.03 | 0.03 | 1.83 |
| | | | AAS (4x4) | 3 | 2.88 | 0.25 | 0.57 | 0.43 | 0.43 | 2.75 |
| | | | Omni | 12 | 1.15 | 0.04 | 0.06 | 0.03 | 0.03 | 0.72 |
| | | | AAS (4x4) | 12 | 2.51 | 0.21 | 0.50 | 0.49 | 0.49 | 2.46 |

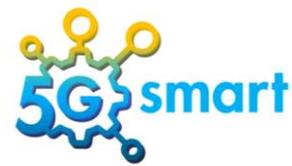

## List of acronyms

| | |
|---|---|
| 3D | Three-Dimensional |
| 3GPP | Third Generation Partnership Project |
| 4G | Fourth Generation Mobile Network |
| 5G | Fifth Generation Mobile Network |
| 5G-ACIA | The 5G Alliance for Connected Industries and Automation |
| 5G-SMART | 5G for Smart Manufacturing |
| AGV | Automated Guided Vehicle |
| AP | Access Point |
| BS | Base Station |
| CDF | Cumulative Distribution Function |
| CG | Configured Grant |
| CHO | Conditional handover |
| CPU | Central Processing Unit |
| CSA | Communication Service Availability |
| CSR | Communication Service Reliability |
| CSI | Channel State Information |
| DAPS | Dual Active Protocol Stack |
| DAS | Distributed Antenna System |
| DL | Downlink |
| DTX | discontinuous transmission |
| E2E | End to end |
| eMBB | Enhanced Mobile Broadband |
| EMC | electromagnetic compatibility |
| FDD | Frequency Division Duplex |
| gNB | gNodeB (5G NR base station) |
| GPU | Graphic Process Unit |
| HARQ | hybrid-automated repeat request |
| IIoT | Industrial Internet of Things |
| IM | Interference Measurement |
| InF-DH | Indoor Factory with Dense clutter and High base station height |
| InF-SH | Indoor Factory with Sparse clutter and High base station height |
| InH | Indoor Hotspot |
| KPI | Key Performance Indicator |
| LDCP | low-density parity-check |
| LTE | 3GPP Long Term Evolution |
| MCS | Modulation and Coding Scheme |
| MIMO | Multiple-Input and Multiple-Output |
| M-MTC | Massive Machine-Type Communication |
| mmWave | Millimiter wave |

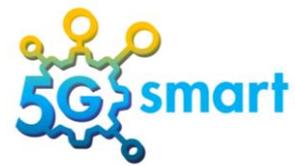

| MNO | Mobile Network Operator |
| MNO | mobile network operator |
| MOCN | Multi-Operator Core Network |
| MTC | Machine-Type Communication |
| NACK | negative acknowledgement |
| NLoS | non line of sight |
| NPN | Non-Public Network |
| NPNs | Non-public networks |
| NR | 3GPP New Radio |
| NR-U | 3GPP New Radio for Unlicensed spectrum |
| OFDM | Orthogonal Frequency Division Multiplexing |
| PDSCH | Physical Downlink Shared Channel |
| PUCCH | Physical Uplink Control Channel |
| QoS | Quality of Service |
| RA | Random Access |
| RACH | Random Access CHannel |
| RAN | Radio Access Network |
| RF | Radio Frequency |
| RRC | Radio Resource Control |
| RS | Reference Signal |
| RSRP | Reference Signal Received Power |
| RTT | Round-Trip Time |
| SA | Service area |
| SCS | Sub-Carrier Spacing |
| SINR | Signal-to-Interference-plus-Noise Ratio |
| SIR | Signal-to-Interference Ratio |
| SLS | System Level Simulation |
| SNR | Signal-to-Noise Ratio |
| SPS | semi-persistent scheduling |
| SR | scheduling request |
| SSB | Synchronization Signal Block |
| TDD | Time Division Duplex |
| TRP | transmission/reception point |
| TTI | Transmission Time Interval |
| UE | User Equipment |
| UL | Uplink |